\documentclass{article}

\usepackage{arxiv}

\usepackage[utf8]{inputenc} 
\usepackage[T1]{fontenc}    
\usepackage{hyperref}       
\hypersetup{
  colorlinks   = true, 
  urlcolor     = ForestGreen, 
  linkcolor    = BrickRed, 
  citecolor   = blue 
}
\usepackage{url}            
\usepackage{booktabs}       
\usepackage{amsfonts}       
\usepackage{nicefrac}       
\usepackage{microtype}      
\usepackage{comment}

\usepackage{subfigure} 

\usepackage{framed,multirow}

\usepackage{amsmath}
\usepackage{algorithm}
\usepackage{algpseudocode}
\usepackage{graphicx}
\usepackage{mathrsfs}

\usepackage{amssymb}
\usepackage{latexsym}
\usepackage{bm}
\usepackage{amssymb}
\usepackage{array}
\usepackage{multirow}
\usepackage[english]{babel}
\usepackage[numbers]{natbib}
\usepackage{footnote}
\usepackage{tabularx}
\makesavenoteenv{tabular}
\makesavenoteenv{table}
\usepackage{relsize}

\usepackage{cleveref}
\Crefname{equation}{Eq.}{Eqs.}
\Crefname{table}{Table}{Tables}
\Crefname{figure}{Fig.}{Figs.}

\Crefname{section}{Section}{Sections}
\Crefname{subsection}{Section}{Secs.}
\Crefname{Algorithm}{Algorihtm}{Algorihtm}
\usepackage{mathtools}
\usepackage{xfrac}
\usepackage{amsthm}

\usepackage[dvipsnames]{xcolor}
\definecolor{newcolor}{rgb}{.8,.349,.1}

\title{Gradient-free Importance Sampling Scheme
for Efficient Reliability Estimation}

\author{
    Elsayed~Eshra\\
The Pennsylvania State University\\
University Park, PA 16802 \\
\texttt{eme5375@psu.edu} \\
\And
 Konstantinos G.~Papakonstantinou \\
The Pennsylvania State University\\
University Park, PA 16802 \\
\texttt{kpapakon@psu.edu} \\
}
\begin{document}
\maketitle
\begin{abstract}
This work presents a novel gradient-free importance sampling-based framework for precisely and efficiently estimating rare event probabilities, often encountered in reliability analyses of engineering systems. The approach is formulated around our foundational Approximate Sampling Target with Post-processing Adjustment (ASTPA) methodology. ASTPA uniquely constructs and directly samples an unnormalized target distribution, relaxing the optimal importance sampling distribution (ISD). The target’s normalizing constant is then estimated using our inverse importance sampling (IIS) scheme, employing an ISD fitted based on the obtained samples. In this work, a gradient-free sampling method within ASTPA is developed through a guided dimension-robust preconditioned Crank-Nicolson (pCN) algorithm, particularly suitable for black-box computational models where analytical gradient information is not available. To boost the sampling efficiency of pCN in our context, a computationally effective, general discovery stage for the rare event domain is devised, providing (multi-modal) rare event samples used in initializing the pCN chains. A series of diverse test functions and engineering problems involving high dimensionality and strong nonlinearity is presented, demonstrating the advantages of the proposed framework compared to several state-of-the-art sampling methods.
\end{abstract}


\section{Introduction}\label{sec: intro}
This work introduces a novel \textit{gradient-free} framework for reliability analysis and rare event probabilities estimation in high-dimensional random variable spaces. Evaluation of system reliability and rare event uncertainty quantification (UQ) play a vital role in modern decision-making across diverse fields. Yet, as our systems and computational models grow increasingly complex, the computational cost of repeatedly evaluating them becomes a major challenge. Additionally, analytically deriving gradients for such models can sometimes be cumbersome or even infeasible, and using numerical gradient estimation methods in high-dimensional settings is impractical due to the high computational cost involved. These challenges are further compounded by the exceptionally low probabilities associated with such rare events, often ranging from $10^{-4}$ to $10^{-9}$ or lower. This necessitates the development of efficient gradient-free approaches capable of accurately estimating rare event probabilities while minimizing the number of model evaluations, a challenge that this work aims to address.

In the context of reliability estimation, failures are often considered rare events; therefore, rare event probability is also commonly referred to as failure probability. Several techniques have been developed over the years to estimate rare event (failure) probabilities, with the sampling-based approaches being able to address this problem in its utmost generality. To overcome the limitations of the crude Monte Carlo approach \cite{bai2023curse}, various advanced sampling techniques have been proposed. Among others, Importance sampling (IS) \citep{rubinstein2016simulation}, a well-known variance-reduction approach, is capable of efficiently estimating rare event probabilities. IS employs an appropriate importance sampling density (ISD), placing greater importance on rare event domains in the random variable space. This results in an increased number of rare event samples, thereby reducing the number of required model calls. Although the theoretically optimal ISD is known \citep{ang1992optimal}, it is challenging to sample from and impractical to implement. Notably, finding an effective near-optimal ISD is an ongoing, challenging research pursuit. The existing approaches can be largely classified into two categories \citep{eshra2024direct}. The first seeks to approximate the optimal ISD by a parametric or non-parametric normalized probability density function (PDF). An early approach in this category utilizes densities centered around one or more design points \citep{schueller1987critical}, a powerful, practical process, yet with certain limitations for high-dimensional, complex spaces, as also discussed in \citep{au2001estimation}. A more recent, popular approach is the cross entropy-based importance sampling (CE-IS). It alternatively constructs a parametric ISD $h$ by minimizing the Kullback-Leibler (KL) divergence
between the optimal ISD and a chosen parametric family of probability distributions \citep{kurtz2013cross, wang2016cross, papaioannou2019improved, uribe2021cross, demange2023variational}. Recent CE-IS variants incorporate information from influential rare event points, e.g., design points, to effectively initialize the cross entropy process \citep{tong2023large,chiron2023failure}. 
Within this first category of methods, several other different noteworthy approaches have also been proposed to approximate the optimal ISD, e.g., \citep{au1999new,au2003important,tabandeh2022review,ehre2023stein, cui2024deep}. The second category of methods employs sequential variants of importance sampling, mainly utilizing a number of intermediate unnormalized distributions \citep{del2006sequential,papaioannou2016sequential,sinha2020neural, xian2024relaxation}. These unnormalized distributions are generally more flexible than the PDFs usually utilized within the first category, thus being capable of providing a better approximation of the optimal ISD. Another sequential sampling method that shares similarities \citep{eshra2024direct} with this second methodological category is the Subset Simulation (SuS) approach, originally presented in \citep{au2001estimation}, and continuously refined \citep{zuev2012bayesian,wang2019hamiltonian,shields2021subset,chen2022riemannian,thaler2024reliability},  due to its potential to handle high-dimensional problems, despite some limitations \citep{breitung2019geometry}. A key to the success of all sequential methods is to design nearby intermediate distributions and to sufficiently sample them to accurately compute the sought rare event probability, a non-trivial and computationally demanding task.

Recently, we developed a novel foundational importance sampling-based approach in a non-sequential manner, termed Approximate Sampling Target with Post-processing Adjustment (ASTPA), for reliability and rare event probabilities estimation \citep{Papakon2023HMCMC,eshra2024direct}. The ASTPA approach innovatively decomposes the challenging rare event probability estimation process into two less demanding estimation problems. First, ASTPA uniquely constructs and directly samples an \textit{unnormalized} target distribution, relaxing the optimal ISD, thereby making it comparatively much easier to sample. The obtained samples are subsequently utilized not only to compute a shifted estimate of the sought probability but also to guide the second ASTPA stage. Post-sampling, the normalizing constant of the approximate sampling target is computed through the inverse importance sampling (IIS) technique \citep{Papakon2023HMCMC}, which utilizes an importance sampling density (ISD) fitted based on the samples drawn in the first stage. The rare event probability is eventually computed by utilizing this computed normalizing constant to correct the shifted estimate of the first stage. Consequently, ASTPA significantly reduces the computational cost typically associated with using multiple intermediate distributions. Sampling the approximate target distribution in ASTPA has been so far performed using gradient-based Hamiltonian Markov Chain Monte Carlo (HMCMC) methods, demonstrating outstanding performance in complex, high-dimensional static and first-passage dynamic problems, both in Gaussian spaces \citep{Papakon2023HMCMC} and directly in non-Gaussian spaces \citep{PapakonICASP2023, eshra2024direct}. However, the competitiveness of this gradient-based approach is hindered when analytical gradients cannot be obtained. 

Gradient-free sampling has been extensively employed in various reliability estimation approaches, such as importance sampling and Subset Simulation, as discussed earlier. One common method employed is the Component-wise Metropolis-Hastings (CWMH) algorithm, often applied within Subset Simulation \citep{au2001estimation} to effectively sample high-dimensional conditional distributions by sequentially updating individual dimensions. To further enhance efficiency, delayed rejection variants of CWMH have been introduced, allowing for multiple proposal attempts within a single iteration when a proposal is rejected, thereby increasing acceptance rates and reducing sample autocorrelation \citep{zuev2011modified, papaioannou2015mcmc, mira2001metropolis, miao2011modified, santoso2011modified}. Another significant development is the preconditioned Crank-Nicolson (pCN) algorithm, a MCMC method designed for high-dimensional probability distributions and infinite-dimensional settings, often encountered in Bayesian inverse problems \citep{Neal1998,cotter2013mcmc}. Inspired by the Crank-Nicolson discretization for solving partial differential equations, the pCN sampler proposes updates by combining the current state with a perturbation drawn from the prior distribution, scaled by a tunable parameter to control the step size. This design maintains stable acceptance probabilities even as dimensionality increases, achieving dimension-robust performance \citep{hairer2014spectral}. To further improve efficiency, some extensions of the standard pCN algorithm incorporate information from the posterior distribution by adapting the proposal covariance using sampling history \citep{hu2017adaptive, carrera2024covariance} or by estimating the posterior covariance through gradient information \citep{rudolf2018generalization}. The pCN algorithm has been applied in various reliability estimation frameworks, including Subset Simulation \citep{au2016rare, papaioannou2015mcmc} and importance sampling \citep{papaioannou2016sequential, xiao2024failure}.\par

In this paper, we introduce a novel gradient-free framework for efficiently and accurately estimating rare event probabilities, building on our foundational ASTPA approach, particularly suitable for black-box computational models where analytical gradient information is not available, for any reason possible. Central to the framework is a guided variant of the dimension-robust pCN algorithm, tailored to sample the approximate target distribution within ASTPA. The standard pCN algorithm encounters challenges with multimodal distributions or when the target is distant from prior information, as often occurs with rare event domains located in the tails of original probability distributions. To overcome these limitations, we propose a computationally efficient discovery stage in this work, properly designed to enhance sampling efficiency by generating (multimodal) rare event samples to serve as initial seeds for the pCN chains. This process begins by sampling from the original distribution or an adjusted version, such as one with a modified covariance matrix, followed by conditional sampling based on the reciprocal of the original distribution, facilitating rapid diffusion toward the rare event domain. The procedure is stopped after generating a specified number of rare event samples, from which a defined number of seeds is randomly selected to initialize the pCN chains. The generated pCN sample set is then employed within the ASTPA approach to eventually estimate the sought failure probability. Consistent with our previous findings \citep{Papakon2023HMCMC, eshra2024direct}, the ASTPA estimator remains unbiased, and the derived analytical coefficient of variation aligns very well with the sampling one, as demonstrated by the numerical experiments presented in this work.\par

The structure of this paper is as follows: \Cref{sec: rare_event_est} defines the rare event probability estimation problem, highlighting its complexities and the considered challenging yet realistic and practical scenarios. The ASTPA framework is presented in \Cref{sec: ASTPA_sec}, followed by a detailed description of the developed gradient-free sampling scheme in \Cref{sec: Sampling}. A summary of the pCN-based ASTPA framework is presented in \Cref{sec: pCN_ASTPA_summary}. To fully evaluate the capabilities of the proposed methodology, \Cref{sec: Numerical_results} demonstrates its performance, successfully comparing it with state-of-the-art Subset Simulation and advanced importance sampling methods across a series of challenging low- and high-dimensional problems. The paper then concludes in \Cref{sec: Conclusion}.

\section{Rare Event Probability Estimation}\label{sec: rare_event_est}

Let $\bm{X}$ be a continuous random vector taking values in $\mathcal{X} \subseteq \mathbb{R}^d$ and having a joint probability density function (PDF) $\pi_{\bm{X}}$. We are interested in rare events $\mathcal{F} \coloneqq \{ \bm{x} \,:\, g(\bm{x}) \leq 0\}$, where $g: \mathcal{X} \rightarrow \mathbb{R}$ is a performance function, also known as the limit-state function, defining the occurrence of a rare event. In this work, we aim to estimate the rare event probability $p_\mathcal{F}$:
\begin{equation}  \label{eq: p_f}
p_\mathcal{F}=  \int_{\mathcal{F}} \pi_{\bm{X}}(\bm{x}) d\bm{x} = 
  \int_{\mathcal{X}} I_{\mathcal{F}} (\bm{x})\,\pi_{\bm{X}}(\bm{x}) d\bm{x} = {\mathop{\mathbb{E}}}_{\pi_{\bm{X}}} [ I_{\mathcal{F}} (\bm{X})] 
\end{equation} 
where $I_{\mathcal{F}}: \mathcal{X} \rightarrow \{0, 1\}$ is the indicator function, i.e., $I_{\mathcal{F}} (\bm{x})=1$ if $\bm{x} \in \mathcal{F}$, and $I_{\mathcal{F}} (\bm{x})=0$ otherwise, and $\mathop{\mathbb{E}}$ is the expectation operator.\par
Our objective in this work is to estimate the described integration in \Cref{eq: p_f} under these challenging yet realistic settings: \textbf{\textit{(i)}} the analytical calculation of \Cref{eq: p_f} is generally intractable; \textbf{\textit{(ii)}} the computational effort for evaluating $I_{\mathcal{F}}$ for each realization $\bm{x}$ is assumed to be quite significant, often relying on computationally intensive models, necessitating the minimization of such function evaluations (model calls); \textbf{\textit{(iii)}} the rare event probability $p_\mathcal{F}$ is extremely low, typically in the range of $10^{-4} - 10^{-9}$ or lower; \textbf{\textit{(iv)}} the random variable space, $\mathcal{X} \subseteq \mathbb{R}^d$, is high-dimensional, with $d$ often as large as $10^2$ or more; \textbf{\textit{(v)}} the analytical gradient of the limit state function is not available, motivating the use of gradient-free approaches;  and \textbf{\textit{(vi)}} the joint probability density, $\pi_{\bm{X}}(\bm{x})$, is assumed to follow an independent standard Gaussian distribution, i.e., $\pi_{\bm{X}}(\bm{x}) = \mathcal{N}(\bm{X};\,\mathbf{0},\, \mathbf{I})$, with $\mathbf{I}$ being an identity matrix, a common assumption in rare event estimation. In most cases, this assumption is not restrictive, as transforming the original random variables to the independent standard Gaussian space can be uncomplicated \citep{hohenbichler1981non, der1986structural, lebrun2009innovating, papaioannou2017efficient}. However, when such Gaussian transformations are not feasible, directly estimating rare event probabilities in complex non-Gaussian spaces may require sophisticated gradient-based samplers, a challenge addressed in our Quasi-Newton mass preconditioned HMCMC (QNp-HMCMC)-based ASTPA framework, as demonstrated in \citep{eshra2024direct, PapakonICOSSAR2022, nikbakht2019HMCMC}. The gradient-free sampling approach developed in this work is specifically tailored for Gaussian spaces, with ongoing efforts to extend these methods in the future to direct estimations in non-Gaussian spaces.\par

Under these settings, several sampling-based methods for estimating rare event probabilities become highly inefficient. For instance, the standard Monte Carlo estimator of \Cref{eq: p_f} using $\{\bm{x}_i\}_1^N$ draws from $\pi_{\bm{X}}$, $\hat{p}_\mathcal{F} = \frac{1}{N} \sum_{i=1}^N  I_{\mathcal{F}} (\bm{x}_i)$, has a coefficient of variation $\sqrt{(1-p_\mathcal{F})/(N p_\mathcal{F})}$, rendering this estimate inefficient by requiring a prohibitively large number of samples to accurately quantify small probabilities; see setting \textbf{\textit{(iii)}} above. As discussed in \Cref{sec: intro}, importance sampling (IS) can efficiently compute the integral in \Cref{eq: p_f}, through sampling from an appropriate importance sampling density (ISD), $h(\boldsymbol{\bm{X}})$. The ISD places greater importance on rare event domain, $\mathcal{F}$, in the random variable space compared to $\pi_{\bm{X}}(\bm{x})$, satisfying the sufficient condition $\text{supp}(I_{\mathcal{F}}\,\pi_{\bm{X}}) \subseteq \text{supp}(h)$, with $\text{supp}(\cdot)$ denoting the support of the function. Eq.~\eqref{eq: p_f} can then be written as:
\begin{equation} \label{eq: p_f_I_S}
p_\mathcal{F}= \int_{\mathcal{X}} I_{\mathcal{F}} (\bm{x})\, \dfrac{\pi_{\bm{X}}(\bm{x})}{h(\bm{x})} h(\bm{x}) d\bm{x} = {\mathop{\mathbb{E}}}_{h} \big[ I_{\mathcal{F}} (\bm{X}) \dfrac{\pi_{\bm{X}}(\bm{X})}{h(\bm{X})}\big] 
\end{equation} 
leading to the unbiased IS estimator: 
\begin{equation} \label{eq: p_f_I_S_estimate}
\hat{p}_\mathcal{F}= \frac{1}{N} \sum_{i=1}^N  I_{\mathcal{F}} (\bm{x}_i)\dfrac{\pi_{\bm{X}}(\bm{x}_i)}{h(\bm{x}_i)}
\end{equation} 
where $\{\bm{x}_i\}_1^N \sim h$. The efficiency of IS relies heavily on the careful selection of both the ISD and the sampling algorithm. A theoretically optimal ISD $h^*$, providing zero-variance IS estimator,  is given by \citep{ang1992optimal}:
\begin{equation}\label{eq: opt_I_S_density}
\begin{aligned}
&h^*(\bm{X}) = \dfrac{1}{p_\mathcal{F}} I_{\mathcal{F}}(\bm{X}) \pi_{\bm{X}}(\bm{X})
\end{aligned}
\end{equation}
However, this optimal ISD cannot be utilized because the normalizing constant in \Cref{eq: opt_I_S_density} is the sought probability $p_\mathcal{F}$. Additionally, $h^*$ can be highly concentrated within the rare event domain, making it challenging to sample, particularly in high dimensions \citep{eshra2024direct}. To address these challenges, we employ our developed importance sampling-based framework, ASTPA \citep{Papakon2023HMCMC,eshra2024direct}, which constructs a relaxed ISD that approximates the optimal one while being much easier to sample. This relaxation allows for more efficient sampling, leading to a more accurate estimation of the sought probability. The primary focus of this work is to introduce an efficient, gradient-free sampling approach, which is especially beneficial when analytical gradients are not available, as detailed in \Cref{sec: Sampling}.

\section{Approximate Sampling Target with Post-processing Adjustment (ASTPA) in Gaussian Spaces}\label{sec: ASTPA_sec}

ASTPA estimates the rare event probability in \Cref{eq: p_f} through an innovative non-sequential, importance sampling variant, as introduced in \citep{Papakon2023HMCMC,eshra2024direct}. In this approach, ASTPA constructs a single unnormalized approximate sampling target $\tilde{h}$, relaxing the optimal ISD in \Cref{eq: opt_I_S_density}. After sampling $\tilde{h}$, its normalizing constant $C_h$ is computed utilizing inverse importance sampling. By employing the unnormalized importance sampling distribution $\tilde{h}$, \Cref{eq: p_f_I_S} can be written as:
\begin{equation} \label{eq: p_f_I_S_ASTPA}
p_\mathcal{F}=  {\mathop{\mathbb{E}}}_{h} \big[ I_{\mathcal{F}} (\bm{X}) \dfrac{\pi_{\bm{X}}(\bm{X})}{h(\bm{X})}\big] = C_{h} \,{\mathop{\mathbb{E}}}_{h} \big[ I_{\mathcal{F}} (\bm{X}) \dfrac{\pi_{\bm{X}}(\bm{X})}{\tilde{h}(\bm{X})}\big] = C_h \, \tilde{p}_\mathcal{F}
\end{equation} 
Thus, computing $p_\mathcal{F}$ is decomposed into two generally simpler tasks. The first involves constructing $\tilde{h}$ and sampling $\{\bm{x}_i\}_{i=1}^N \sim h$\footnote{$\{\bm{x}_i\}_{i=1}^N \sim h$ is equivalent to $\{\bm{x}_i\}_{i=1}^N \sim \tilde{h}$.},  to compute the unbiased expectation of the weighted indicator function (shifted probability estimate) as:
\begin{equation} \label{eq: p_f_tilde_ASTPA}
\tilde{p}_\mathcal{F} = \,{\mathop{\mathbb{E}}}_{h} \big[ I_{\mathcal{F}} (\bm{X}) \dfrac{\pi_{\bm{X}}(\bm{X})}{\tilde{h}(\bm{X})}\big] \approx \hat{\tilde{p}}_\mathcal{F} = \dfrac{1}{N} \sum_{i=1}^{N} I_{\mathcal{F}} (\bm{x}_i) \dfrac{\pi_{\bm{X}}(\bm{x}_i)}{\tilde{h}(\bm{x}_i)} 
\end{equation} 
The normalizing constant, $\hat{C}_h$, is then estimated using inverse importance sampling, as discussed in \Cref{sec: IIS}. The ASTPA estimator for the sought rare event probability can then be computed as:  
\begin{equation} \label{eq: p_f_ASTPA}
\hat{p}_\mathcal{F} = \hat{\tilde{p}}_\mathcal{F} \,\, \hat{C}_h 
\end{equation} 

\begin{figure}[t!]
 \centering
 \vspace*{-0.7in}
  \begin{tabular}{ccccc}
   \vspace*{-0.2in}
    \hspace*{-0.1in}
   &\includegraphics[width=.245\textwidth,keepaspectratio]{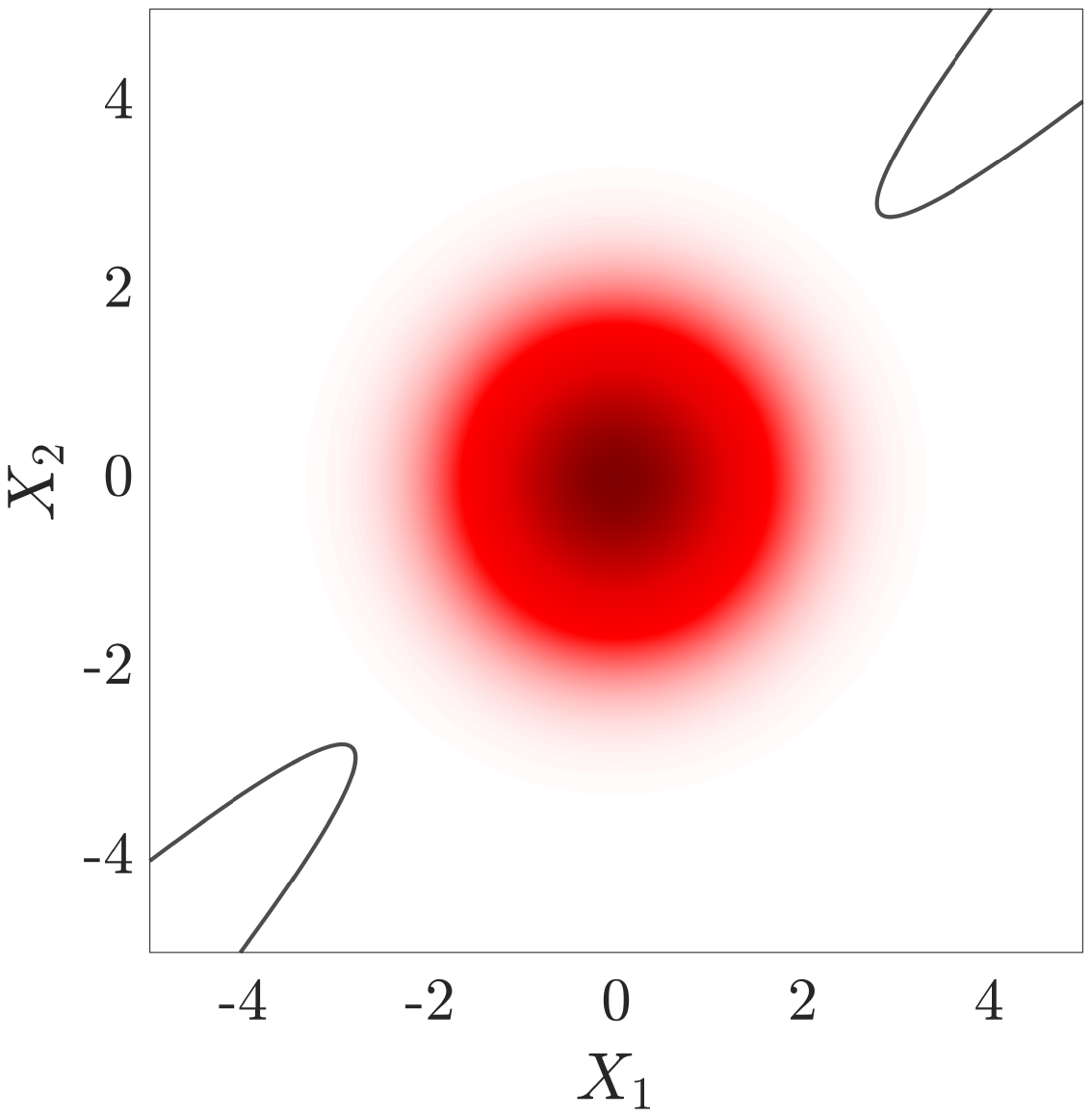}&    \hspace*{-0.2in}
   \includegraphics[width=.245\textwidth,keepaspectratio]{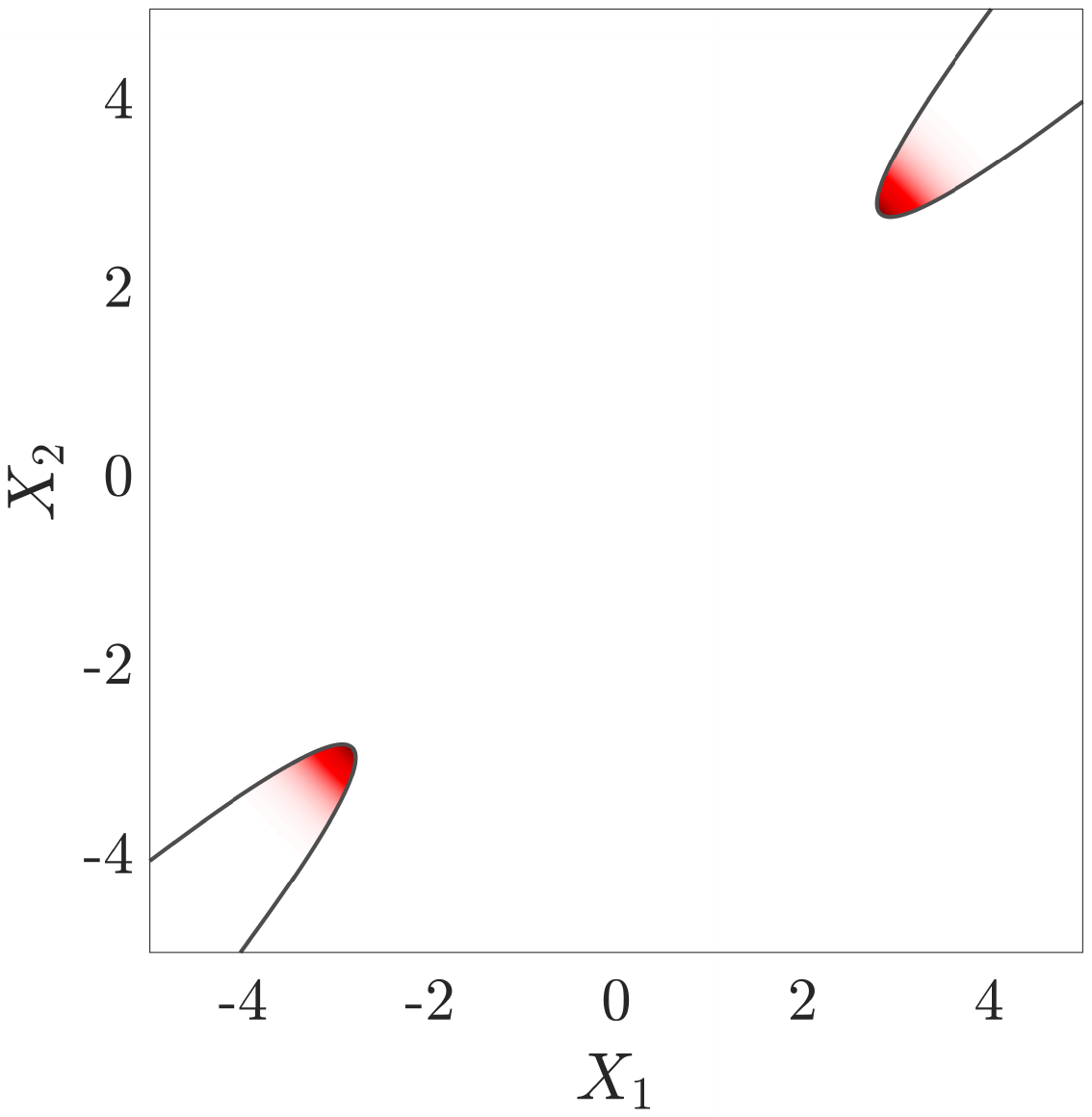}&  &\\&
   \hspace*{-0.2in}{\fontsize{8.5pt}{7.2} (a) $\pi_{\bm{X}}$}&\hspace*{-0.05in}{\fontsize{8.5pt}{7.2} (b) $h^*(\bm{X})$}&\vspace*{0.1in}\\
    \cline{1-4}\vspace*{-0.35in}\\
   \hspace*{-0.2in}
   \includegraphics[width=.245\textwidth,keepaspectratio]{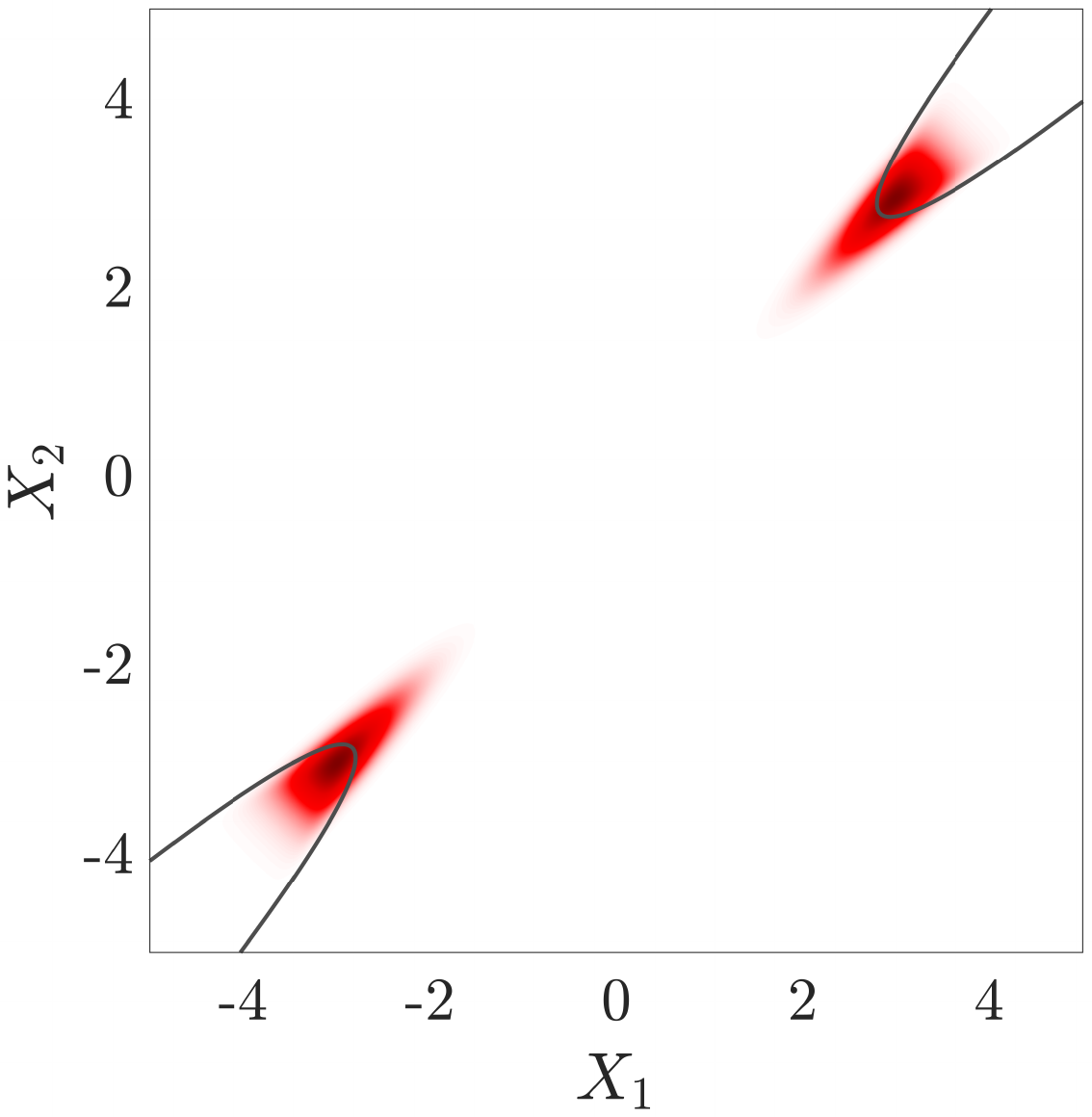}&
   \hspace*{-0.2in}
      \includegraphics[width=.245\textwidth,keepaspectratio]{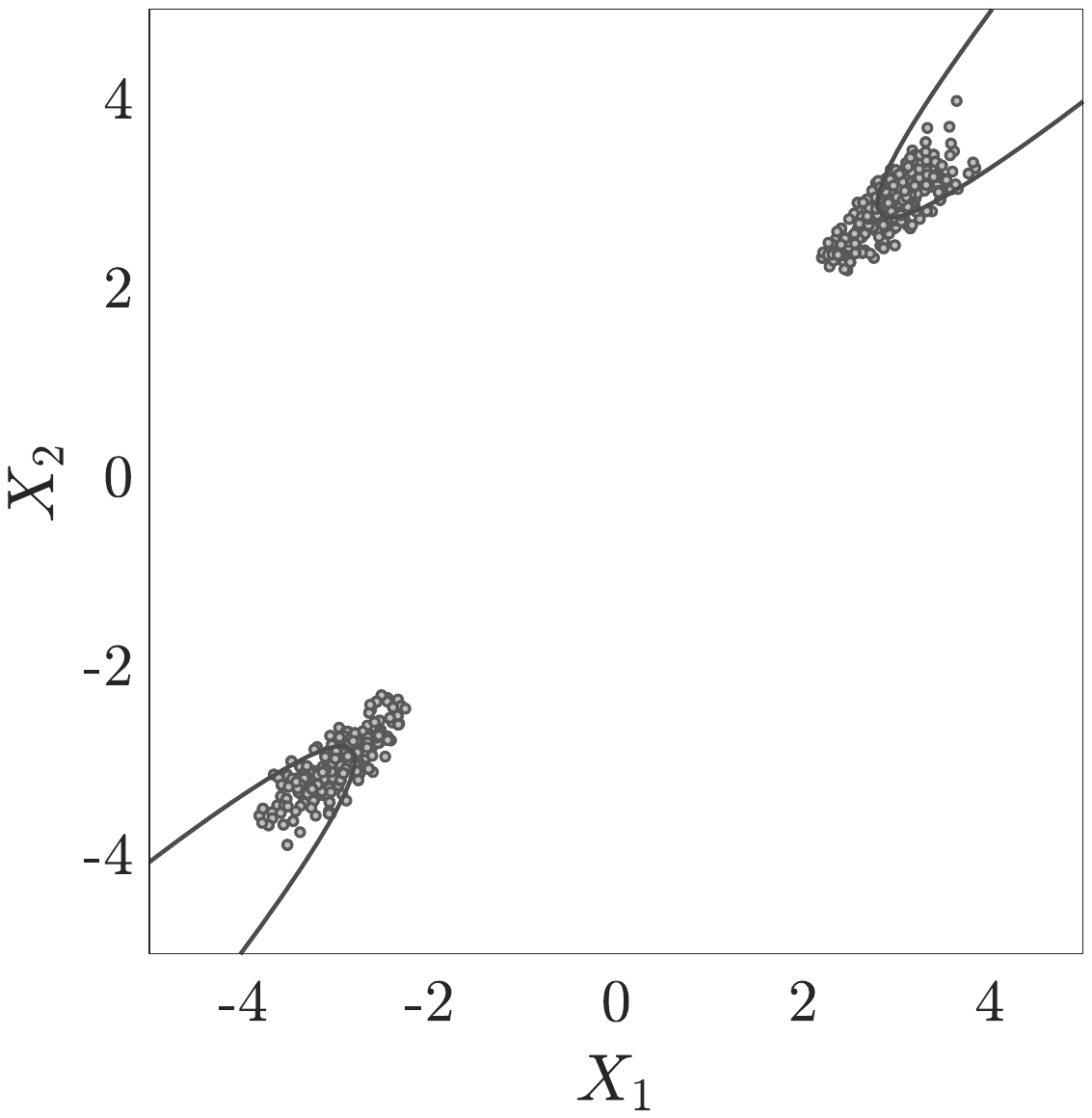}& \hspace*{-0.2in} 
         \includegraphics[width=.245\textwidth,keepaspectratio]{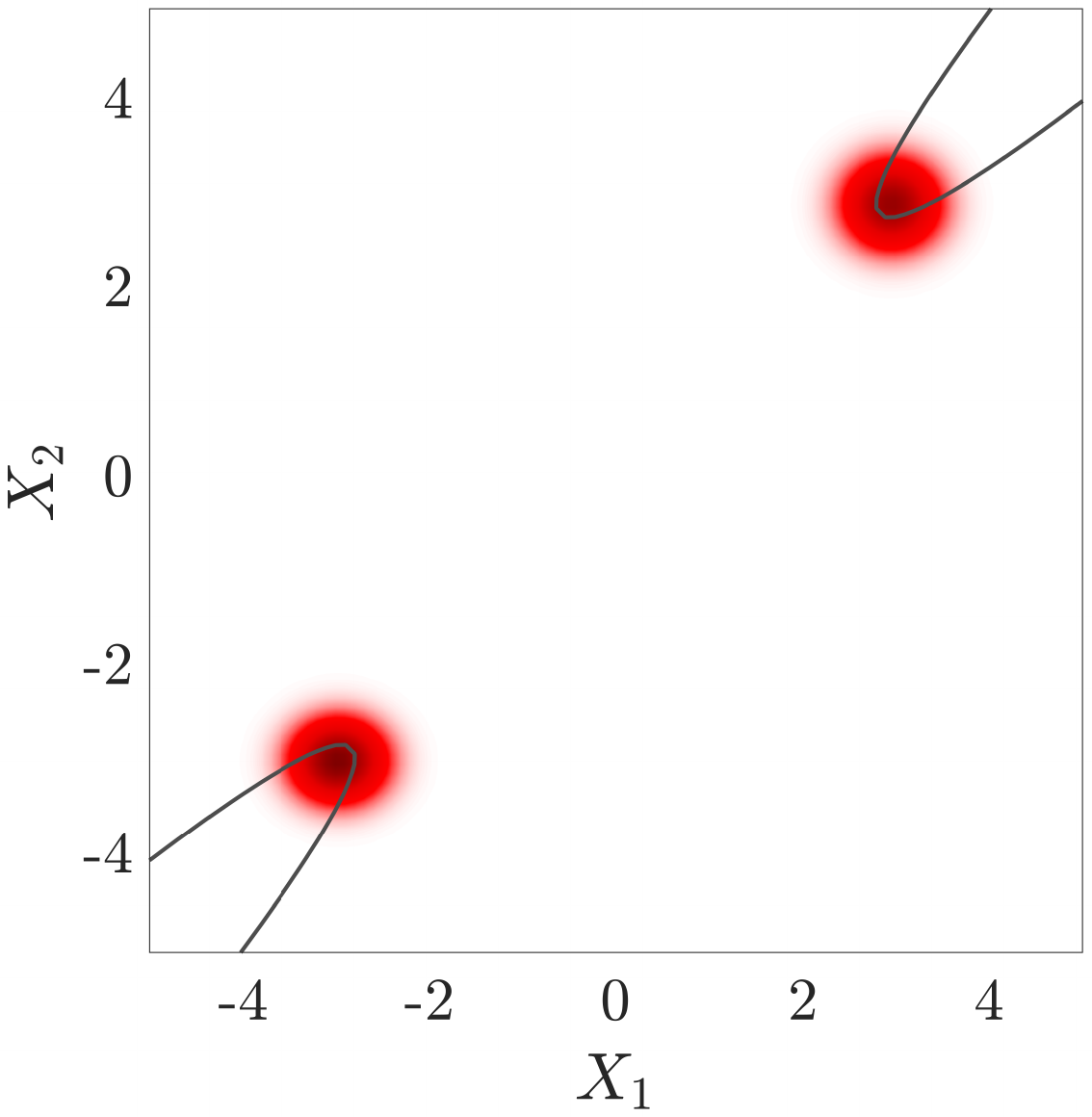}& \hspace*{-0.2in}
   \includegraphics[width=.245\textwidth,keepaspectratio]{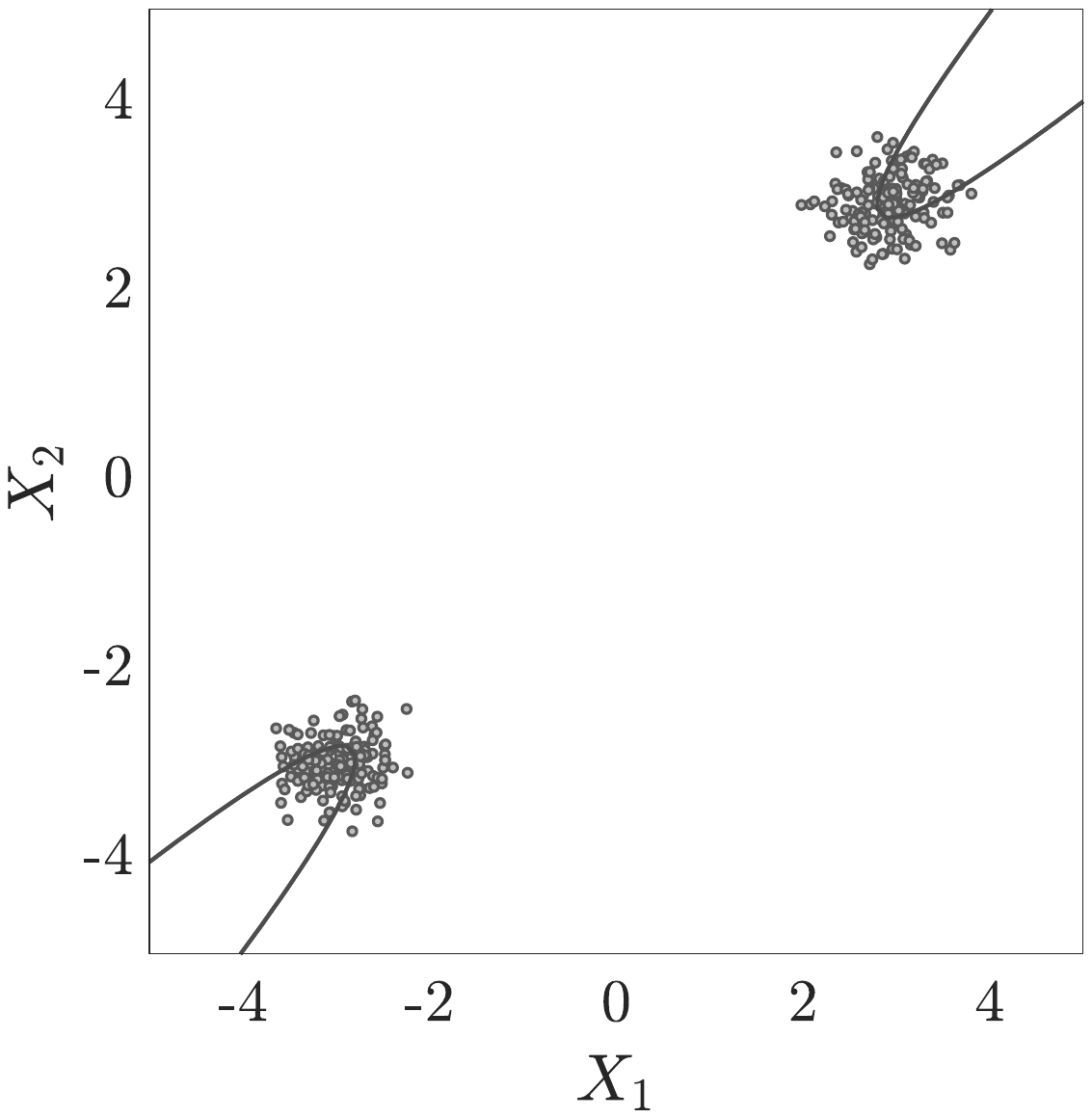}
\vspace*{-0.25in}\\
    \hspace*{-0.25in}
 \vspace*{+0.05in}
    {\fontsize{8.5pt}{7.2} (c) $\tilde{h}(\bm{X})$}&
    \hspace*{-0.25in}
    {\fontsize{8.5pt}{7.2}  (d) Samples $\sim \tilde{h}(\bm{X})$}&
    \hspace*{-0.25in}
   {\fontsize{8.5pt}{7.2}  (e) $Q(\bm{X})$}& \hspace*{-0.25in}
   {\fontsize{8.5pt}{7.2}  (f) Samples $\sim Q(\bm{X})$}\\
    \vspace*{-0.1in}\\
   \vspace*{-0.2in}
 \end{tabular}
 \caption{Outlining the ASTPA framework; (a) shows an original bivariate independent standard Gaussian distribution $\pi_{\bm{X}}$ in red, and a limit state function at $g(\bm{X})=0$ shown using a grey line, with a bimodal rare event domain being inside $g(\bm{X})=0$, as detailed in \Cref{sec: Bimodal_nonlinear}, 
 (b) visualizes the optimal importance sampling density, $h^*$, in \Cref{eq: opt_I_S_density}, (c) depicts the constructed approximate sampling target, $\tilde{h}$, in \Cref{eq: Tar_ASTPA}, (d) showcases samples from this target, subsequently used to compute $\hat{\tilde{p}}_\mathcal{F}$ in \Cref{eq: p_f_tilde_ASTPA}, and (e) and (f) demonstrate the inverse importance sampling procedure to compute the normalizing constant $\hat{C}_h$ in \Cref{eq: IIS_est_hat}. Specifically, (e) and (f) present a crudely fitted $Q(\bm{X})$ and its samples, respectively. The sought probability is eventually computed using \Cref{eq: p_f_ASTPA}.}
 \label{fig: ASTPA_framework}
\end{figure}

\Cref{fig: ASTPA_framework} provides a concise illustration of the ASTPA framework using a nonlinear bimodal  limit state function, as detailed in \Cref{sec: Bimodal_nonlinear}, with $p_{\mathcal{F}}\sim 9.47 \times10^{-6}$. The construction of the approximate sampling target in ASTPA, along with recommended parameters, is discussed in \Cref{sec: Targ_form}. We then present the inverse importance sampling technique in \Cref{sec: IIS}. Subsequently, \Cref{sec: AnalCOV} shows the statistical properties of the ASTPA estimator presented in \Cref{eq: p_f_ASTPA}.

\subsection{Target distribution formulation}\label{sec: Targ_form}

\begin{figure}[t!]
 \vspace*{-0.7in}
 \centering
  \begin{tabular}{ccccc}
   \vspace*{-0.7in}
    \hspace*{-0.12in}\includegraphics[width=.205\textwidth,keepaspectratio]{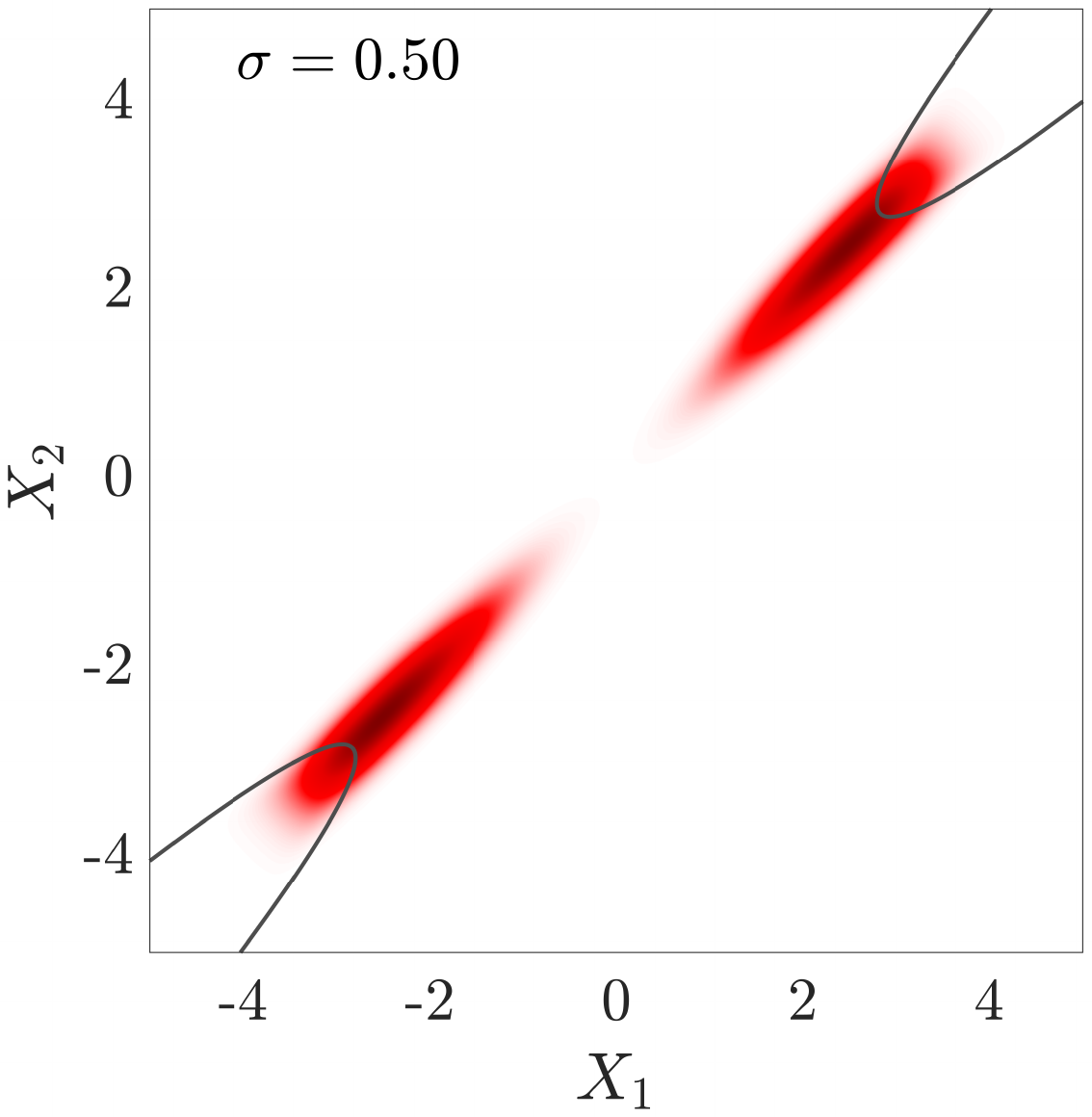}& \hspace*{-0.2in}
   \includegraphics[width=.205\textwidth,keepaspectratio]{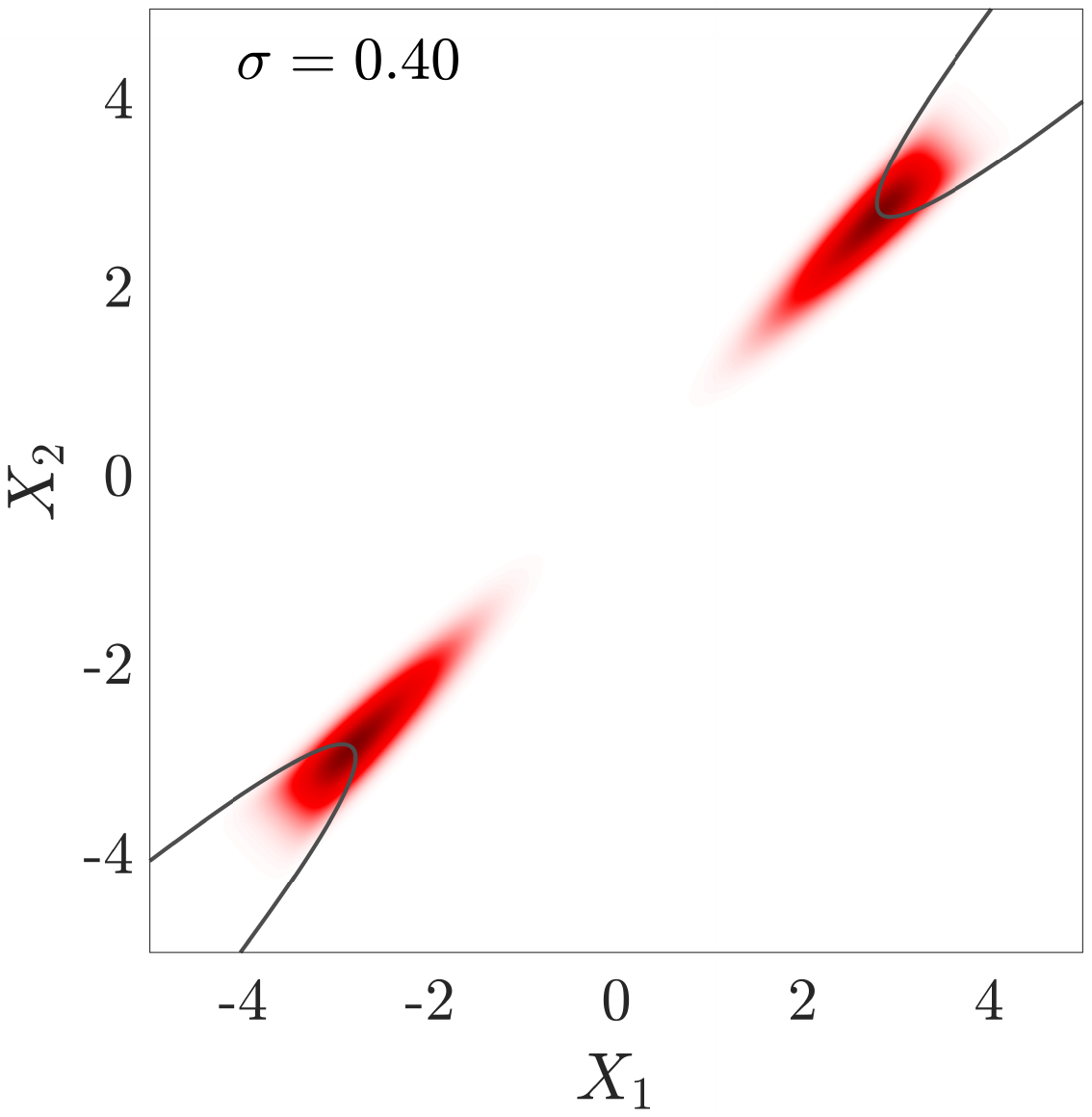}& \hspace*{-0.2in}
      \includegraphics[width=.205\textwidth,keepaspectratio]{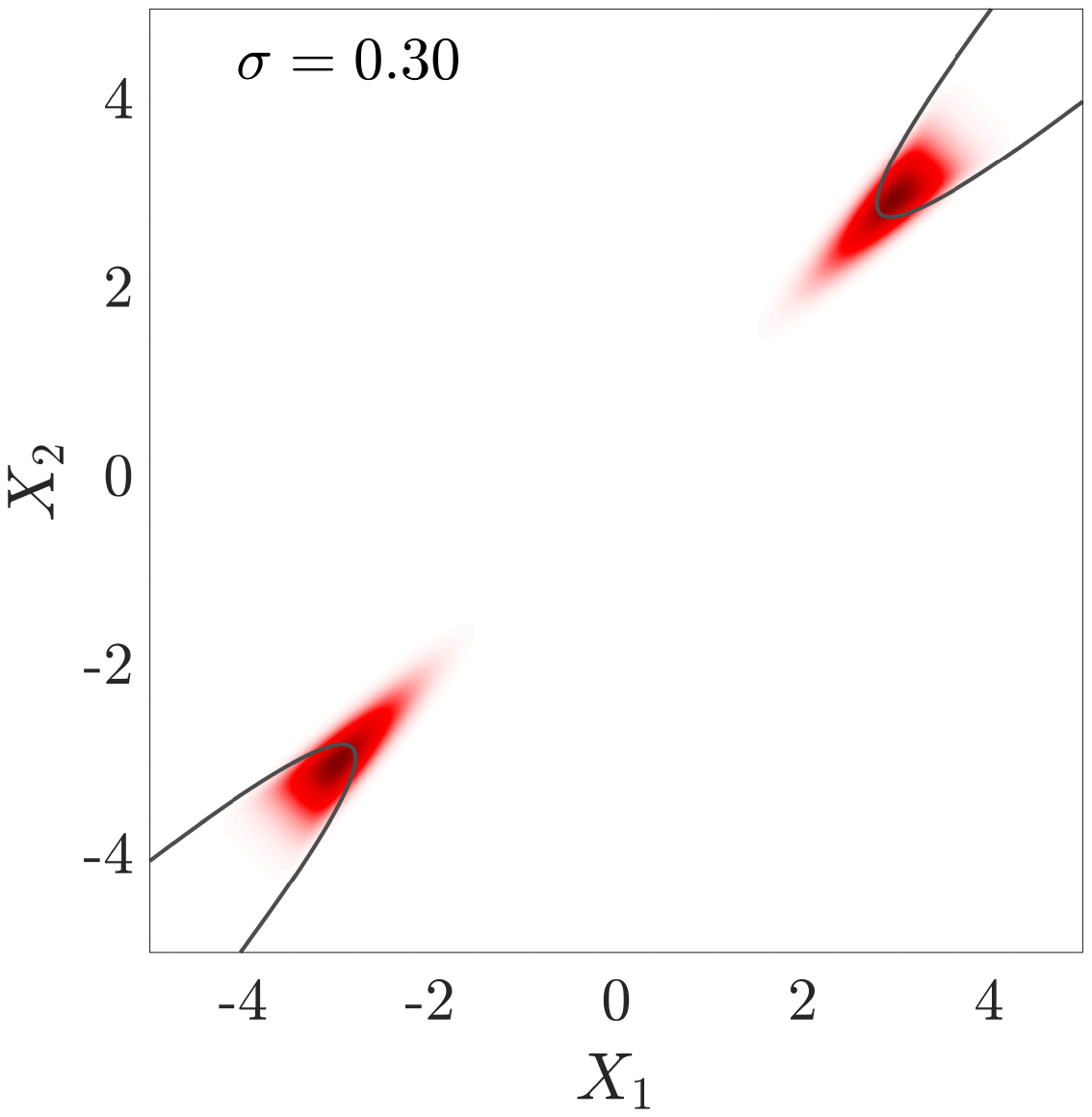}& \hspace*{-0.2in} \includegraphics[width=.205\textwidth,keepaspectratio]{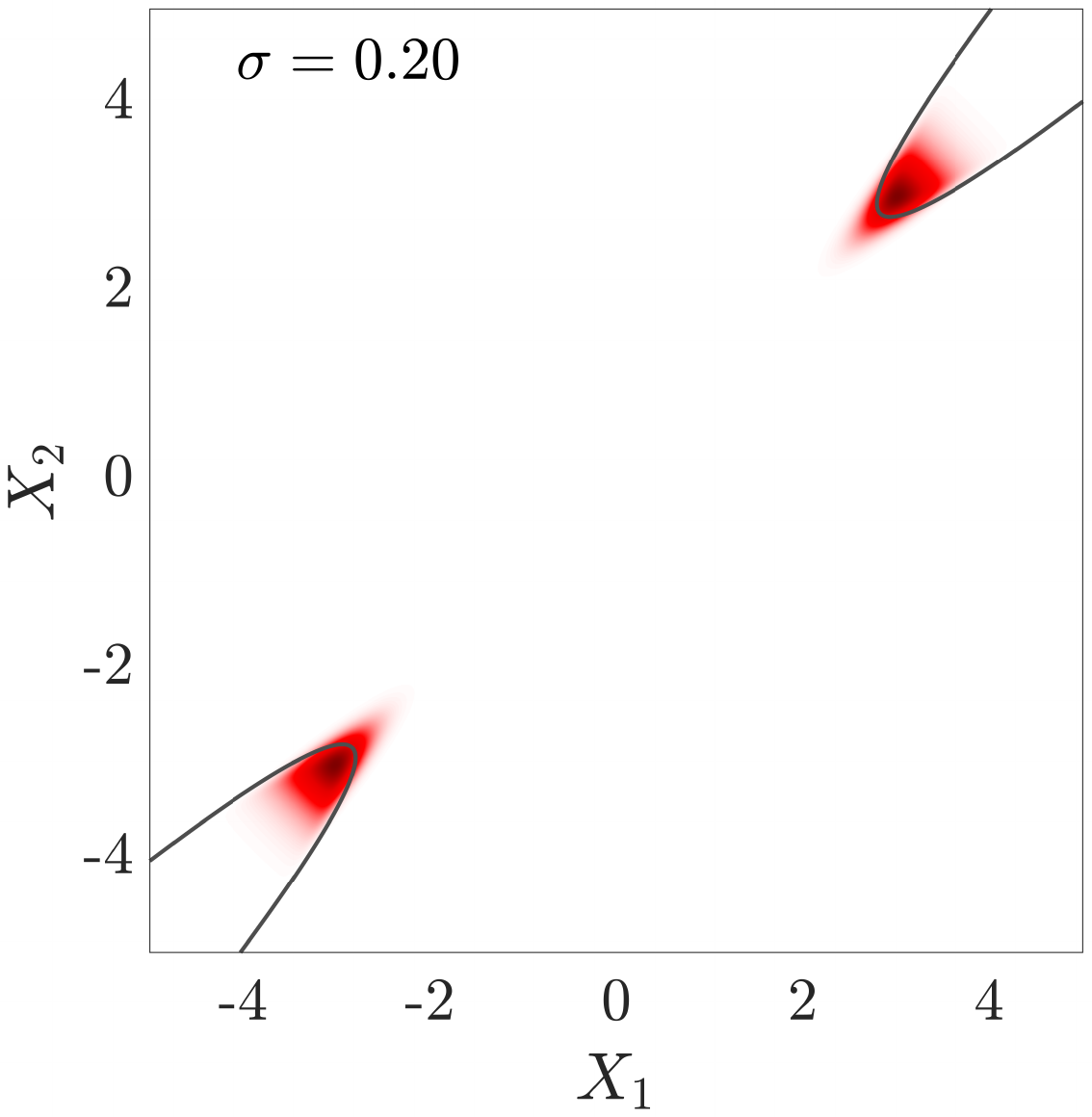}& \hspace*{-0.2in} \includegraphics[width=.205\textwidth,keepaspectratio]{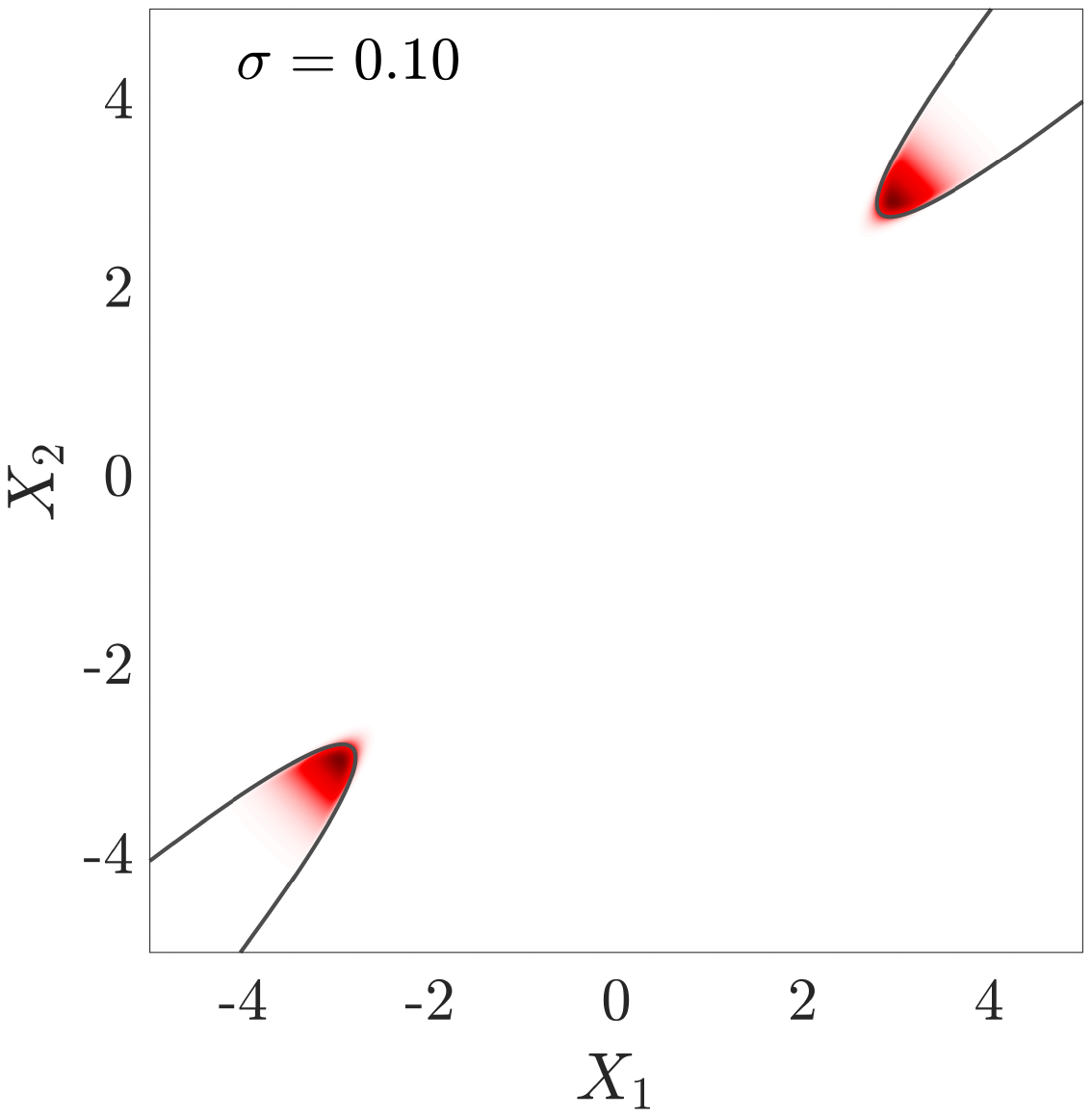}
 \vspace{0.4in}
 \end{tabular}
 \caption{Effect of the likelihood dispersion factor $\sigma$ on the constructed target distribution for a nonlinear bimodal  limit-state function, shown in gray at $g(\bm{X})=0$, as detailed in \Cref{sec: Bimodal_nonlinear}.}
 \label{fig: Effect_on_target_sigma}
\end{figure}

As discussed earlier, the basic idea is to construct a near-optimal sampling target distribution $\tilde{h}$, placing higher importance on the rare event domain $\mathcal{F}$. In ASTPA, this is achieved by approximating the indicator function $I_\mathcal{F}$ in \Cref{eq: opt_I_S_density} with a smooth likelihood function $\ell_{g_{\bm{X}}}$, chosen as a logistic cumulative distribution function (CDF) of the negative of a scaled limit state function, $F_{\text{CDF}}\left(\nicefrac{-g(\bm{X})}{g_c}\right)$, with $g_c$ being a scaling constant:
\begin{equation}
\begin{aligned}
\ell_{g_{\bm{X}}} =  \, F_{\text{CDF}} \bigg(\dfrac{-g(\bm{X})}{g_c} \mid\ \mu_{g} , \ \sigma\bigg ) =  \, \Bigg (1 + \exp\bigg (\dfrac{(\frac{g(\bm{X})}{g_{c}})+\mu_{g}}{(\frac{\sqrt{3}}{\pi})\sigma} \bigg ) \Bigg )^{-1}
\label{eq: likelihood}
\end{aligned}
\end{equation}
where $\mu_{g}$ and $\sigma$ are the mean and dispersion factor of the logistic CDF, respectively. The approximate sampling target is ASTPA is then expressed as: 
\begin{equation}
\begin{aligned}
\tilde{h}(\bm{X}) = \ell_{g_{\bm{X}}} \, \pi_{\bm{X}}(\bm{X}) =  \, \Bigg (1 + \exp\bigg (\dfrac{(\frac{g(\bm{X})}{g_{c}})+\mu_{g}}{(\frac{\sqrt{3}}{\pi})\sigma} \bigg ) \Bigg )^{-1} \, \pi_{\bm{X}}(\bm{X})
\label{eq: Tar_ASTPA}
\end{aligned}
\end{equation}
The dispersion factor $\sigma$ controls the spread of the likelihood function and thus influences the shape of the constructed target $\tilde{h}$. Recommended values for $\sigma$ lie in the range of $[0.1, \, 0.6]$, with lower values (typically between $0.1$ and $0.4$) generally working efficiently.  Higher values within the recommended range may, however, be needed for high-dimensional cases involving strongly nonlinear effects. The mean parameter, $\mu_{g}$, of the logistic CDF is typically chosen as $\mu_{g} = 1.21\sigma$, generally placing $\mu_{g}$ inside the rare event domain, as derived in \citep{eshra2024direct}. In \Cref{fig: Effect_on_target_sigma}, we illustrate how the dispersion factor, $\sigma$, influences the shape of the target distribution, based on a nonlinear bimodal limit state function, as detailed in \Cref{sec: Bimodal_nonlinear}. As shown, decreasing $\sigma$ results in a more concentrated target distribution inside the rare event domain. Although pushing the target further inside the rare event domain, i.e., closer to the optimal distribution $(I_{\mathcal{F}} \pi_{\bm{X}})$, can generate more rare event samples, it may also complicate the sampling process and, thus, the computation of the normalizing constant $\hat{C}_h$, as discussed in \citep{eshra2024direct}.

The scaling $\nicefrac{g(\bm{X})}{g_c}$ accounts for the varying magnitudes of limit-state functions, ensuring that the input to the logistic CDF $\left(\nicefrac{g(\bm{X})}{g_{c}}\right)$, evaluated at the mean of the original distribution, consistently falls within a predefined range. 
This process standardizes the input while keeping the recommended values for $\sigma$ unchanged. To ensure the input at $\bm{x} = \bm{0}$ lies within a desired range of $[3, 7]$, $g_c$ is defined as $g_c = \nicefrac{g(\bm{0})}{q}$, where $\, q \in [3, 7] \text{ if } g(\bm{0}) \notin [3, 7]$, and $g_c = 1$ otherwise. Our empirical analysis confirms the effectiveness of this scaling technique, demonstrating that the recommended range for $q$ is practical for the studied Gaussian problems. 

\begin{figure}[t!]
 \centering
  \begin{tabular}{ccc}
   \hspace*{-.15in}
	\includegraphics[trim=0cm 3.5cm 0cm 3cm,width=0.21\textwidth,keepaspectratio]{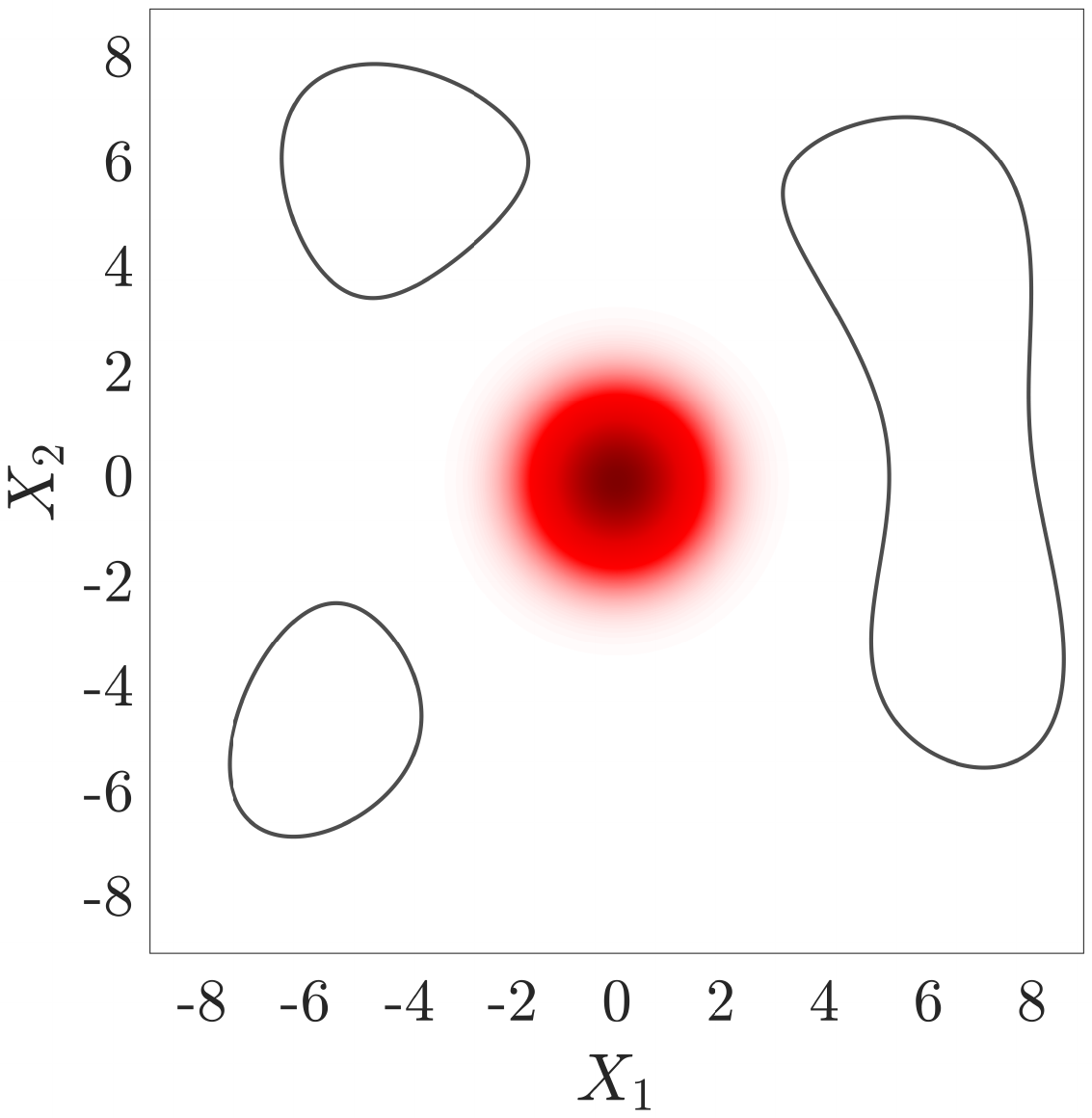}&\hspace{-0.2in}
			\quad
	\includegraphics[trim=0cm 3.5cm 0cm 3cm,width=0.21\textwidth,keepaspectratio]{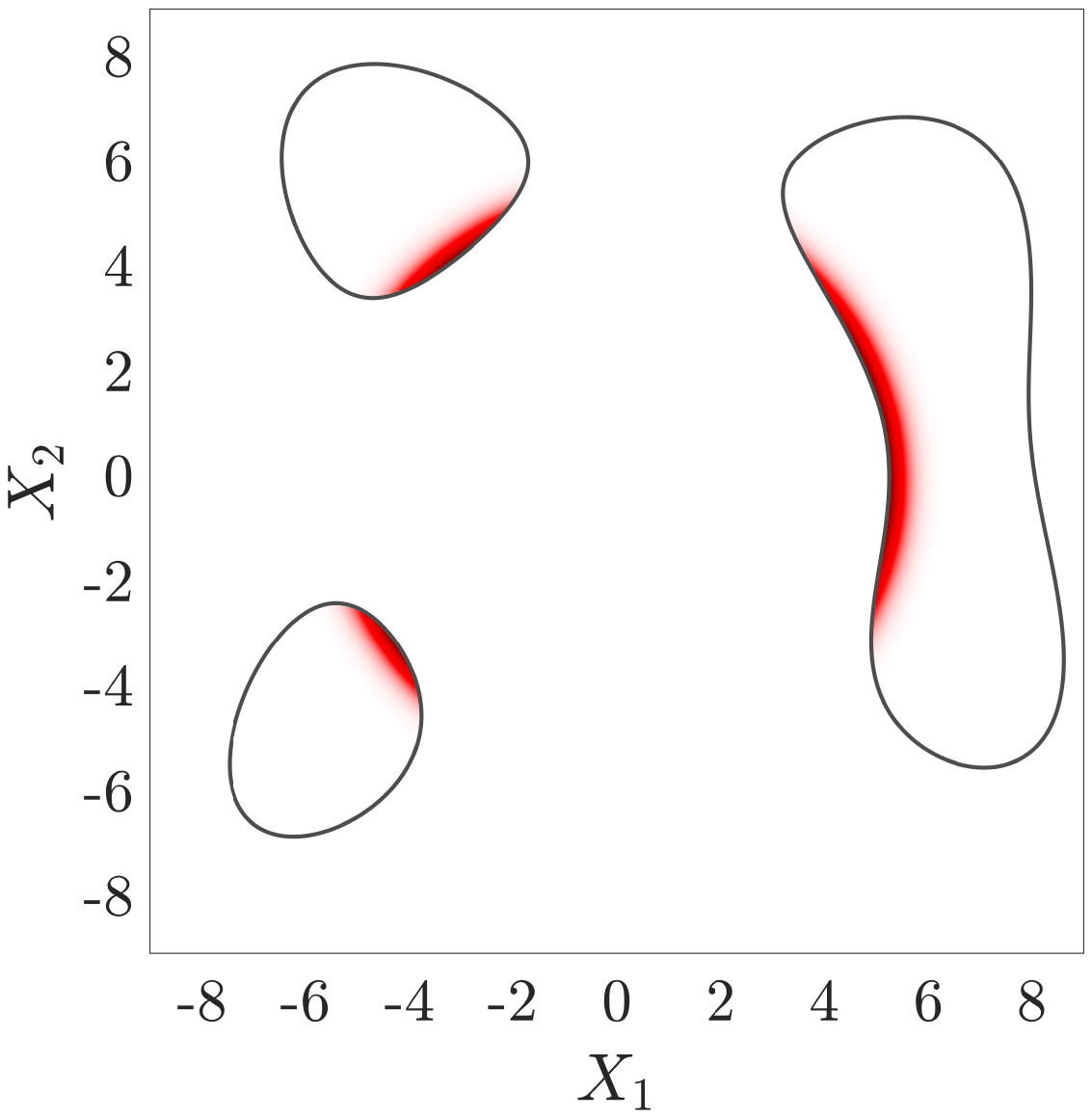}& \hspace{-0.2in}
   			\quad
	\includegraphics[trim=0cm 3.5cm 0cm 3cm,width=0.21\textwidth,keepaspectratio]{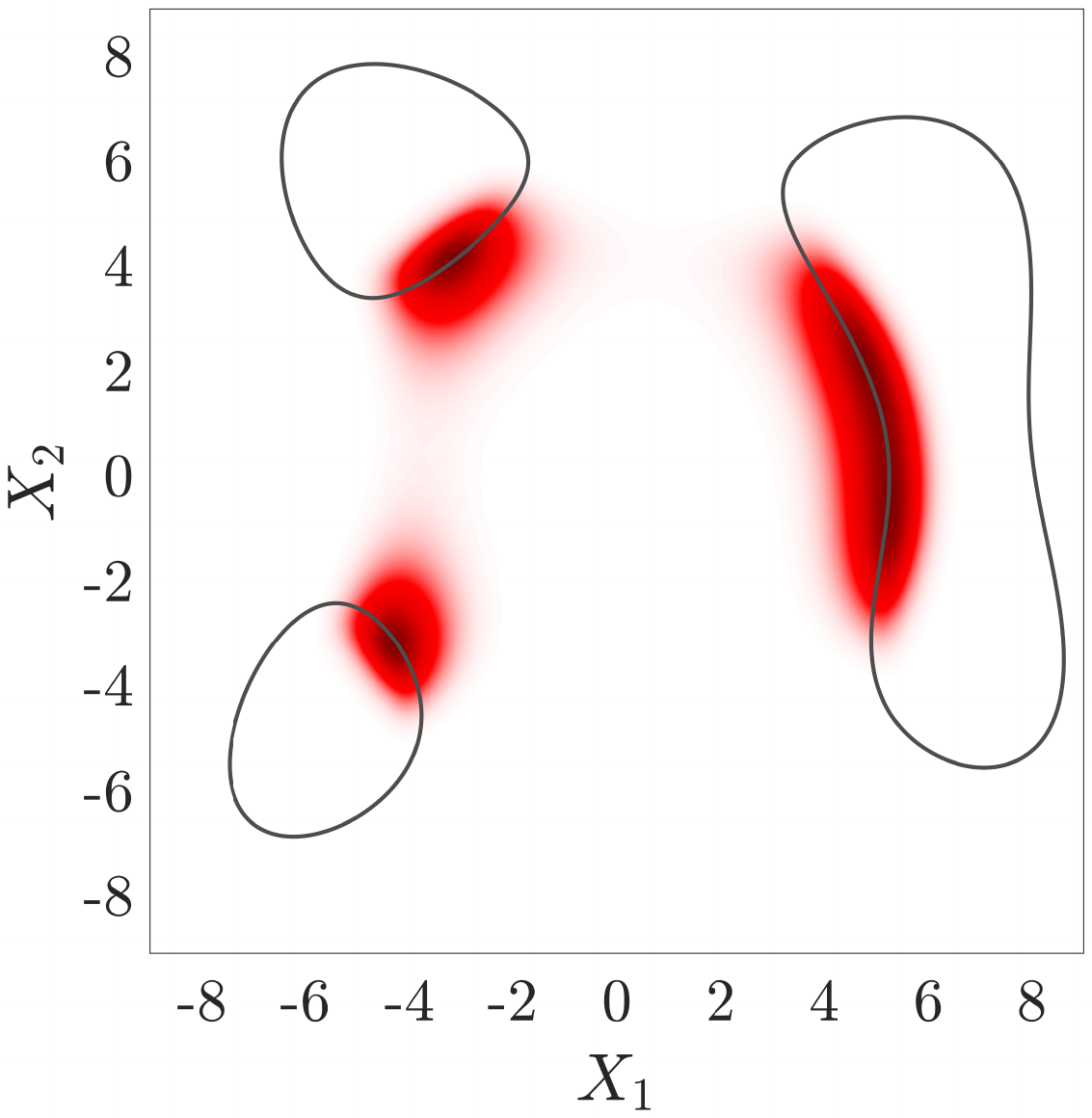}\\
    \hspace*{-0.25in}
     {\fontsize{8.5pt}{7.2} (a) $\pi_{\bm{X}}$}&
    \hspace*{-0.25in}
    {\fontsize{8.5pt}{7.2} (b) $\tilde{h}(\bm{X}); g_{c}=1 $} &
    \hspace*{-0.25in}
    {\fontsize{8.5pt}{7.2} (c) $\tilde{h}(\bm{X}); {g_{c}=\dfrac{g(\textbf{0})}{4}}$} 
  \end{tabular}
    \caption{Illustrating the impact of the scaling constant $g_c$ on constructing of the target distribution $\tilde{h}(\bm{X})$. The gray island-shaped curve represents Himmelblau's multi-modal limit-state function $g(\bm{X})$, with the failure domain inside the curves and $p_\mathcal{F} \sim 2.81 \times 10^{-7}$, as detailed in \Cref{sec: Him}. (a) Bivariate Gaussian distribution. (b) Constructed target distribution  $\tilde{h}(\bm{X})$ without the proper value of the scaling constant. (c) The corrected target $\tilde{h}(\bm{X})$ after applying an appropriate scaling constant $g_c = \nicefrac{g(\bm{0})}{4}$.}\label{fig: Effect_on_target_scale}
\end{figure}

\Cref{fig: Effect_on_target_scale} demonstrates the effect of the scaling constant $g_{c}$ on constructing the target distribution $\tilde{h}(\bm{X})$. Since the limit state function considered in this problem provides $g(\bm{0}) > 7$, the scaling constant $g_c$ can be defined as $\nicefrac{g(\bm{0})}{4}$. \Cref{fig: Effect_on_target_scale}(b) shows the unnormalized target distribution $\tilde{h}(\bm{X})$ constructed without applying the proper scaling constant. As a result, the input $\left(\nicefrac{g(\bm{X})}{g_c}\right)$ at $\bm{x} = 0$ does not lie within the desired range, leading to a highly concentrated target. In contrast, \Cref{fig: Effect_on_target_scale}(c) presents the corrected target distribution after applying the appropriate scaling factor $g_c = \nicefrac{g(\mathbf{0})}{4}$. These plots clearly demonstrate how scaling the limit-state function helps maintain an appropriately relaxed target, with the recommended parameter values remaining unchanged, thus enhancing sampling efficiency.

\subsection{Inverse importance sampling}\label{sec: IIS} 
Inverse importance sampling (IIS) is a general technique to compute normalizing constants, demonstrating exceptional performance in the context of rare event probability estimation \citep{Papakon2023HMCMC}. Given an unnormalized distribution $\tilde{h}$, and samples $\{\bm{x}_i\}_{i=1}^N \sim h$, inverse importance sampling first fits an ISD $Q(\bm{x})$ based on the samples $\{\bm{x}_i\}_{i=1}^N \sim h$. IIS then estimates the normalizing constant $C_h$ as:
\begin{equation} \label{eq: IIS_est}
C_h = \int_{\mathcal{X}} \tilde{h}(\bm{x}) d\bm{x} = \int_{\mathcal{X}} \dfrac{\tilde{h}(\bm{x})}{Q(\bm{x})}Q(\bm{x}) d\bm{x}\,=\mathbb{E}_{Q} \big[ \dfrac{\tilde{h}(\bm{X})}{Q(\bm{X})}] 
\end{equation} 
By drawing $\{\bm{x}^\prime\}_{i=1}^M$ i.i.d. samples from $Q$, the unbiased IIS estimator can be computed as:
\begin{equation} \label{eq: IIS_est_hat}
\hat{C}_h = \dfrac{1}{M}\sum_{i=1}^{M} \dfrac{\tilde{h}(\bm{x}_i^\prime)}{Q(\bm{x}_i^\prime)}
\end{equation}
In this work, a Gaussian Mixture Model (GMM), based on the available samples $\{\bm{x}_i \}_{i=1}^N \sim h$, is employed as the ISD $Q(\bm{X})$. In low-dimensional spaces, where $d<20$, we employ a GMM with a large number $(\sim 10)$ of Gaussian components that have full covariance matrices. This approach aims to closely approximate the target distribution $\tilde{h}$ with an ISD $Q(.)$, thus improving the IIS estimator in \cref{eq: IIS_est_hat}. To address the challenges posed by high-dimensional spaces, GMMs with diagonal covariance matrices are used in high-dimensional examples in this work. This choice reduces the number of GMM parameters to be estimated, thereby mitigating the scalability issues typically associated with density estimation methods. Notably, IIS has been shown to work effectively even with a crudely fitted $Q$ \citep{eshra2024direct, Papakon2023HMCMC}. 

The IIS technique is further enhanced using the splitting approach introduced in \citep{eshra2024direct} to improve the robustness of the IIS estimator against the impact of rarely observed outliers or anomalous samples. In this approach, the sample set $\{\bm{x}_i^\prime\}_{i=1}^M \sim Q$ is split into two equal-sized subsets, and the IIS estimate $\hat{C}_h$ is computed separately for each subset. If the two estimates are within a reasonable range of each other (specifically, $\frac{1}{3} \leq \nicefrac{\hat{C}_h^1}{\hat{C}_h^2} \leq 3$), the final estimate is taken as their average. Otherwise, the final estimate is conservatively set to the minimum of the two. This method has proven effective across all examples presented in this work.

\subsection{Statistical properties of the ASTPA estimator} \label{sec: AnalCOV}
The ASTPA estimator of the rare event probability can finally be computed as:
\begin{equation} \label{eq: p_f_ASTPA_final}
\begin{aligned}
p_\mathcal{F}&= \,{\mathop{\mathbb{E}}}_{h} \big[ I_{\mathcal{F}} (\bm{X}) \dfrac{\pi_{\bm{X}}(\bm{X})}{\tilde{h}(\bm{X})}\big] \,\mathbb{E}_{Q} \big[ \dfrac{\tilde{h}(\bm{X})}{Q(\bm{X})}] 
\approx \hat{p}_\mathcal{F} = \hat{\tilde{p}}_\mathcal{F} \, \hat{C}_h  =  \bigg(\dfrac{1}{N} \sum_{i=1}^{N} I_{\mathcal{F}} (\bm{x}_i) \dfrac{\pi_{\bm{X}}(\bm{x}_i)}{\tilde{h}(\bm{x}_i)}\bigg) \bigg( \dfrac{1}{M}\sum_{i=1}^{M} \dfrac{\tilde{h}(\bm{x}_i^\prime)}{Q(\bm{x}_i^\prime)}\bigg)
\end{aligned}
\end{equation} 
where $\{\bm{x}_i\}_{i=1}^N$ and $\{\bm{x}_i^\prime\}_{i=1}^M$ are samples from $h$ and $Q$, respectively. 

As shown in \citep{Papakon2023HMCMC,eshra2024direct}, the ASTPA estimator is unbiased, i.e., $\mathop{\mathbb{E}}[\hat{p}_{\mathcal{F}}] = \mathop{\mathbb{E}}[\hat{\tilde{p}}_\mathcal{F} \, \hat{C}_h] = \tilde{p}_\mathcal{F}\, C_h$. The analytical Coefficient of Variation (C.o.V) of the ASTPA estimator can also be computed as $\textrm{C.o.V} \approx \nicefrac{\sqrt{\widehat{\text{Var}}(\hat{p}_{\mathcal{F}})}} {\hat{p}_\mathcal{F}}$, where the variance ($\widehat{\text{Var}}(\hat{p}_{\mathcal{F}})$) is estimated by:
\begin{equation}\label{eq: var_p_f_c_h}
\widehat{\text{Var}}(\hat{p}_{\mathcal{F}}) = (\hat{\tilde{p}}_{\mathcal{F}})^2 \,\widehat{\text{Var}}(\hat{C}_h) + (\hat{C}_h)^2 \widehat{\text{Var}}(\hat{\tilde{p}}_{\mathcal{F}}) + \widehat{\text{Var}}(\hat{\tilde{p}}_{\mathcal{F}})\widehat{\text{Var}}(\hat{C}_h)
\end{equation}
Here,  $\hat{\tilde{p}}_{\mathcal{F}}$ and $\hat{C}_h$ can be computed according to \Cref{eq: p_f_tilde_ASTPA,eq: IIS_est_hat}, respectively. The variances $\widehat{\text{Var}}(\hat{\tilde{p}}_{\mathcal{F}})$ and $\widehat{\text{Var}}(\hat{C}_h)$ can thus be estimated as follows:
\begin{equation}\label{eq: var_p_f_c_h_each}
\begin{aligned}
\widehat{\text{Var}}(\hat{\tilde{p}}_{\mathcal{F}}) = \frac{1}{N_{s}(N_{s}-1)} \sum_{i=1}^{N_{s}} \bigg ( \dfrac{I_{\mathcal{F}} (\bm{x}_i)\,\pi_{\bm{X}}(\bm{x}_i)}{\tilde{h}(\bm{x}_i)}  - \hat{\tilde{p}}_\mathcal{F} \bigg )^{2},  \quad \quad \widehat{\text{Var}}(\hat{C}_h) = \frac{1}{M(M-1)} \sum_{i=1}^{M} \Bigg (\dfrac{\tilde{h}(\bm{x}_i^\prime)}{Q(\bm{x}_i^\prime)} -\hat{C}_{h} \Bigg )^{2}
\end{aligned}
\end{equation}
where $N_{s}$ denotes the number of used Markov chain samples, accounting for the fact that the samples are not independent and identically distributed (i.i.d) in this case. To address this, a thinning process is adopted to reduce the sample size from $N$ to $N_s$ based on the effective sample size (ESS) of the sample set $\{\bm{x}_i\}_{i=1}^{N}$. We select every $j^{th}$ sample, where $j$, an integer, is determined by:
\begin{equation}
    j=\left\lfloor\dfrac{N}{4 \cdot \text{ESS}_{\text{min}}}\right\rfloor
    \label{thining}
\end{equation}
where $\text{ESS}_{\text{min}}$ represents the minimum effective sample size across all dimensions. To ensure $j$ remains within practical bounds, it is constrained to the set $\{3, 4, 5, \ldots, 30\}$. Any calculated value of $j$ outside this range is adjusted to the nearest bound within it. The analytical C.o.V computed according to this approach showed very good agreement with the sample C.o.V in all our numerical examples.

\section{Gradient-free Sampling of the Approximate Target (\texorpdfstring{$\tilde{h}$}{h}) in ASTPA}\label{sec:  Sampling}

In this section, we introduce a novel gradient-free approach for sampling the approximate target, $\tilde{h} = \ell_{g_{\bm{X}}} \, \pi_{\bm{X}}$, in ASTPA. Rare event probability estimation typically faces two fundamental sampling challenges. First, efficiently sampling high-dimensional distributions is difficult due to the curse of dimensionality, where the volume of the space expands exponentially with the number of dimensions \citep{betancourt2017conceptual}. Second, the rare event domain, $\mathcal{F} \coloneqq \{ \bm{x} \,:\, g(\bm{x}) \leq 0\}$, often exists randomly in the tails of the probability distribution $\pi_{\bm{X}}$, making it challenging to effectively guide samples toward these regions of interest, which may also exhibit multi-modality; see \Cref{fig: ASTPA_framework}(a) for an example. In addressing the first challenge, we adopt the dimension-robust preconditioned Crank-Nicolson (pCN) algorithm within our approach, as discussed in \Cref{sec: pCN}. For the second challenge, we develop an effective discovery method, aimed at precisely locating rare event domains and appropriately initializing the pCN chains, as detailed in \Cref{sec: Discovery}. Our developed algorithm, termed guided-pCN, thus enhances sampling for $\tilde{h}$ by overcoming the limitations of the standard pCN sampler in the context of rare event probability estimation.

\subsection{The preconditioned Crank-Nicolson (pCN) algorithm }\label{sec:  pCN}

The preconditioned Crank-Nicolson (pCN) algorithm is a gradient-free Markov Chain Monte Carlo (MCMC) method designed for efficient sampling from high-dimensional probability distributions \citep{Neal1998, cotter2013mcmc}. To sample from a target distribution $\tilde{h} = \ell_{g_{\bm{X}}} \, \pi_{\bm{X}}$, the pCN sampler proposes an update, $\tilde{\bm{x}}$, at each iteration, $t$, by combining the current sample, $\bm{x}_{t}$,  with a perturbation drawn from the original (prior) Gaussian distribution, $\pi_{\bm{X}}=\mathcal{N}(\bm{X};\,\bm{0},\mathbf{C})$, as follows:
\begin{equation}
    \tilde{\bm{x}} = \sqrt{1-\beta^2}\,\bm{x}_{t}+\beta \,\boldsymbol{\xi} , \quad \bm{\xi} \sim \mathcal{N}(\mathbf{0}, \mathbf{C})
    \label{eq: pNC_proposal}
\end{equation}
where $\beta \in (0,1]$ is a tunable scaling parameter, discussed below, and $\mathbf{C}$ is a covariance matrix. This update is equivalent to using a Gaussian proposal distribution $\mathcal{N}(\sqrt{1-\beta^2}\,\bm{x}_{t}, \, \beta^2\mathbf{C})$, yielding the acceptance probability:  
\begin{equation}
\alpha = \min\bigg\{1,\dfrac{\ell_{g_{\bm{X}}}(\tilde{\bm{x}})}{\ell_{g_{\bm{X}}}(\bm{x}_{t})}\bigg\}
\label{eq: MH_prob_pCN}
\end{equation}
where $\ell_{g_{\bm{X}}}$ is the likelihood function in \Cref{eq: likelihood}. This formula illustrates a fundamental characteristic of the pCN algorithm: its invariance to the underlying Gaussian distribution $\mathcal{N}(\bm{X};\,\bm{0},\mathbf{C})$. In essence, this ensures that the proposed sample is always accepted when sampling from $\pi_{\bm{X}}$, making the pCN highly suitable for efficiently exploring high-dimensional Gaussian spaces. This invariance can be understood by examining the Metropolis-Hastings (MH) ratio when using the pCN algorithm to sample $\mathcal{N}(\bm{X};\,\bm{0},\mathbf{C})$, as:
\begin{equation}
\alpha = \min\bigg\{1,\dfrac{ \mathcal{N}(\tilde{\bm{x}};\,\bm{0},\mathbf{C})\,\mathcal{N}(\bm{x}_t\mid \tilde{\bm{x}};\, \sqrt{1-\beta^2}\,\tilde{\bm{x}}, \, \beta^2\mathbf{C})} {\mathcal{N}(\bm{x}_t;\,\bm{0},\mathbf{C})\,\mathcal{N}(\tilde{\bm{x}} \mid \bm{x}_t;\,\sqrt{1-\beta^2}\,\bm{x}_{t}, \, \beta^2\mathbf{C})}\bigg\}
\label{eq: MH_prior}
\end{equation}

\begin{algorithm}[t!]
\caption{The preconditioned Crank-Nicolson (pCN) Algorithm}\label{alg: pCN}
\begin{algorithmic}[1]
\Procedure{pCN}{$\bm{x}_{0}$, $\beta_{0}$, $\alpha^*$, $L_{\text{chain}}$, $\ell_{g_{\bm{X}}}(\bm{X})$, $\mathbf{C}$} 
\For{$t=1$ $to$ ${L_\text{Chain}}$}
\State $\bm{x}_{t}$ $\gets$ $\bm{x}_{t-1}$
\State $\tilde{\bm{x}} = \sqrt{1-\beta_{t-1}^2}\,\bm{x}_{t-1}+\beta_{t-1} \,\boldsymbol{\xi} , \quad \bm{\xi} \sim \mathcal{N}(\mathbf{0}, \mathbf{C})$ \Comment{Propose candidate sample}
\State $with$ $probability$:\Comment{Accept or reject candidate sample}\\ 
       \hspace{2cm} 
       $\alpha_{t} = \min\bigg\{1,\dfrac{\ell_{g_{\bm{X}}}(\tilde{\bm{x}})}{\ell_{g_{\bm{X}}}(\bm{x}_{t-1})}\bigg\}$ \\
       \hspace{2cm} $\bm{x}_{t}$ $\gets$ $\tilde{\bm{x}}$  
\State $\log \beta_{t} = \log \beta_{t-1} + t^{\nicefrac{-1}{2}} (\alpha_{t} - \alpha^{*})$ 
\Comment{Adapt $\beta$ using Robbins-Monro algorithm (\Cref{eq:robbins_monro_beta})}   
\EndFor
\EndProcedure
\end{algorithmic}
\end{algorithm}

Considring the marginal distribution $\mathcal{N}(\tilde{\bm{X}};\,\bm{0},\mathbf{C})$ and the conditional distribution $\mathcal{N}(\bm{X}_t\mid \tilde{\bm{X}};\, \sqrt{1-\beta^2}\,\tilde{\bm{x}}, \, \beta^2\mathbf{C})$, the joint random vector $[\tilde{\bm{X}} \,\, \bm{X}_t]^\text{T}$ is therefore Gaussian with a mean vector of $[\bm{0} \, \, \bm{0}]^\text{T}$ and a covariance matrix of $\begin{bmatrix}  \mathbf{C} & \sqrt{1-\beta^2}\, \mathbf{C} \\ \sqrt{1-\beta^2} \,\mathbf{C} & \mathbf{C}\end{bmatrix}$. The conditional distribution of  $\tilde{\bm{X}}$ given $\bm{X}_t$ is hence Gaussian with a conditional mean $\boldsymbol{\mu}_{\tilde{\bm{X}}\mid\bm{X}_t} = \sqrt{1-\beta^2}\bm{X}_t$ and a covariance $\mathbf{\Sigma}_{\tilde{\bm{X}}\mid\bm{X}_t} = \beta^2 \mathbf{C}$. We therefore get:
\begin{equation}\begin{aligned}
    \mathcal{N}(\tilde{\bm{X}};\,\bm{0},\mathbf{C}) \, \mathcal{N}(\bm{X}_t\mid \tilde{\bm{X}};\, \sqrt{1-\beta^2}\,\tilde{\bm{x}}, \, \beta^2\mathbf{C}) &= \mathcal{N}\bigg(\begin{bmatrix} \tilde{\bm{X}} \\ \bm{X}_t \end{bmatrix}; \,  \begin{bmatrix} \bm{0} \\ \bm{0} \end{bmatrix}, \begin{bmatrix}  \mathbf{C} & \sqrt{1-\beta^2}\, \mathbf{C} \\ \sqrt{1-\beta^2} \,\mathbf{C} & \mathbf{C}\end{bmatrix} \bigg)   \\& = \mathcal{N}(\bm{X}_t;\,\bm{0},\mathbf{C})\, \mathcal{N}(\tilde{\bm{X}} \mid \bm{X}_t ;\, \sqrt{1-\beta^2}\,\bm{x}_t, \, \beta^2\mathbf{C})
    \end{aligned}
\end{equation}
Consequently, the MH ratio in \Cref{eq: MH_prior} simplifies to one, confirming that the pCN proposal is indeed invariant under the original Gaussian distribution $\pi_{\bm{X}}$. As a result, when applying the pCN algorithm to sample from the target distribution $\tilde{h} = \ell_{g_{\bm{X}}} \, \pi_{\bm{X}}$, the acceptance probability reduces to the ratio between the likelihood function for the proposed and the current samples:    
\begin{equation}
\alpha = \min\bigg\{1,\dfrac{ \ell_{g_{\bm{X}}}(\tilde{\bm{x}})\,\mathcal{N}(\tilde{\bm{x}};\,\bm{0},\mathbf{C})\,\mathcal{N}(\bm{x}_t\mid \tilde{\bm{x}};\, \sqrt{1-\beta^2}\,\tilde{\bm{x}}, \, \beta^2\mathbf{C})} {\ell_{g_{\bm{X}}}(\bm{x}_t) \,\mathcal{N}(\bm{x}_t;\,\bm{0},\mathbf{C})\,\mathcal{N}(\tilde{\bm{x}} \mid \bm{x}_t;\,\sqrt{1-\beta^2}\,\bm{x}_{t}, \, \beta^2\mathbf{C})}\bigg\}= \min\bigg\{1,\dfrac{\ell_{g_{\bm{X}}}(\tilde{\bm{x}})}{\ell_{g_{\bm{X}}}(\bm{x}_{t})}\bigg\}
\label{eq: MH_proof}
\end{equation}
The scaling parameter $\beta$ determines the balance between retaining information from the current sample and exploring new regions of the random variable space. When $\beta$ is close to zero, we have $\sqrt{1 - \beta^2} \approx 1$, which keeps the proposed sample $\tilde{\bm{x}}$ close to the current one. This proximity leads to high acceptance rates but potentially slows the mixing of the Markov chain due to limited exploration, thus reducing the effective sample size. Conversely, as $\beta$ approaches one, we have $\sqrt{1 - \beta^2} \approx 0$, making $\tilde{\bm{x}}$ largely independent of $\bm{x}_t$. This allows the sampler to explore the random variable space more broadly, but the proposed samples are more likely to land in low-probability regions, leading to an increase in rejections. To enhance the sampling efficiency, it is essential to find an optimal $\beta$ balancing these trade-offs, which is typically achieved by tuning $\beta$ to target a desired average acceptance ratio $\alpha^*$. The Robbins--Monro stochastic approximation algorithm provides a method for such adaptation as \citep{robbins1951stochastic,andrieu2008tutorial}:
\begin{equation}
\log \beta_{t+1} = \log \beta_{t} + \gamma_t \left( \alpha_t - \alpha^{*} \right)
\label{eq:robbins_monro_beta}
\end{equation}
where $\alpha_t$ is the acceptance probability at iteration $t$, computed by \Cref{eq: MH_prob_pCN}, and $\gamma_t$ is a diminishing step size sequence, commonly chosen as $\gamma_t = t^{-0.5}$ \citep{papaioannou2015mcmc}. In this work, we target an average acceptance probability in the range of $20$--$40\%$, with the lower values applied to high-dimensional distributions and the higher values to low-dimensional ones. This range is consistent with the values typically employed in the literature \citep{cotter2013mcmc,rudolf2018generalization} and has proven effective across all numerical examples considered in this work. \Cref{alg: pCN} describes the pCN algorithm, where $\bm{x}_0$ is the initial sample, $\beta_0$ is the initial scaling parameter ($\beta_0 = 0.5$ in this work), $\alpha^*$ is the desired average acceptance ratio, and $L_\text{chain}$ indicates the number of steps in the pCN chain.

\begin{figure}[t!]
 \vspace*{-0.9in}
 \centering
  \begin{tabular}{ccccc}
   \hspace*{-1.3in} \includegraphics[width=.3\textwidth,keepaspectratio]{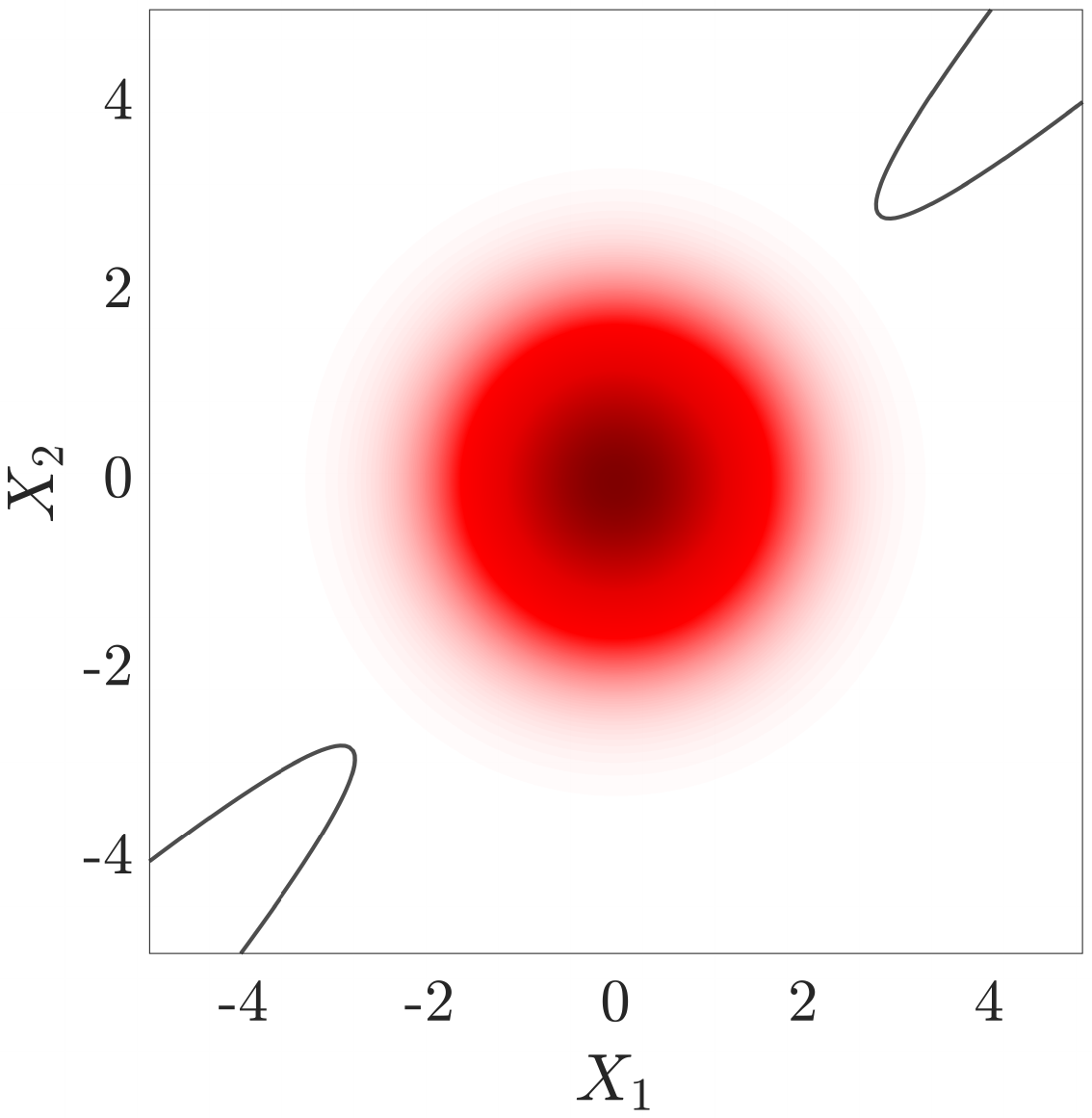}& \hspace*{-2.7in} \includegraphics[width=.3\textwidth,keepaspectratio]{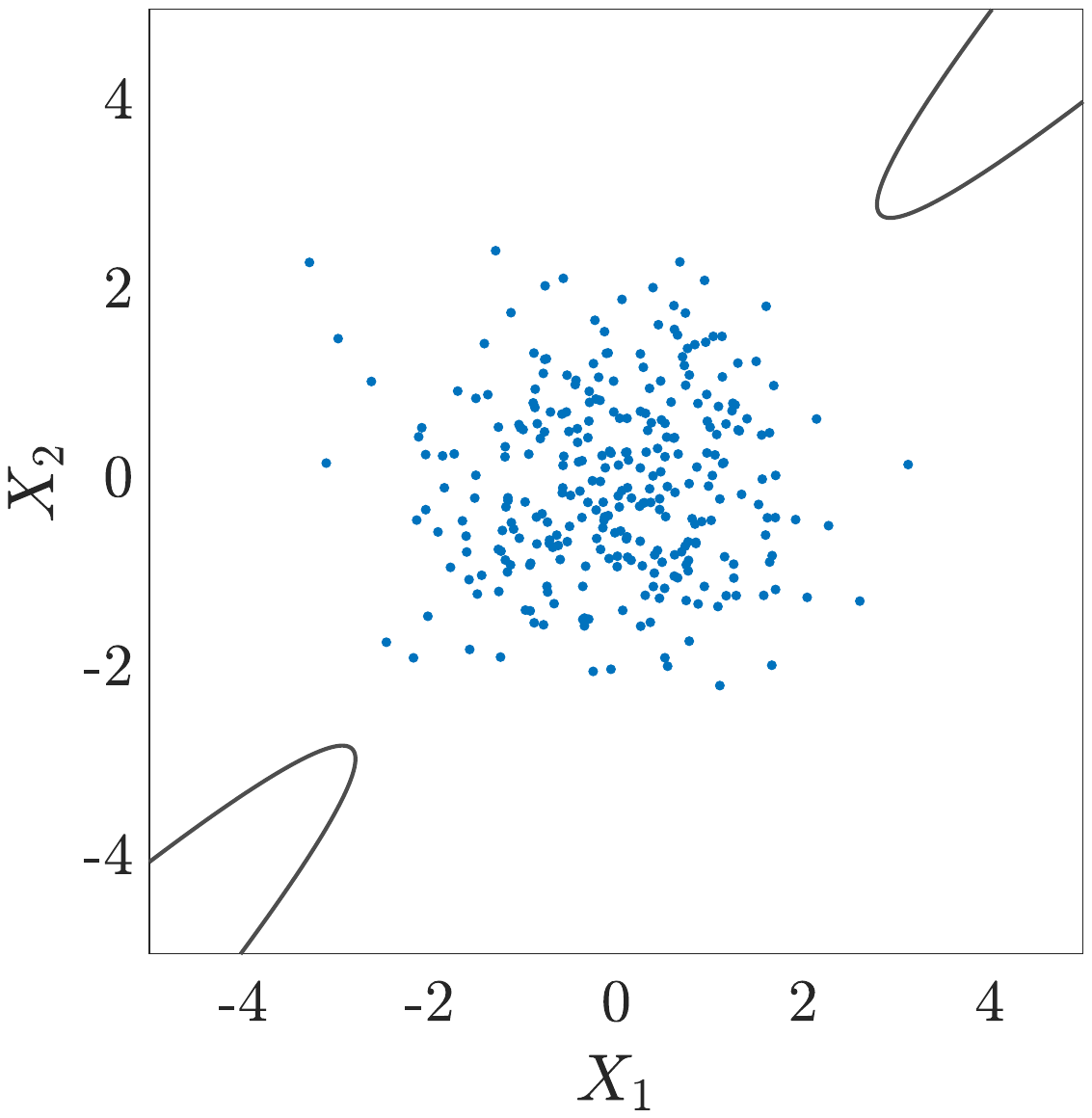}& \hspace*{-1.5in} 
         \includegraphics[width=.3\textwidth,keepaspectratio]{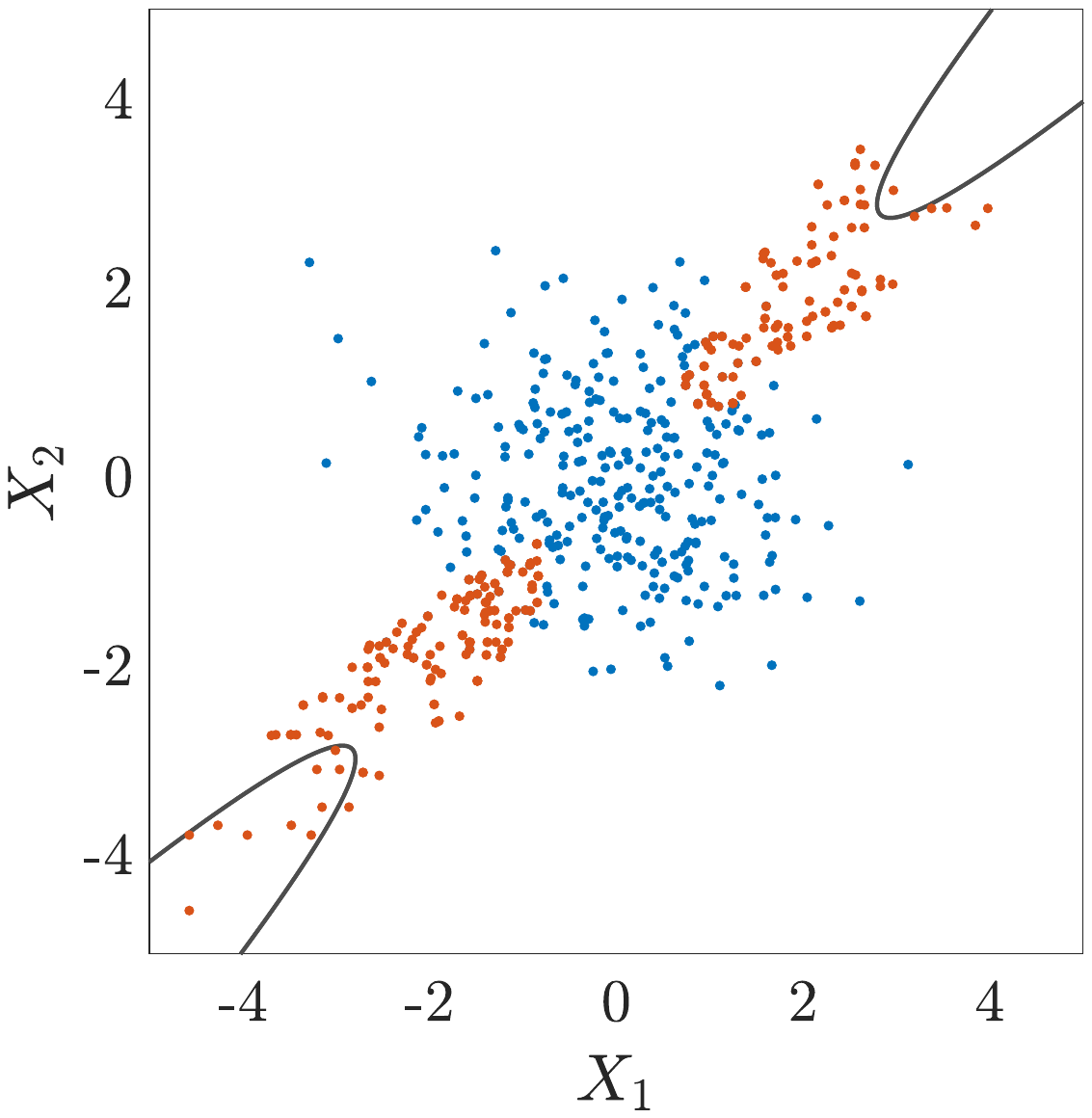}\vspace*{-0.3in}\\ \hspace*{-1.3in}
     {\fontsize{8.5pt}{7.2} (a) $\pi_{\bm{X}}=\mathcal{N}(\bm{X};\,\bm{0},\mathbf{I})$}&
    \hspace*{-2.7in}
    {\fontsize{8.5pt}{7.2} (b) $\{\bm{x}_i^0\}_{i=1}^{300} \sim\pi_{\bm{X}}$} &
    \hspace*{-1.5in}
    {\fontsize{8.5pt}{7.2} (c) $\{\bm{x}_i^1\}_{i=1}^{300} \sim \dfrac{1}{\pi_{\bm{X} \mid \mathcal{F}_1}}$} \vspace*{-0.3in}\\ \hspace*{1.2in}
    \includegraphics[width=.3\textwidth,keepaspectratio]{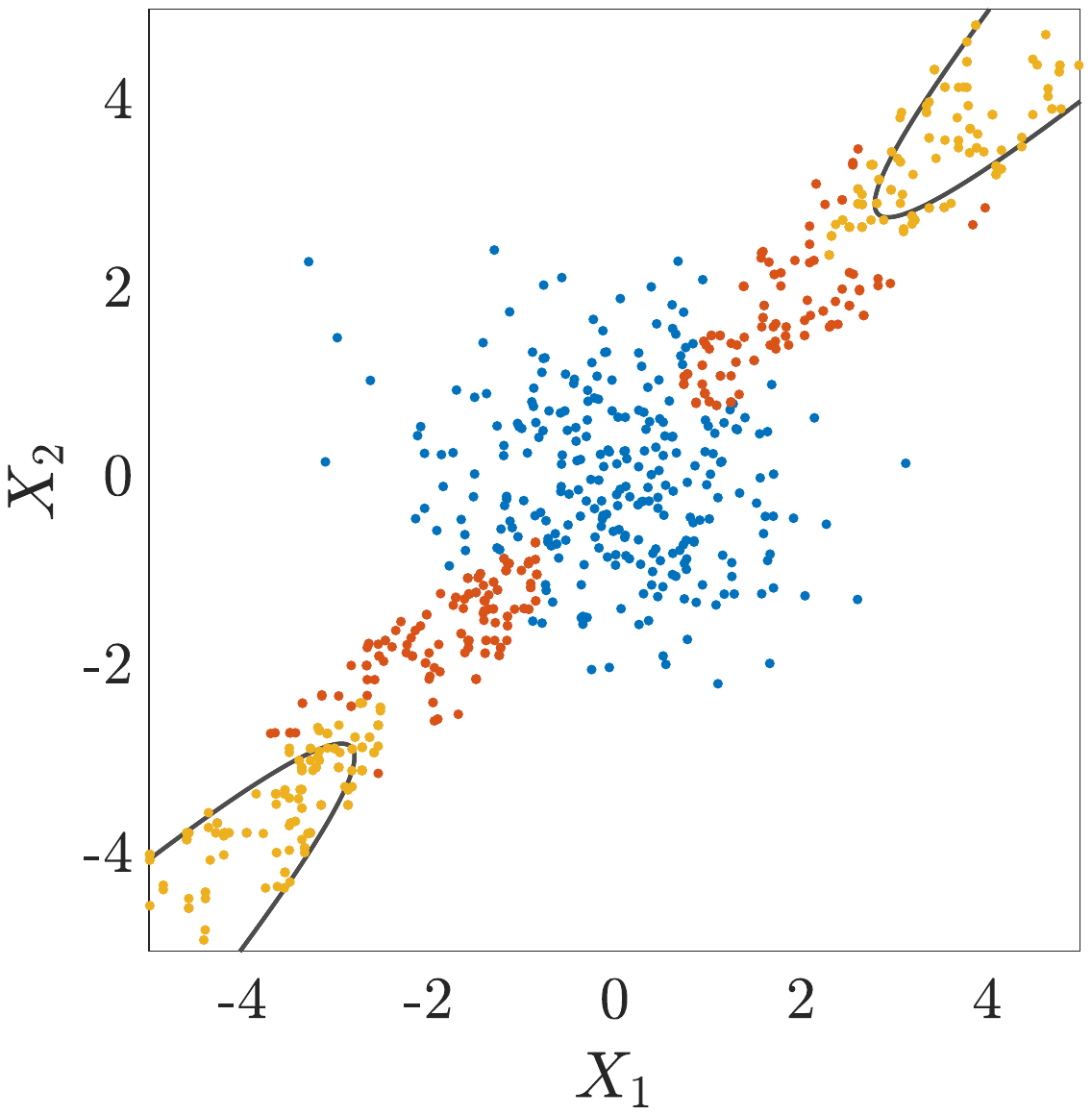}& \hspace*{-0.3in}\includegraphics[width=.3\textwidth,keepaspectratio]{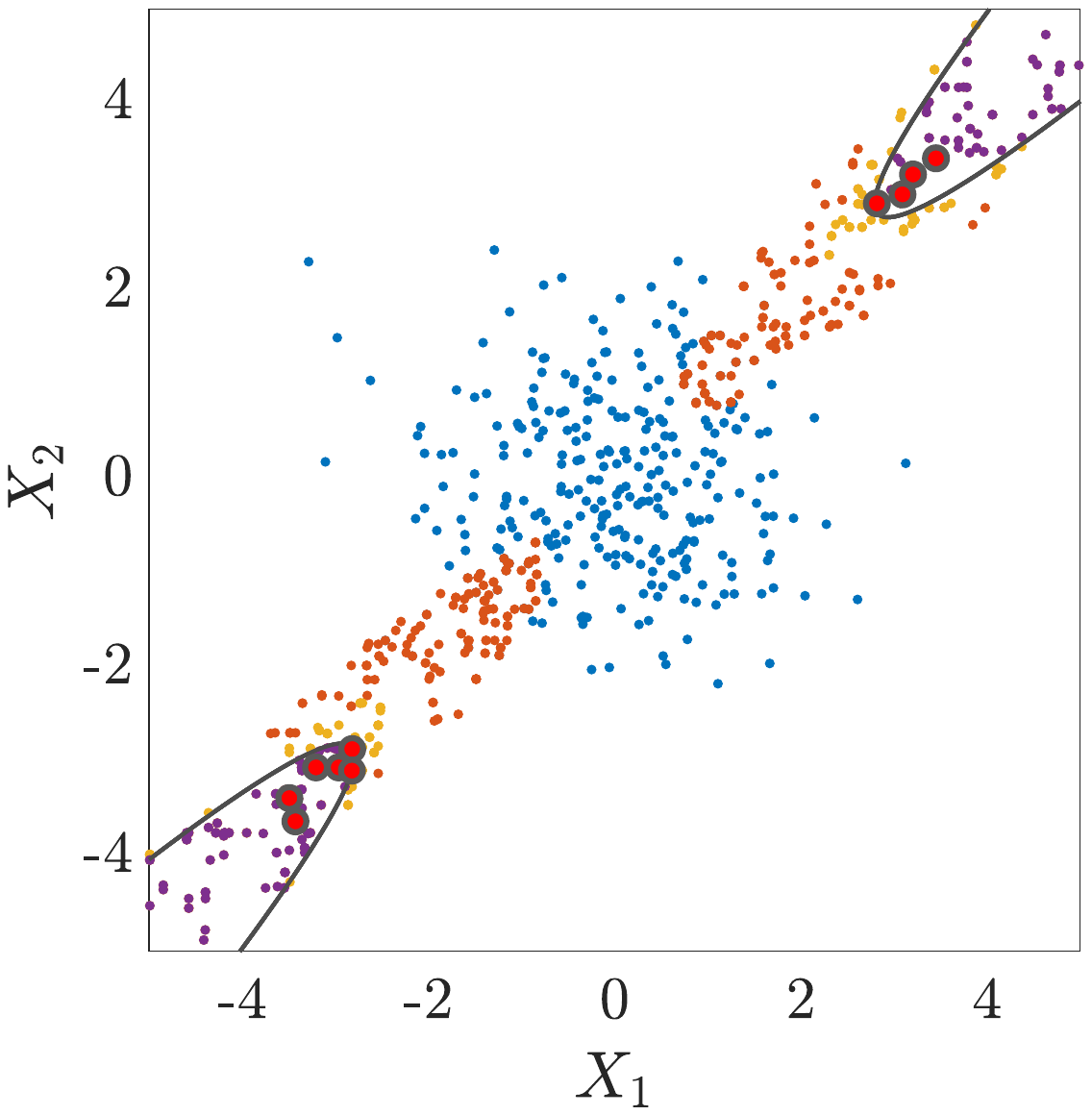}
         \vspace*{-0.3in}\\
    \hspace*{1.3in}
    {\fontsize{8.5pt}{7.2} (d) $\{\bm{x}_i^2\}_{i=1}^{300} \sim \dfrac{1}{\pi_{\bm{X} \mid \mathcal{F}_2}}$} &
    \hspace*{-0.25in}
    {\fontsize{8.5pt}{7.2} (e) Selected seeds} 
 \end{tabular}
  \caption{Outlining the developed discovery approach of the rare event domain discussed in \Cref{sec: Discovery}; (a) shows a bivariate independent standard Gaussian distribution in red, and a limit state function at $g(\bm{X})=0$ shown using a grey line, (b) presents the first sample set drawn from the original distribution $\mathcal{N}(\bm{X};\,\bm{0},\mathbf{I})$, (c) visualizes an additional conditional sample set drawn from $\nicefrac{1}{\pi_{\bm{X} \mid \mathcal{F}_1}}$, (d) adds a second conditional sample set drawn from $\nicefrac{1}{\pi_{\bm{X} \mid \mathcal{F}_2}}$, which provided at least $p_0 N_{\text{level}}$ rare event samples, thus this procedure was stopped, and (e) depicts the pCN seeds in red, selected according to the computed probabilities in \Cref{eq: seeds_weight}. These seeds are used to initialize the pCN chains but are not utilized in estimating the sought probability.}
 \label{fig: Discovery_sigma}
\end{figure}

\subsection{Rare event domain discovery}\label{sec: Discovery}
In this section, we introduce a computationally efficient discovery approach for rare event domains, designed to enhance sampling efficiency by generating (multimodal) rare event samples that serve as initial seeds for the pCN chains. This approach addresses the limitations the pCN algorithm may encounter in scenarios involving multimodality and target distributions, $\tilde{h}$, that are significantly distant from the original (prior) distribution $\pi_{\bm{X}}$.

\begin{figure}[t!]
 \vspace*{-0.7in}
 \centering
  \begin{tabular}{ccccc}
   \hspace*{-0.22in} \includegraphics[width=.28\textwidth,keepaspectratio]{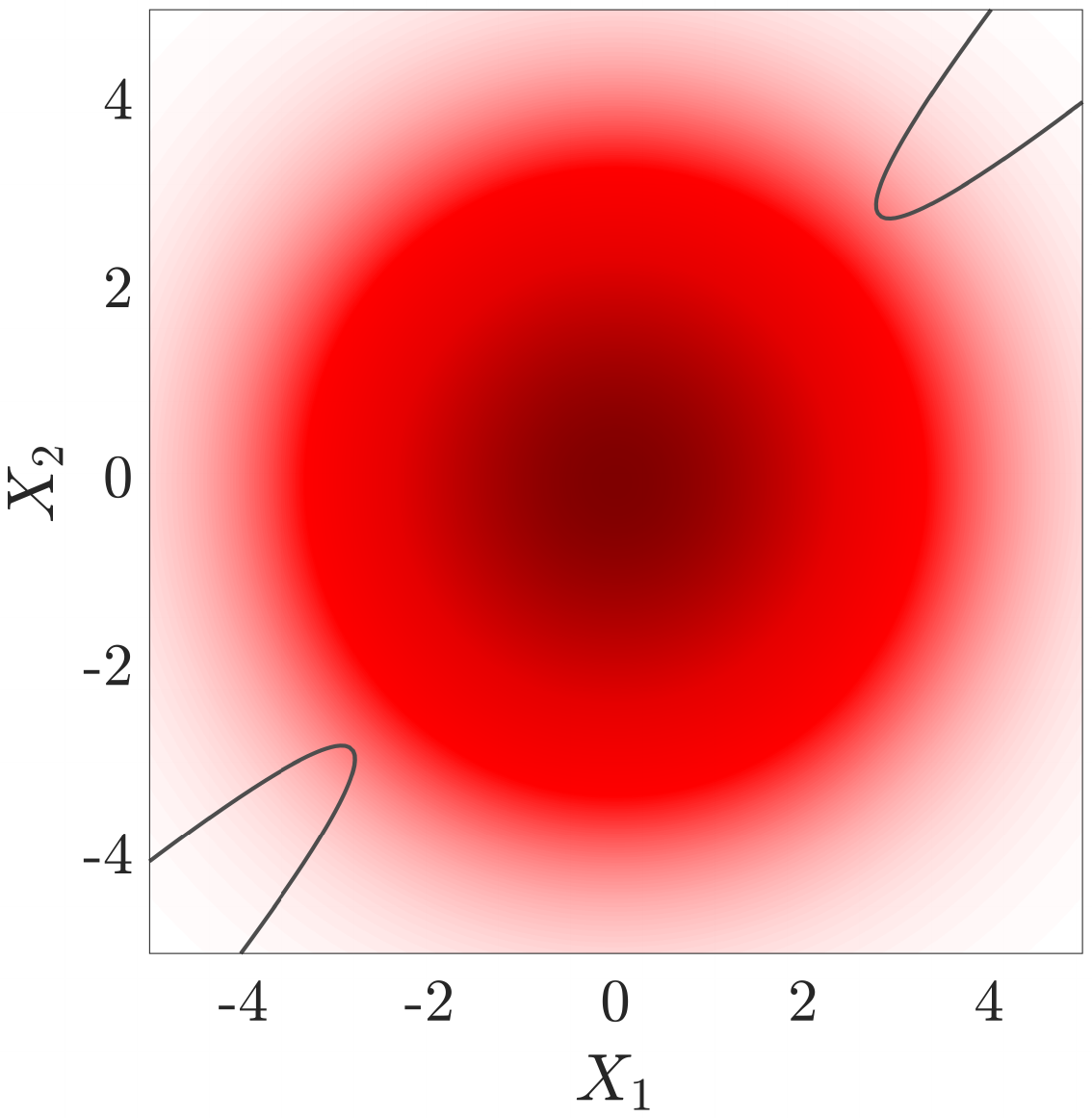}& \hspace*{-0.35in} \includegraphics[width=.28\textwidth,keepaspectratio]{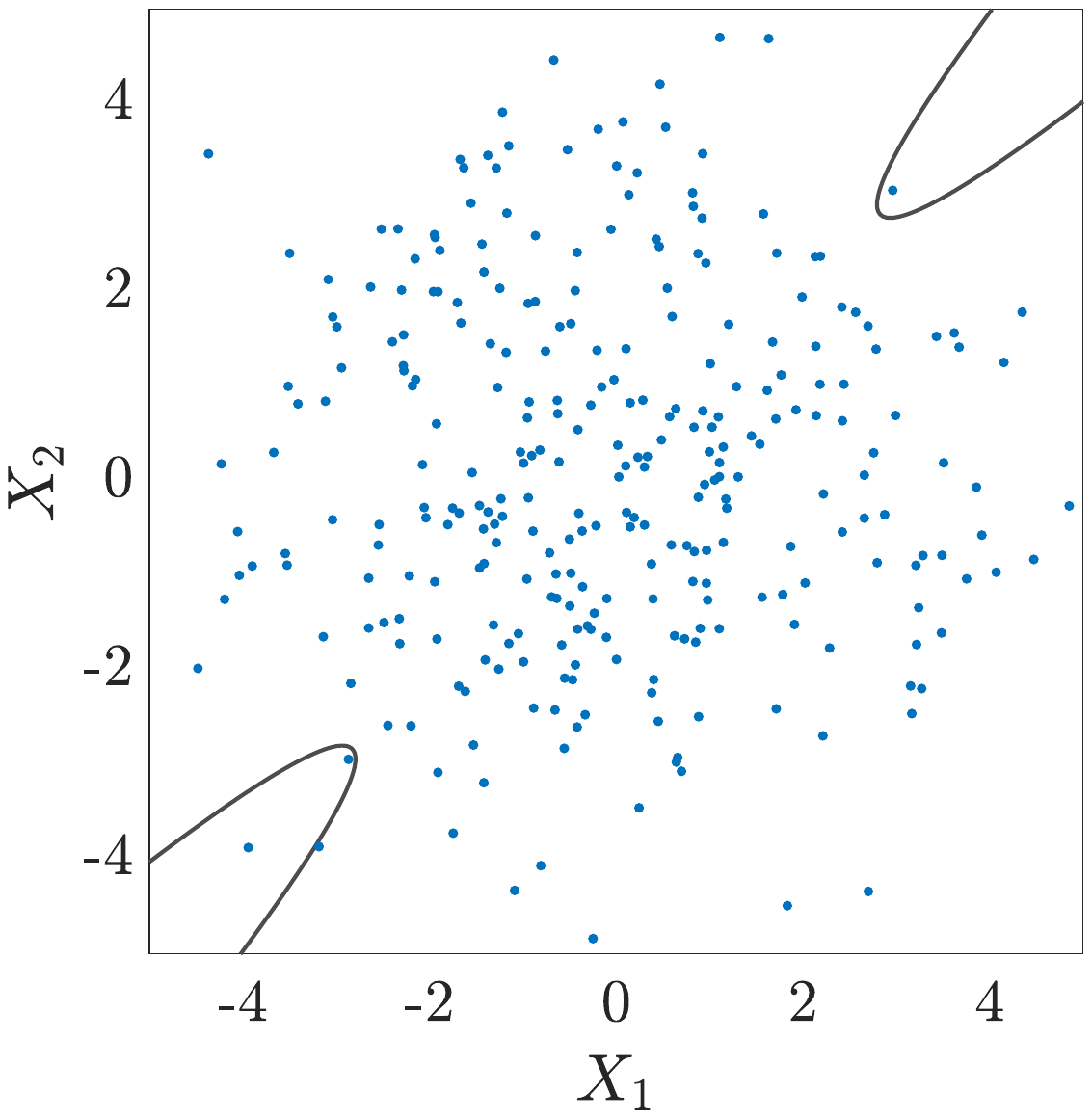}& \hspace*{-0.35in} 
         \includegraphics[width=.28\textwidth,keepaspectratio]{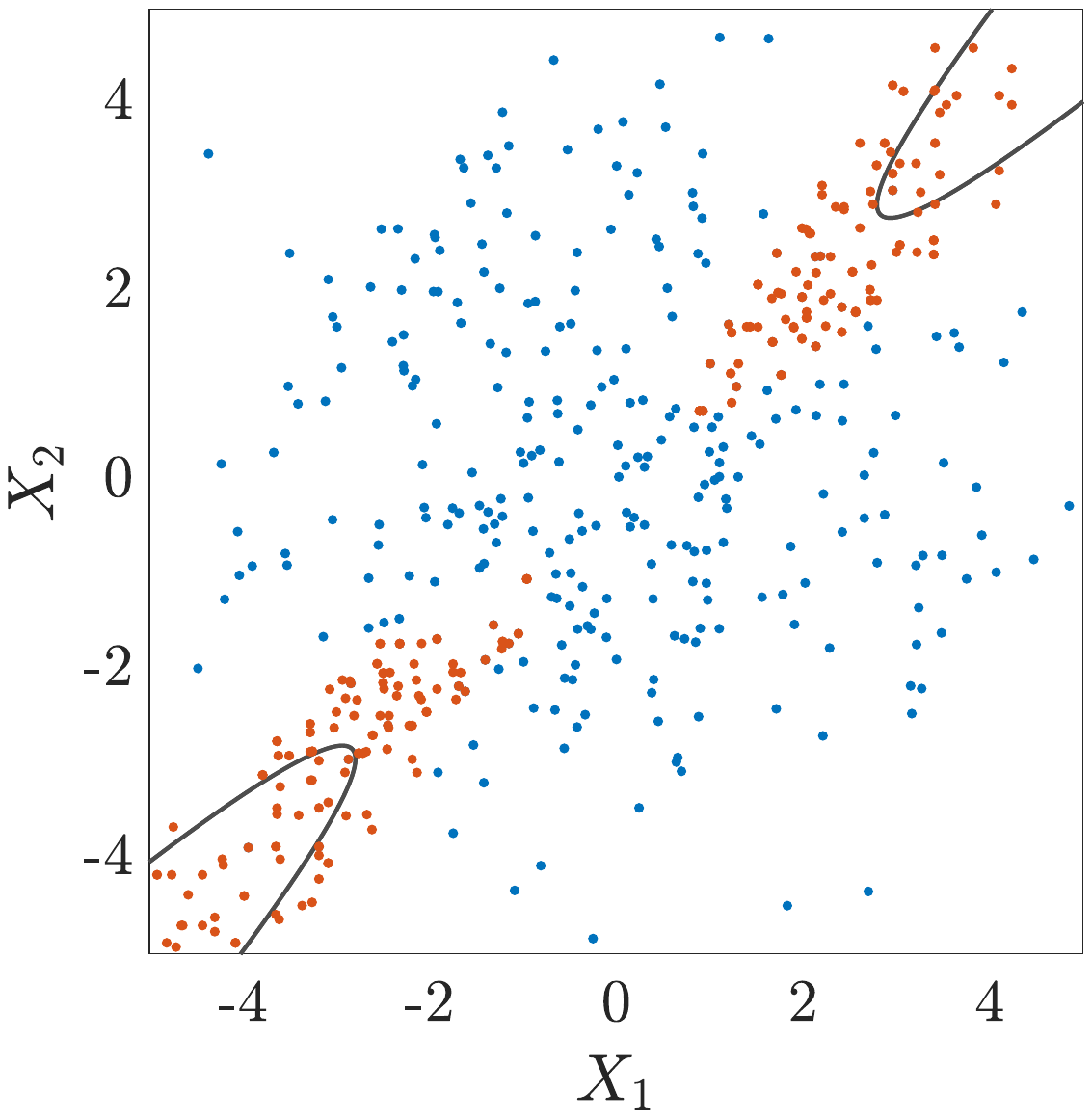}& \hspace*{-0.35in}\includegraphics[width=.28\textwidth,keepaspectratio]{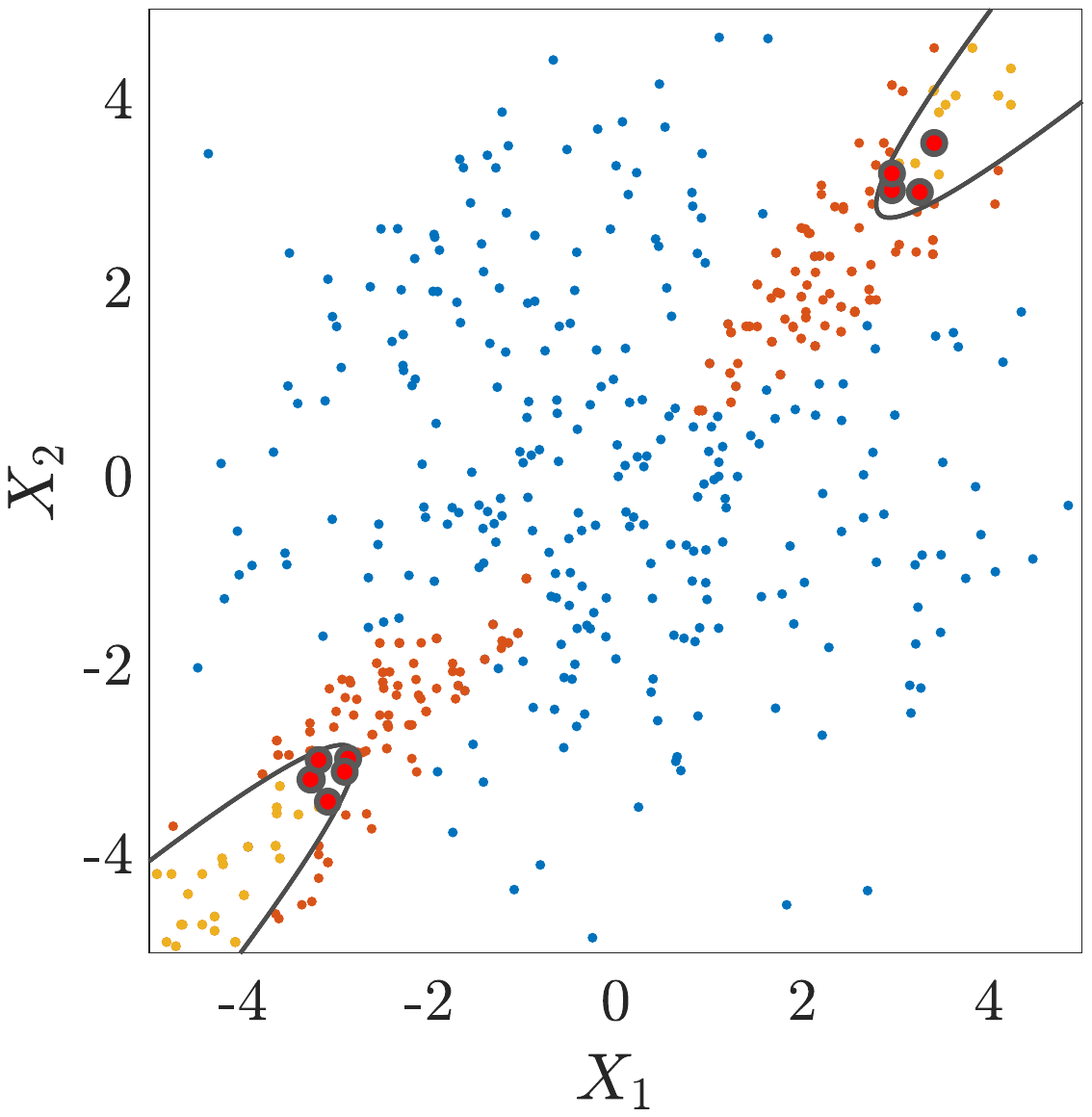}
\vspace*{-0.25in}\\\hspace*{-0.1in}
     {\fontsize{8.5pt}{7.2} (a) $\mathcal{N}(\bm{X};\,\bm{0},4\mathbf{I})$}&
    \hspace*{-0.25in}
    {\fontsize{8.5pt}{7.2} (b) $\{\bm{x}_i^0\}_{i=1}^{300} \sim\mathcal{N}(\bm{X};\,\bm{0},4\mathbf{I})$} &
    \hspace*{-0.25in}
    {\fontsize{8.5pt}{7.2} (c) $\{\bm{x}_i^1\}_{i=1}^{300} \sim \dfrac{1}{\pi_{\bm{X} \mid \mathcal{F}_1}}$} &
    \hspace*{-0.25in}
    {\fontsize{8.5pt}{7.2} (d) Selected seeds} 
 \end{tabular}
   \caption{Illustrating the benefits of utilizing a more dispersed initial distribution during the rare event domain discovery, using the same example as in \Cref{fig: Discovery_sigma}; (a) shows a bivariate independent Gaussian distribution with a covariance matrix of $4\mathbf{I}$, (b) presents the first sample set drawn from $\mathcal{N}(\bm{X};\,\bm{0},4\mathbf{I})$, (c) visualizes an additional conditional sample set drawn from $\nicefrac{1}{\pi_{\bm{X} \mid \mathcal{F}_1}}=\nicefrac{1}{\mathcal{N}(\bm{X} \mid \mathcal{F};\,\bm{0},\,\mathbf{I})}$, which provides at least $p_0 N_{\text{level}}$ rare event samples. As a result, this procedure is stopped one step earlier compared to \Cref{fig: Discovery_sigma}, highlighting greater computational efficiency. (d) depicts the pCN seeds in red, selected according to the computed probabilities in \Cref{eq: seeds_weight}. These seeds are used to initialize the pCN chains but are not utilized in estimating the sought probability. }
 \label{fig: Discovery_2sigma}
\end{figure}

\begin{algorithm}[t!]
\caption{Rare Event Domain Discovery Algorithm}\label{alg: Discovery}
\begin{algorithmic}[1]
\Procedure{RareEventDiscovery}{$N_{\text{level}}$, $p_0$, $\epsilon$, $g(\bm{X})$, $N_{\text{chain}}$} 
    \State Generate an initial sample set $\{\bm{x}_i^0\}_{i=1}^{N_{\text{level}}}$ from $\mathcal{N}(\bm{0},\epsilon \mathbf{I})$.
    
    \State Sort the samples $\{\bm{x}_i^0\}_{i=1}^{N_{\text{level}}}$ in increasing order of their limit-state values $\{g(\bm{x}_i^0)\}_{i=1}^{N_{\text{level}}}$.
    
    \State Find the threshold $\lambda_1$ as the $p_0$-percentile of the ordered $\{g(\bm{x}_i^0)\}_{i=1}^{N_{\text{level}}}$, and set $\mathcal{F}_1 = \{\bm{x} : g(\bm{x}) \leq \lambda_1\}$.
    
    \State Set the intermediate events counter $j \gets 1$.

    \While{$\lambda_j > 0$}
        \State  Find $N_S = p_0 N_{\text{level}}$ seeds, $\{\bm{x}_i^{(j-1)}\}_{i=1}^{N_S}$, where $\bm{x}_i^{(j-1)} \in \mathcal{F}_j$.
        
        \State Generate $N_{\text{level}}$ samples $\{\bm{x}_i^j\}_{i=1}^{N_{\text{level}}}$ from $\nicefrac{1}{\pi_{\bm{X} \mid \mathcal{F}_j}}$, by drawing $(\nicefrac{1}{p_0} - 1)$ additional samples, starting from  \hspace*{1cm}  each seed $\bm{x}_i^{(j-1)} \in \mathcal{F}_j$, using an MCMC sampler.\Comment{$\pi_{\bm{X}\mid \mathcal{F}_j} =  \mathcal{N}(\bm{X} \mid \mathcal{F}_j;\,\bm{0},\mathbf{I})$}

        \State Find $\lambda_{j+1}$ as the $p_0$-percentile of  the ascendly ordered $\{g(\bm{x}_i^j)\}_{i=1}^{N_{\text{level}}}$, and set $\mathcal{F}_{j+1} = \{\bm{x} : g(\bm{x}) \leq \lambda_{j+1}\}$.

        \State $j \gets j + 1$
    \EndWhile
    
    \State Sample $N_{\text{chain}}$ pCN seeds from the final rare event sample set $\{\bm{x}_i^{(j-1)}\}_{i=1}^{N_S}$, for which $\bm{x}_i^{(j-1)} \in \mathcal{F}$.
\EndProcedure
\end{algorithmic}
\end{algorithm}

This process begins by sampling from the original distribution, $\pi_{\bm{X}}=\mathcal{N}(\bm{X};\,\bm{0},\mathbf{I})$, or a more dispersed version, $\mathcal{N}(\bm{X};\,\bm{0},\epsilon \mathbf{I})$, with the dispersion parameter $\epsilon\geq1$, to facilitate broader exploration of the random variable space. This step is followed by conditional sampling based on the reciprocal of the conditional distribution $\nicefrac{1}{\pi_{\bm{X} \mid \mathcal{F}_j}}$, where $\mathcal{F}_j$ is an intermediate event defined as $\mathcal{F}_{j} \coloneqq \{ \bm{x} \,:\, g(\bm{x}) \leq \lambda_j\}$, with $\mathcal{F}_1 \supset \mathcal{F}_2 \supset \dots \supset \mathcal{F}_M$, where $\mathcal{F}_M = \mathcal{F}$ and $\lambda_1 \geq \lambda_2 \geq \dots \geq \lambda_M = 0$, chosen adaptively as discussed below. The use of $\nicefrac{1}{\pi_{\bm{X} \mid \mathcal{F}_j}}$ significantly facilitates rapid diffusion toward the rare event domain, compared to sampling from the conditional distribution $\pi_{\bm{X} \mid \mathcal{F}_j}$. While the latter is necessary for methods sequentially computing the sought probability, our method focuses solely on generating samples in the rare event domain (without utilizing any sequential information). The procedure stops after generating a specified number of rare event samples from the synthetic distribution  $\nicefrac{1}{\pi_{\bm{X} \mid \mathcal{F}}}$, from which a defined number of seeds, $N_{\text{chain}}$, is randomly selected to initialize the pCN chains, ensuring that the sampler is initialized within the rare event domain. 

The detailed steps of this discovery process are outlined in Algorithm \ref{alg: Discovery}, where  $N_{\text{level}}$ denotes the number of samples of each intermediate event $\mathcal{F}_j$, $p_0$ is the conditional probability for intermediate events, controlling the selection of threshold levels, $\lambda_j$, $\epsilon$ is a dispersion parameter for the initial distribution, $\mathcal{N}(\bm{0}, \epsilon \mathbf{I})$, and $N_{\text{chain}}$ represents the number of the desired pCN chains. \Cref{fig: Discovery_sigma} outlines the proposed rare event domain discovery stage, showing the evolution of samples and seed selection, while the computational benefits of using a more dispersed initial distribution are demonstrated in \Cref{fig: Discovery_2sigma}. In this work, we use $\epsilon = 4$ for all numerical examples to facilitate broader exploration of the random variable space by the initial sample set $\{\bm{x}_i^0\}_{i=1}^{N_{\text{level}}}$. 

\begin{figure}[t!]
   \vspace*{-1in}
 \centering
  \begin{tabular}{ccccc}
   \includegraphics[width=.3\textwidth,keepaspectratio]{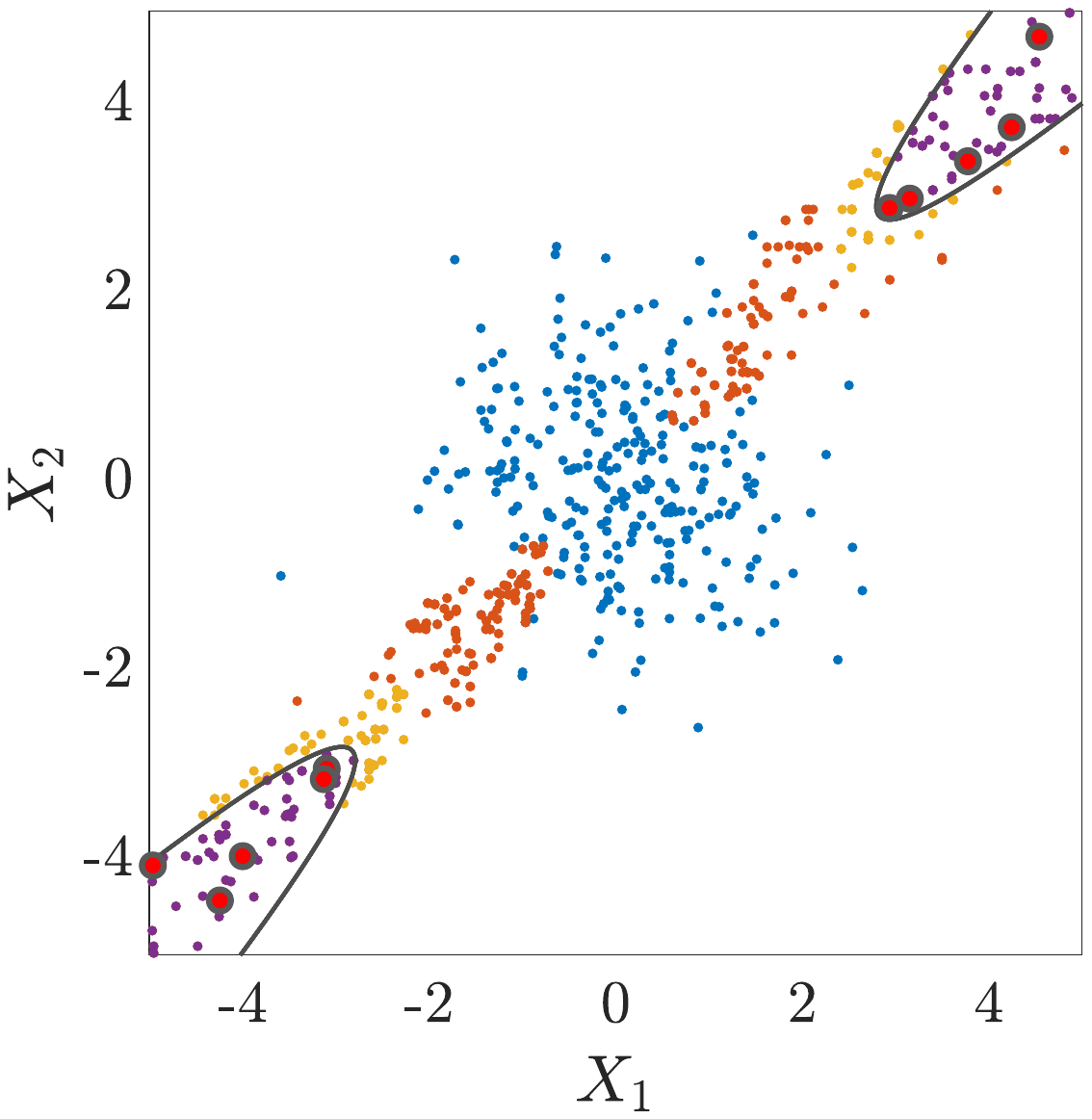}& \hspace*{-0.1in} 
         \includegraphics[width=.3\textwidth,keepaspectratio]{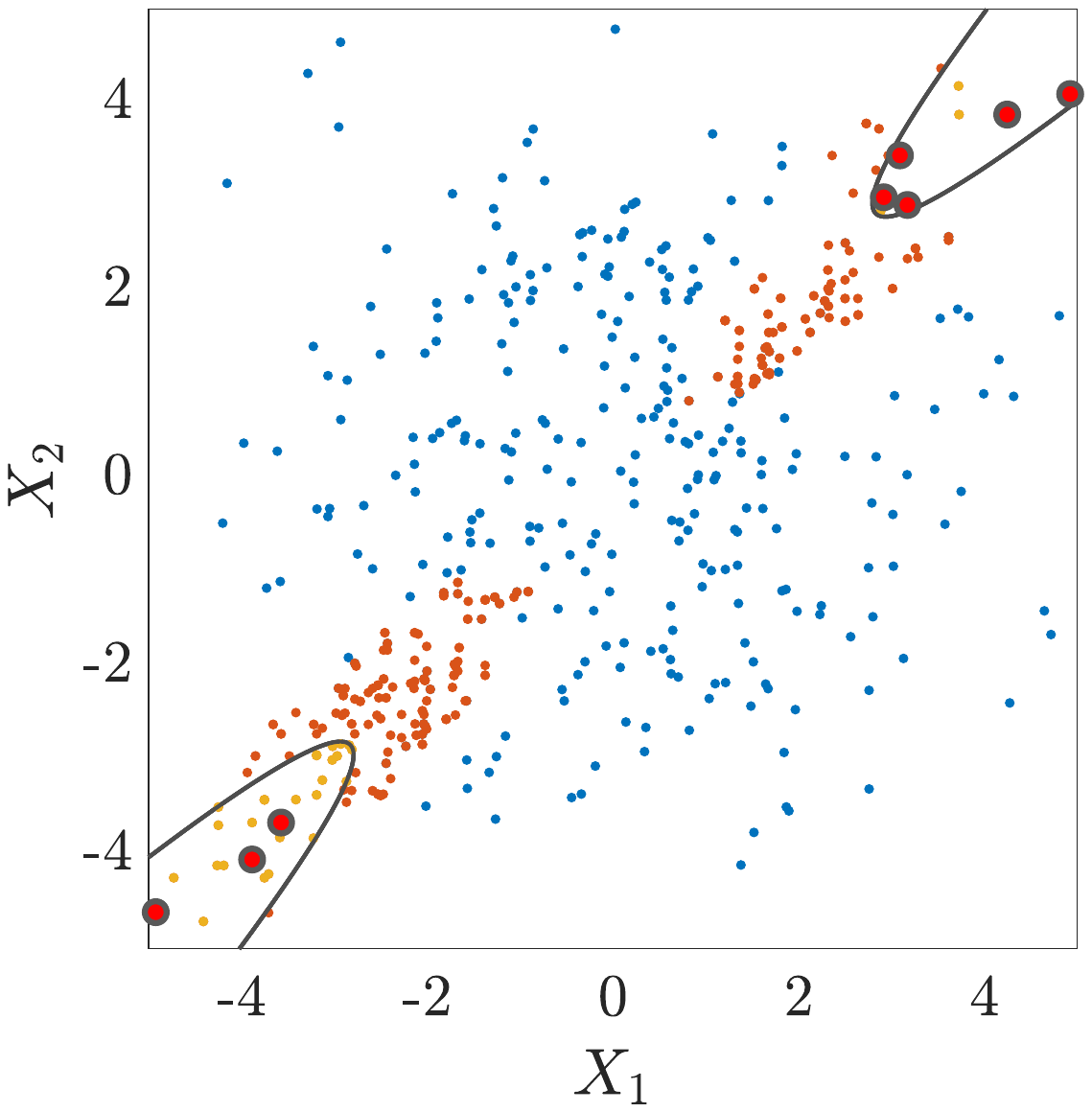}
         & \hspace*{-0.2in} 
         \includegraphics[width=.3\textwidth,keepaspectratio]{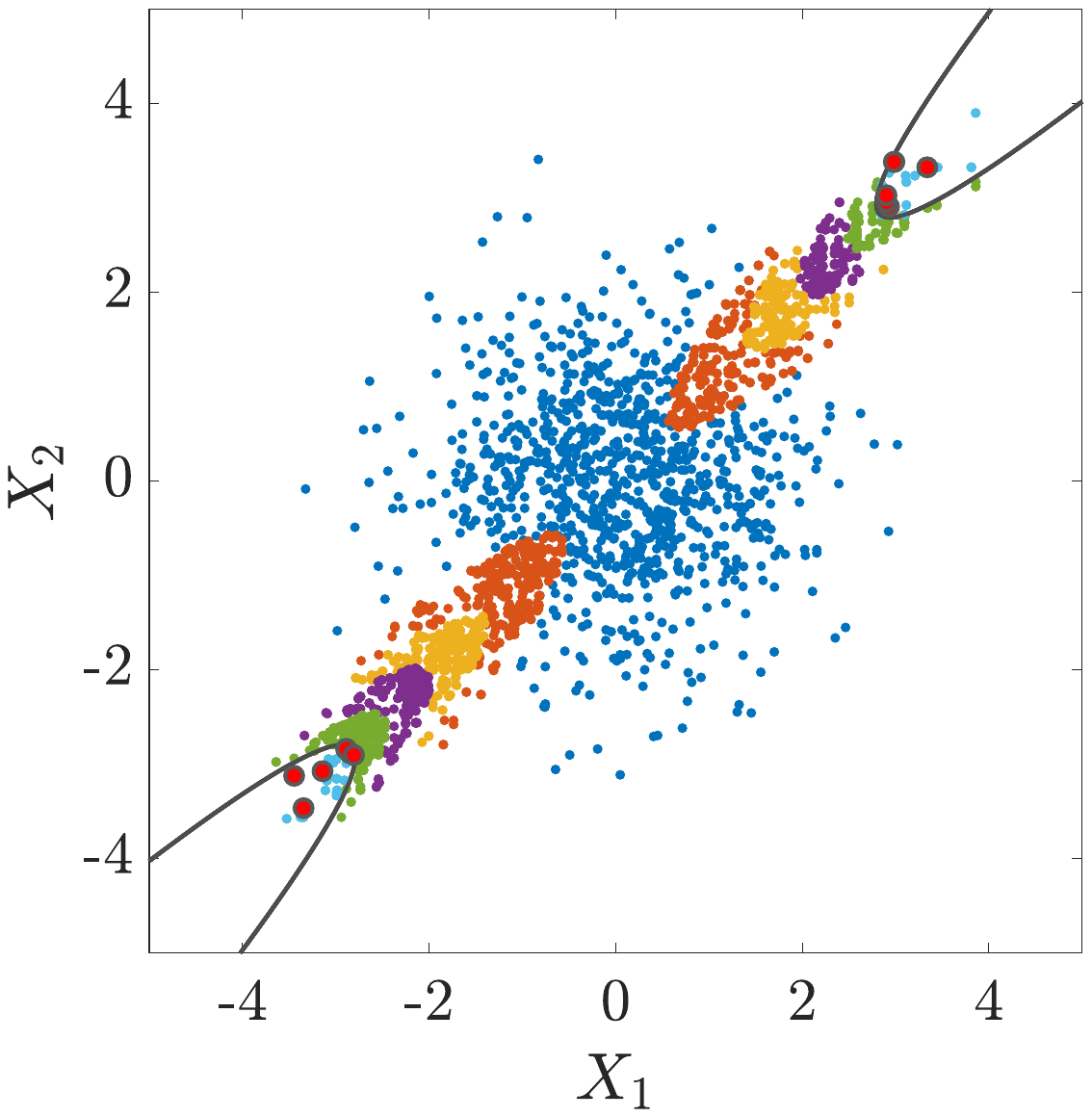}
\vspace*{-0.25in}\\\hspace*{0.02in}
     {\fontsize{8.5pt}{7.2} (a) }&
    \hspace*{-0.02in}
    {\fontsize{8.5pt}{7.2} (b) } &
    \hspace*{-0.13in}
    {\fontsize{8.5pt}{7.2} (c) } 
 \end{tabular}
      \caption{(a) and (b) visualize uniformly selected seeds (shown in red) for the two cases depicted in \Cref{fig: Discovery_sigma} and \Cref{fig: Discovery_2sigma}, respectively. These seeds are used to initialize the pCN chains but, unlike Subset Simulation (SuS), are not utilized in estimating the sought probability.  (c) shows the sampling evolution of the SuS method, emphasizing the computational efficiency of the proposed discovery stage, marked by utilizing significantly fewer samples just in order to discover the rare event domain.}
 \label{fig: Discovery_Uniform}\vspace{-0.15in}
\end{figure}

Let $\mathbf{X}^\mathcal{F} = \{\bm{x}_1, \bm{x}_2, \ldots, \bm{x}_{N_\mathcal{F}}\}$, with $N_\mathcal{F} \geq N_{S}$, be the rare event sample set obtained in \Cref{alg: Discovery}. The probability of selecting $\bm{x}_i \in \mathbf{X}^\mathcal{F}$ as a pCN seed can be given by:
\begin{equation}\vspace{-0.05in}
    \text{Pr}(\bm{x}_i)=\dfrac{\tilde{h}(\bm{x}_i)}{\sum_{j}^{N_\mathcal{F}}\tilde{h}(\bm{x}_j)}
    \label{eq: seeds_weight}
\vspace{-0.05in}\end{equation}
where \( \tilde{h}(\bm{x}_i) \) is the value of the target distribution in ASTPA at sample $\bm{x}_i$.  Then, $N_{\text{chain}} $ pCN seeds can be sampled from the set $\mathbf{X}^\mathcal{F}$  without replacement, using the selection probability $\text{Pr}(\bm{x}_i)$. This approach ensures that pCN seeds are more likely to be selected from different rare event modes, based on their statistical weights. \Cref{fig: Discovery_sigma}(e) and \Cref{fig: Discovery_2sigma}(d) illustrate the selected pCN seeds using this weighted selection method. While this technique is effective for generating multimodal seeds in low-dimensional cases, as demonstrated in the low-dimensional examples in this work, it may not scale well in high-dimensional, multimodal spaces due to the sensitivity of the high-dimensional target distribution values, even for relatively small perturbations. In such cases, we recommend using uniform sampling to select pCN seeds, which ensures a broader exploration of high-dimensional, multimodal rare event domains. This uniform sampling method is successfully adopted in the high-dimensional examples in this work. Two examples of this uniform sampling are shown in \Cref{fig: Discovery_Uniform}(a) and \Cref{fig: Discovery_Uniform}(b), illustrating that the selected seeds are located in the rare event domain but do not follow any relevant target distribution ($\pi_{\bm{X}}$ or $\tilde{h}$). The computational efficiency of the proposed discovery approach, which requires significantly fewer model evaluations, just in order to identify the relevant domains, compared to conventional sequential methods utilized for probability estimation, is highlighted by comparing its performance in \Cref{fig: Discovery_Uniform}(a) and \Cref{fig: Discovery_Uniform}(b) with the sampling evolution of the Subset Simulation (SuS) method, as shown in \Cref{fig: Discovery_Uniform}(c).

\section{Summary of Gradient-free ASTPA}\label{sec: pCN_ASTPA_summary}
\noindent The proposed guided pCN-based ASTPA framework can be applied as follows:
\begin{enumerate}
    \item Construct the approximate sampling target $\tilde{h}$ in \Cref{eq: Tar_ASTPA}, where its two parameters $\sigma$ and $g_c$ are determined as recommended in \Cref{sec: Targ_form}.

    \item Run the rare event discovery stage as described in \Cref{alg: Discovery} to obtain $N_{\text{chain}}$ initial seeds for the pCN sampler. $N_\text{level} = 300$, $p_0=\{0.1, 0.2\}$, and $\epsilon = 4$ typically work effectively in \Cref{alg: Discovery}, with $p_0 = 0.2$ being more suitable for high-dimensional problems. Denote the total number of model calls in this stage as $N_\text{discovery}$.

    \item Generate $N_{\text{chain}}$ pCN chains, as described in \Cref{alg: pCN}, with a number of steps, $L_{\text{chain}}$, starting from the seeds obtained in the previous step, to get a sample set $\{\bm{x}_{i}\}_{i=1}^N \sim h$. A burn-in phase with a length of $10\%\, \text{of}\, L_\text{chain}$ is considered, with $N_{\text{burn}}$ denoting the total number of burn-in samples, i.e, $N = \left\lceil0.9 \, N_{\text{chain}} \cdot L_{\text{chain}} \right\rceil$, and  $N_{\text{burn}} = \left\lfloor  0.1 \, N_{\text{chain}} \cdot L_{\text{chain}} \right\rfloor$. The required number of chains, $N_{\text{chain}}$, typically ranges between 5 and 25, with higher numbers recommended for multimodal cases. 

    \item Compute the expected value of the weighted indicator function, $\hat{\tilde{p}}_\mathcal{F}$, using \Cref{eq: p_f_tilde_ASTPA}, based on $\{\bm{x}_{i}\}_{i=1}^{N}$.

    \item Apply Inverse Importance Sampling (IIS) to compute the normalizing constant of $\hat{C}_h$. This process initiates by fitting a GMM, $Q(\boldsymbol{\bm{X}})$, based on $\{\bm{x}_{i}\}_{i=1}^{N}$, with the $Q(\boldsymbol{\bm{X}})$ structured as recommended in \Cref{sec: IIS}. An i.i.d. sample set $\{\bm{x}_i^\prime\}_{i=1}^{M}$ is then drawn from $Q(\boldsymbol{\bm{X}})$ to compute $\hat{C}_h$ using \Cref{eq: IIS_est_hat}. $M$ can be generally around $30\%$ of $N$.

    \item Compute the sought rare event probability as $\hat{p}_\mathcal{F}= \hat{\tilde{p}}_\mathcal{F} \,\, \hat{C}_h $.
\end{enumerate}
A target total computational effort of model calls, $N_{\text{total}}=N_{\text{discovery}}+N_{\text{burn}}+N+M$, can be attributed to the different stages according to all the aforementioned steps. The convergence of the estimator and/or any additional needed model calls can then be checked through \Cref{eq: var_p_f_c_h_each}. 

\section{Numerical Results}\label{sec: Numerical_results}

This section studies several numerical examples to examine the performance and efficiency of the proposed methods. In all examples, the guided pCN-based ASTPA is implemented as summarized in \Cref{sec: pCN_ASTPA_summary}. The results are compared with state-of-the-art gradient-free algorithms, including the improved Cross Entropy-based Importance Sampling (iCE-IS) algorithm \citep{papaioannou2019improved}, the Sequential Importance Sampling (SIS) method \citep{papaioannou2016sequential}, and the adaptive Conditional Sampling Subset Simulation variant (aCS-SuS) \citep{papaioannou2015mcmc}, all of which have well documented, publicly available code implementations \citep{PapaMatlab}. Comparisons are made in terms of accuracy and computational cost, based on $500$ independent simulations. Specifically, for each simulation $j$, we record the target probability estimate $(\hat{p}_\mathcal{F})^j$, the total number of model calls $(N_{\text{total}})^j$, and the analytical Coefficient of Variation of ASTPA, $(\text{C.o.V-Anal})^j$, computed as discussed in \Cref{sec: AnalCOV}. $N_{\text{total}}$ is defined for ASTPA in \Cref{sec: pCN_ASTPA_summary}. Then, we report the mean of the rare event probability estimates, $\mathop{\mathbb{E}}[\hat{p}_\mathcal{F}]$, the mean of the total number of model calls $\mathop{\mathbb{E}}[N_{\text{tota}l}]$, and the sampling C.o.V, computed as:
\begin{equation}
    \textrm{C.o.V}=\dfrac{\sqrt{\widehat{\text{Var}}(\hat{p}_\mathcal{F})}}{\mathop{\mathbb{E}}[\hat{p}_\mathcal{F}]},\quad \widehat{\text{Var}}(\hat{p}_{\mathcal{F}})=\dfrac{1}{500-1}\,\sum_{j=1}^{500}\big((\hat{p}_\mathcal{F})^j-\mathop{\mathbb{E}}[\hat{p}_\mathcal{F}]\big)^2
\end{equation}
The mean of the analytical C.o.V, $\mathbb{E}[\text{C.o.V-Anal}]$, is also reported in parentheses. The total number of limit-state function evaluations  $N_{\text{total}}$ for ASTPA has been determined based on achieving C.o.V values $ \in [0.1,\, 0.35]$. Reference rare event probabilities are computed based on $100$ independent standard Monte Carlo simulations (MCs), as described in \Cref{sec: rare_event_est}, using $10^{7}$-$10^{9}$ samples, as appropriate. The problem dimensionality is denoted by $d$, and all ASTPA parameters are carefully chosen for all examples but are not optimized. 

\subsection{Example 1: Nonlinear bimodal convex limit-state function} \label{sec: Bimodal_nonlinear}
The first example involves a bimodal nonlinear  limit-state function characterized by two independent standard normal random variables:
\begin{equation}
\begin{aligned}
g(\bm{X}) = \min \left[4 - \frac{1}{\sqrt{2}}\ (X_{1}+X_{2}) + 2.5\ (X_{1}-X_{2})^{2}, \,\, 4 + \frac{1}{\sqrt{2}}\ (X_{1}+X_{2}) + 2.5\ (X_{1}-X_{2})^{2}\right]
\end{aligned}\label{eq: Bimodal_}
\end{equation}
The approximate sampling target, $\tilde{h}$, in ASTPA is constructed using a likelihood dispersion factor $\sigma= 0.3$ and a scaling constant $g_c=1$, as discussed in \Cref{sec: Targ_form}. The discovery stage then starts using $N_{\text{level}}=300$ and $p_0=0.1$, yielding 10 rare event seeds for the pCN sampler ($N_{\text{chain}} = 10$), with each chain running for $L_{\text{chain}}=150$ iterations. Subsequently, inverse importance sampling (IIS) is applied using $M=300$ samples. \Cref{Table: EX1} compares the performance of the guided pCN-ASTPA with iCE-IS, SIS, and aCS-SuS. As shown in \cref{Table: EX1}, the guided pCN-based ASTPA provides an $\mathbb{E}[\hat{p}_{\mathcal{F}}]$ estimate, closely matching the reference probability obtained through standard Monte Carlo simulation (MCS). Compared to the other methods considered, the guided pCN-based ASTPA demonstrates improved performance, achieving a lower C.o.V using fewer total model calls, $\mathbb{E}[N_{\text{total}}]$. Notably, iCE-IS also demonstrates strong efficiency, outperforming both SIS and aCS-SuS in this 2D example. The excellent agreement between the sampling C.o.V and analytical C.o.V, as reported in parentheses, confirms the accuracy and effectiveness of the ASTPA analytical C.o.V expression. Since this example was introduced earlier in the paper to illustrate the ASTPA framework and the discovery stage, the steps and performance of the proposed framework are visualized in \Cref{fig: ASTPA_framework,fig: Discovery_2sigma}. 

\begin{table}[t!]
\captionsetup{justification=centering}
\vspace*{-0.5in}
\caption{Example 1: Performance of various methods for the nonlinear bimodal limit-state function.}
\centering
\footnotesize
\setlength\tabcolsep{4pt}
\begin{tabular}{p{1.0cm}ccccccccc}
  \toprule[1.5pt]
  \multirow{2}{*}{\Cref{eq: Bimodal_}}& \multirow{2}{*}{\textbf{500 Simulations}} & \multirow{2}{*}{\textbf{MCS}}& \textbf{pCN-ASTPA}& \multirow{2}{*}{\textbf{iCE-IS}}& \multirow{2}{*}{\textbf{SIS}}& \multirow{2}{*}{\textbf{aCS-SuS}}\\ 
  \cline{4-4}
    \addlinespace[2pt]
  & && \multicolumn{1}{c}{($\sigma = 0.3, \, q = $ n.a.)} & &  &\\ 
 \cmidrule(lr){1-7}
         \multirow{3}{*}{\shortstack[l]{$d=2$}}\rule{0pt}{2.5ex}    &$\mathop{\mathbb{E}}[N_{\text{total}}]$ &1.00E9 & 2,373& 2,972 & 5,525& 5,741\\
      &C.o.V &0.01& 0.16$\color{ForestGreen}($0.16$\color{ForestGreen})$  &0.18 & 1.18 & 0.72 \\
       &$\mathop{\mathbb{E}}[\hat{p}_{\mathcal{F}}]$    &9.47E-6& 9.46E-6& 9.52E-6  & 9.89E-6 & 9.66E-6 \\
      \bottomrule[1.5pt]
\end{tabular}\label{Table: EX1}
\end{table}

\newpage
\subsection{Example 2: Quartic bimodal limit-state function}
In the second example, we consider a quartic bimodal limit-state function, defined in the standard Gaussian space as:
\begin{equation}
\begin{aligned}
g(\bm{X}) = 6.5 - \frac{1}{\sqrt{2}}\ (X_{1}+X_{2}) - 2.5\ (X_{1}-X_{2})^{2}+\ (X_{1}-X_{2})^{4}\label{eq: EX2}
\end{aligned} 
\end{equation}

\begin{figure}[t!]

\centerline{\subfigure[]{\includegraphics[trim=0cm 3.5cm 0cm 3cm,width=0.31\textwidth]{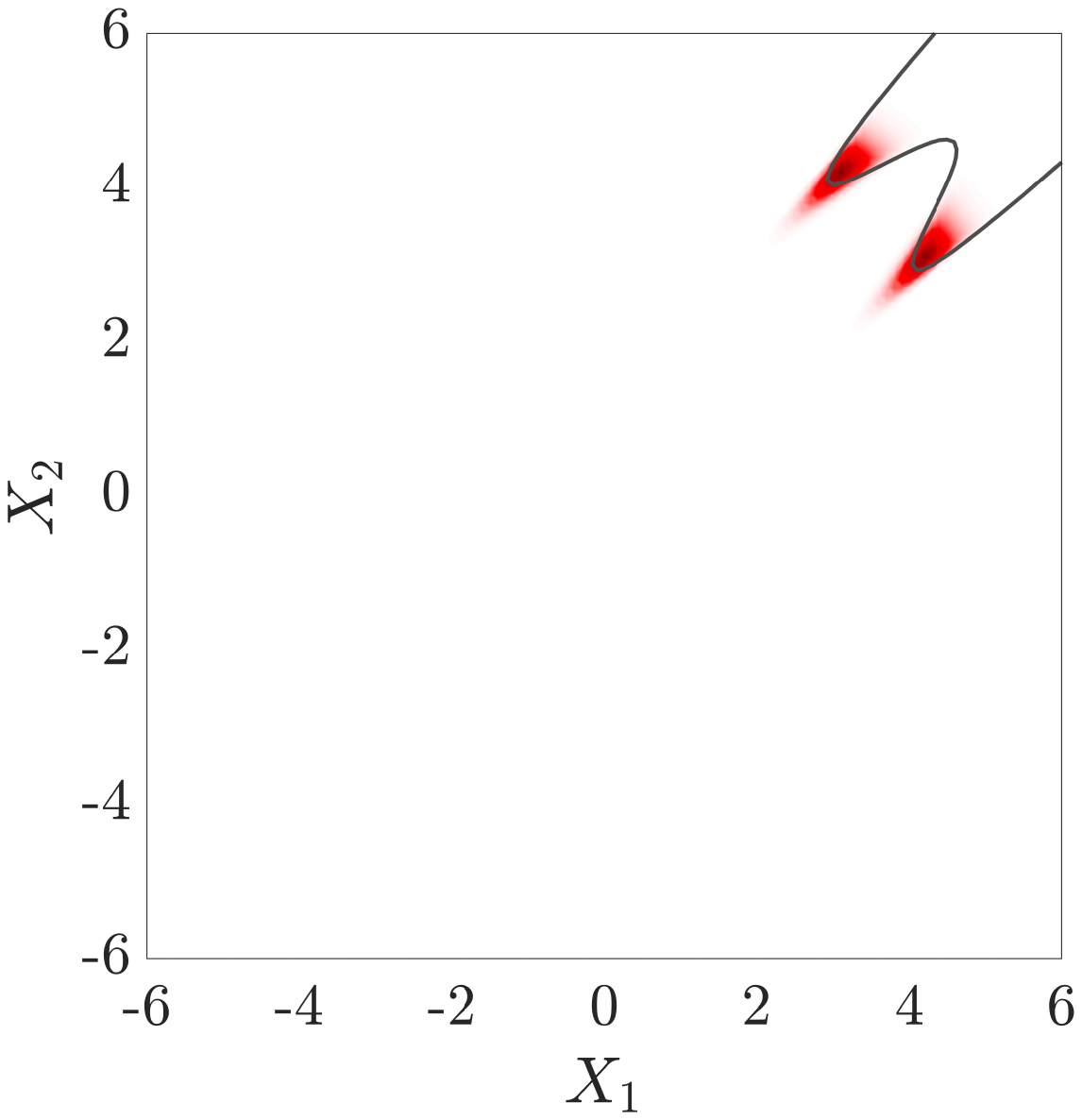}}\hspace{-0.2in}
\quad
\subfigure[]{\includegraphics[trim=0cm 3.5cm 0cm 3cm,width=0.31\textwidth]{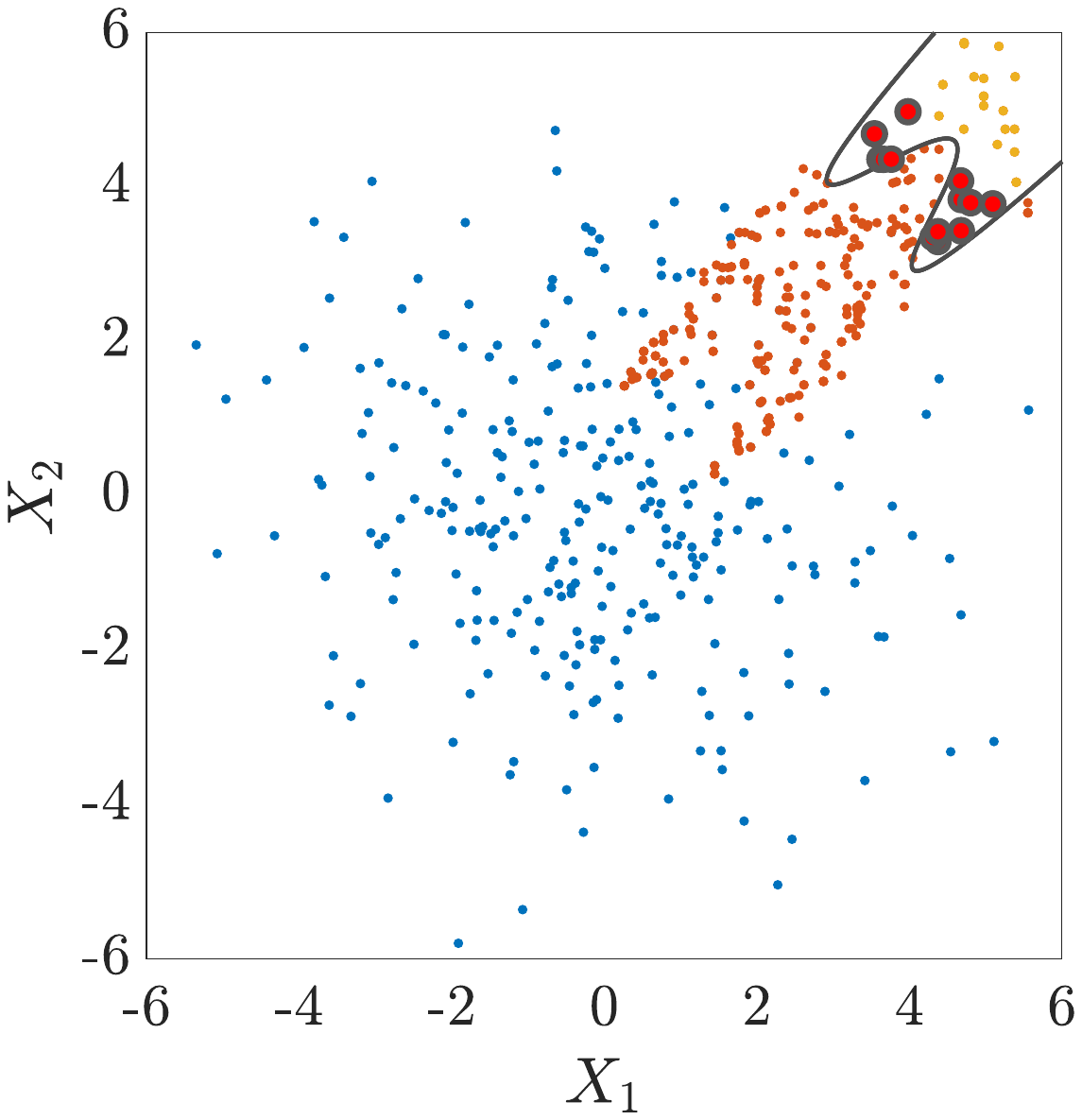}}\hspace{-0.2in}
\quad
\subfigure[]{\includegraphics[trim=0cm 3.5cm 0cm 3cm,width=0.31\textwidth]{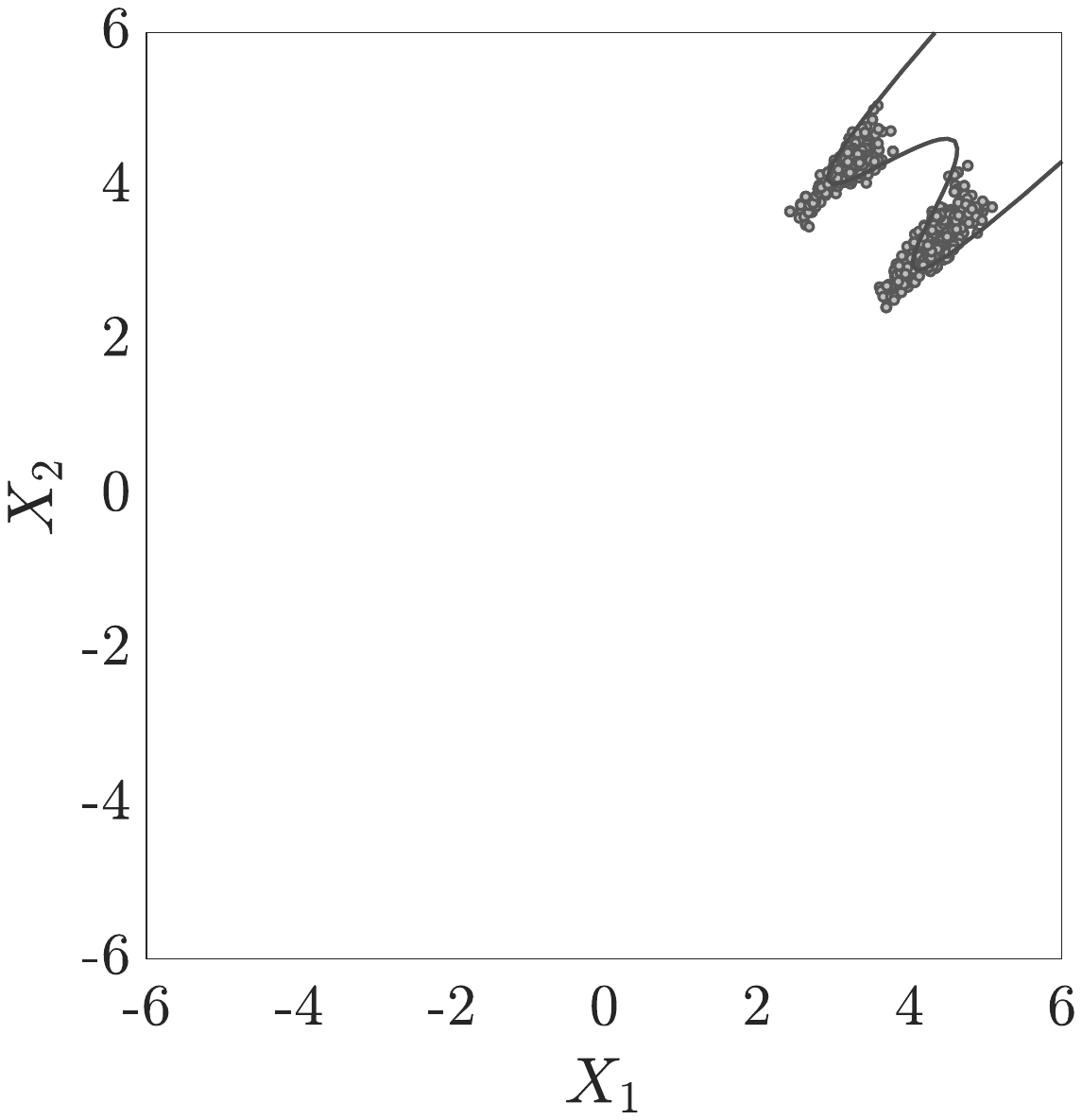}}}
  \captionsetup{labelfont={color=Black}}
\caption{Example 2: (a) The constructed approximate sampling target $\tilde{h}$, (b) the sampling evolution of the rare event domain discovery stage, with red points being the selected pCN seeds, and (c) samples from $\tilde{h}$, subsequently used to compute $\hat{\tilde{p}}_\mathcal{F}$ in \Cref{eq: p_f_tilde_ASTPA}. \vspace{-0.1in}}\label{fig: EX2}
\end{figure}

The approximate target $\tilde{h}$ is constructed with a likelihood dispersion factor of $\sigma = 0.2$ and a scaling constant $g_c = 1$. The discovery stage is applied using $N_{\text{level}} = 300$ and $p_0 = 0.1$, generating 18 rare event seeds for the pCN sampler ($N_{\text{chain}} = 18$), with each chain running for $L_{\text{chain}} = 120$ steps. As in the previous example, inverse importance sampling (IIS) is used with $M = 300$ samples. \Cref{fig: EX2} shows the evolution of the rare event domain discovery stage, highlighting the effectiveness of the proposed algorithm in accurately identifying and sampling rare event regions. 

\begin{table}[t!]
\captionsetup{justification=centering}
\caption{Example 2: Performance of various methods for the quartic bimodal limit-state function.}
\centering
\footnotesize
\setlength\tabcolsep{4pt}
\begin{tabular}{p{1.0cm}cccccccccc}
  \toprule[1.5pt]
  \multirow{2}{*}{\Cref{eq: EX2}}& \multirow{2}{*}{\textbf{500 Simulations}} & \multirow{2}{*}{\textbf{MCS}}& \textbf{pCN-ASTPA}& \multirow{2}{*}{\textbf{iCE-IS}}& \multirow{2}{*}{\textbf{SIS}}& \multirow{2}{*}{\textbf{aCS-SuS}}\\ 
  \cline{4-4}
    \addlinespace[2pt]
  & && \multicolumn{1}{c}{($\sigma = 0.2, \, q =$ n.a.)}  & &  &\\ 
 \cmidrule(lr){1-7}
         \multirow{3}{*}{\shortstack[l]{$d=2$}}\rule{0pt}{2.5ex}    &$\mathop{\mathbb{E}}[N_{\text{total}}]$ &1.00E9 & 3,165 & 5,000 &  10,750&7,945 \\
      &C.o.V &0.30& 0.12$\color{ForestGreen}($0.11$\color{ForestGreen})$ & 0.15 & 2.12 & 2.07 \\
       &$\mathop{\mathbb{E}}[\hat{p}_{\mathcal{F}}]$   & 5.91E-8 & 5.92E-8 & 5.80E-8 & 4.28E-8  & 5.79E-8 \\
      \bottomrule[1.5pt]
\end{tabular}\label{Table: Ex2}
\end{table}

The results, as summarized in \Cref{Table: Ex2},  show that the guided pCN-ASTPA significantly outperforms the other methods, achieving the lowest C.o.V,  while using the fewest model calls ($\mathop{\mathbb{E}}[N_{\text{total}}]$). This demonstrates the robustness and efficiency of the guided pCN-ASTPA framework, even in the presence of increased nonlinearity introduced through higher-order terms in the limit-state function. The guided pCN-based ASTPA also provides $\mathop{\mathbb{E}}[\hat{p}_{\mathcal{F}}]$ estimates in close agreement with the reference probability computed by Monte Carlo simulation (MCS). The very good agreement between the sampling and analytical C.o.V values (reported in parentheses) further confirms the reliability of the analytical C.o.V expression for ASTPA in this more challenging problem.

\subsection{Example 3: The Himmelblau function}\label{sec: Him}
\begin{figure}[t!]
\vspace*{-0.7in}
    \centerline{\subfigure[]{\includegraphics[trim=0cm 3.5cm 0cm 3cm,width=0.31\textwidth]{Figures/Him_target.pdf}}\hspace{-0.2in}
            \quad
        \subfigure[]{\includegraphics[trim=0cm 3.5cm 0cm 3cm,width=0.31\textwidth]{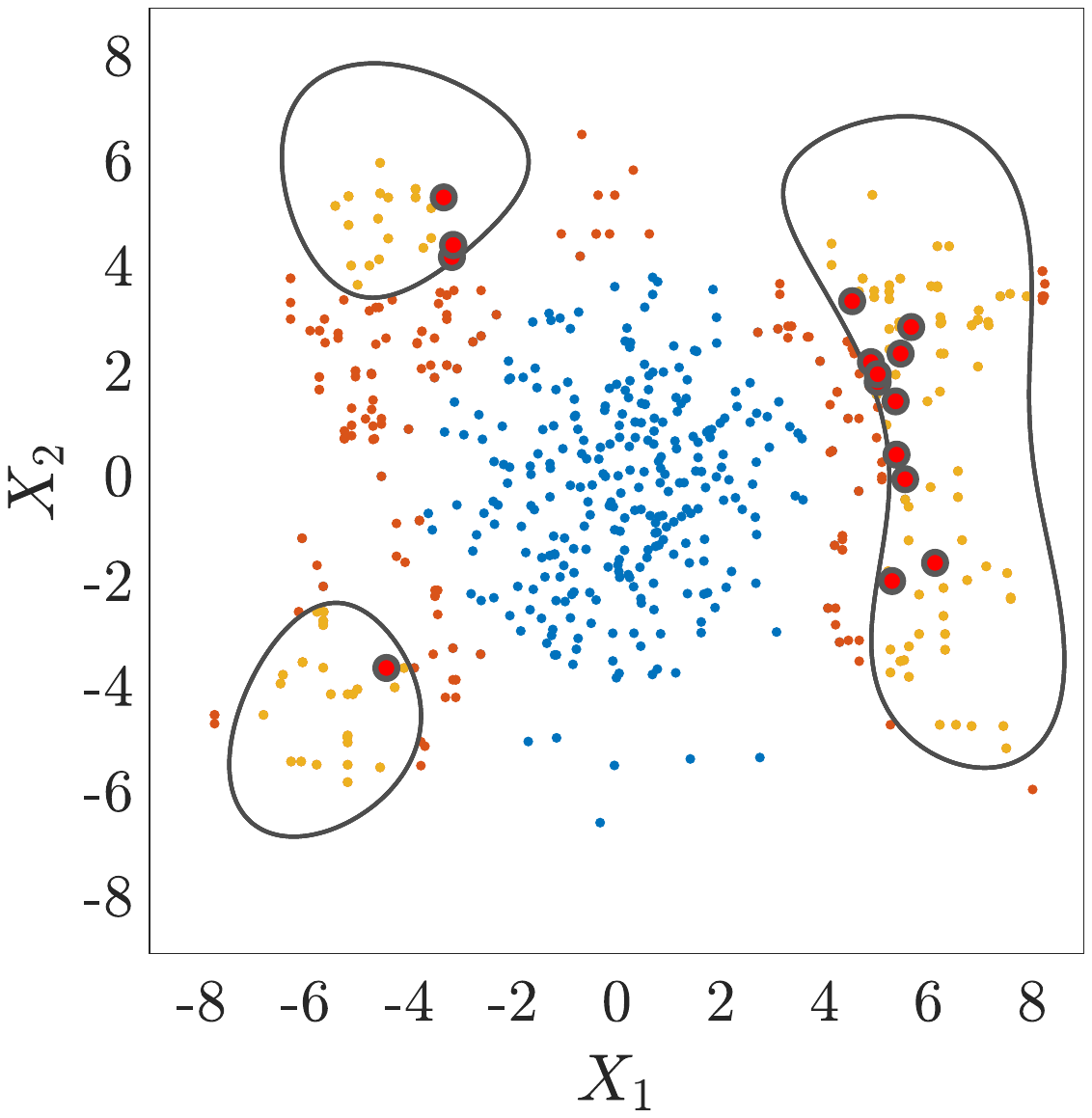}}\hspace{-0.2in}
            \quad
        \subfigure[]{\includegraphics[trim=0cm 3.5cm 0cm 3cm,width=0.31\textwidth]{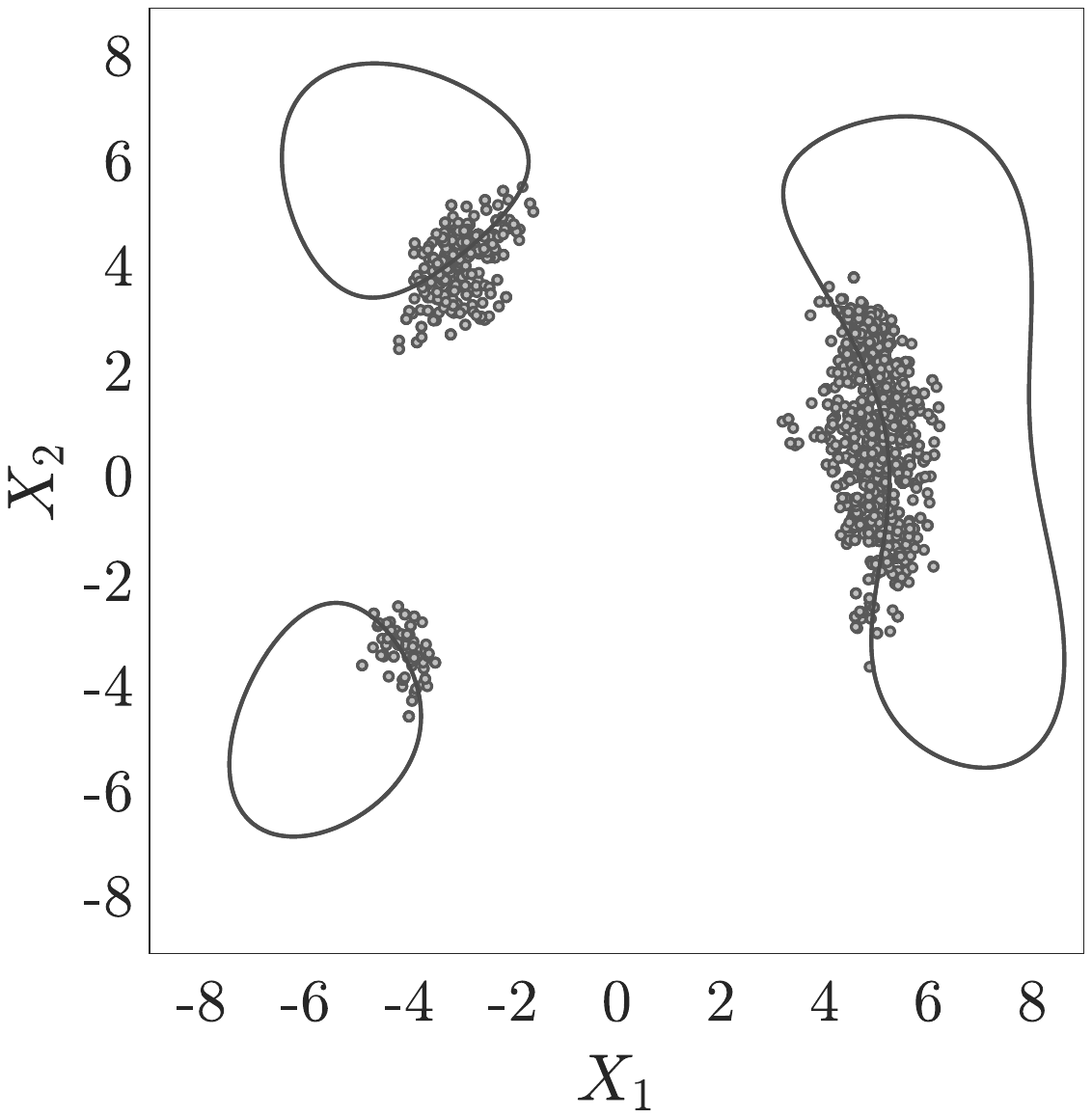}}}
\caption{Example 3: (a) The constructed approximate sampling target $\tilde{h}$, (b) the sampling evolution of the rare event domain discovery stage, with red points representing the selected pCN seeds, and (c) samples from $\tilde{h}$.}\label{fig: EX3}
\end{figure}

The Himmelblau function \citep{himmelblau1972applied} is a commonly used fourth-order polynomial test function in nonlinear optimization. In this example, we modify the Himmelblau function to suit rare event scenarios, particularly those with multiple separated rare event domains. The modified limit-state function is defined in the standard Gaussian space as:
\begin{equation}
    g(\bm{X}) =  \bigg (\dfrac{(0.75X_{1} - 0.5)^{2}}{1.81} + \dfrac{(0.75X_{2} - 0.5)}{1.81} - 11 \bigg )^{2} + \bigg (\dfrac{(0.75X_{1} - 1)}{1.81} + \dfrac{(0.75X_{2} - 0.5)^{2}}{1.81} - 7 \bigg )^{2} -50               
    \label{eq: EX3}
\end{equation}
where $X_{1}$ and $X_{2}$ are independent standard normal random variables. The approximate target $\tilde{h}$ is constructed using a likelihood dispersion factor $\sigma = 0.3$ and a scaling constant $g_c = \nicefrac{g(\bm{0})}{4}$, as described in \Cref{sec: Targ_form} and \Cref{fig: Effect_on_target_scale}. The discovery stage is applied using $N_{\text{level}} = 300$ and $p_0 = 0.1$, generating 16 rare event seeds for the pCN sampler, with each chain subsequently running for $L_{\text{chain}} = 160$ iterations. Similar to the previous examples, inverse importance sampling (IIS) is applied using $M = 300$ samples. \Cref{fig: EX3} illustrates the constructed target distribution, the rare event domain discovery process, and the pCN samples, effectively exploring all rare event modes. The results in \Cref{Table: EX3} show that the guided pCN-ASTPA outperforms the other methods, achieving a lower C.o.V and fewer total model evaluations. This demonstrates the robustness and efficiency of the guided pCN-ASTPA framework, particularly in handling complex, multi-modal rare event domains. The $\mathop{\mathbb{E}}[\hat{p}_{\mathcal{F}}]$ estimate produced by the guided pCN-based ASTPA closely matches the reference probability, further confirming the method's accuracy.

\begin{table}[t!]
\captionsetup{justification=centering}
\caption{Example 3: Performance of various methods for the Himmelblau limit-state function.}
\centering
\footnotesize
\setlength\tabcolsep{4pt}
\begin{tabular}{p{1.0cm}cccccccccc}
  \toprule[1.5pt]
  \multirow{2}{*}{\Cref{eq: EX3}}& \multirow{2}{*}{\textbf{500 Simulations}} & \multirow{2}{*}{\textbf{MCS}}& \textbf{pCN-ASTPA}& \multirow{2}{*}{\textbf{iCE-IS}}& \multirow{2}{*}{\textbf{SIS}}& \multirow{2}{*}{\textbf{aCS-SuS}}\\ 
  \cline{4-4}
    \addlinespace[2pt]
  & && \multicolumn{1}{c}{($\sigma = 0.3, \, q =4$)}& &  &\\ 
 \cmidrule(lr){1-7}           
    \multirow{3}{*}{\shortstack[l]{$d=2$}}\rule{0pt}{2.5ex}  &$\mathop{\mathbb{E}}[N_{\text{total}}]$ &1.00E9 &3,430 &6,466 & 6,216& 7,101 \\
       & C.o.V & 0.05&  0.18$\color{ForestGreen}($0.17$\color{ForestGreen})$& 0.37 & 0.30 & 0.53\\
       &$\mathop{\mathbb{E}}[\hat{p}_{\mathcal{F}}]$   &2.81E-7& 2.82E-7 &2.36E-7 & 2.21E-7 & 3.02E-7  \\
      \bottomrule[1.5pt]
\end{tabular}\label{Table: EX3}
\end{table}


\subsection{Example 4: Limit-state surface with changing topological parameter space.} \label{sec: Changing topo}
\begin{figure}[t!]
\vspace*{-0.7in}
\centerline{\subfigure[]{\includegraphics[trim=0cm 3.5cm 0cm 3cm,width=0.3\textwidth]{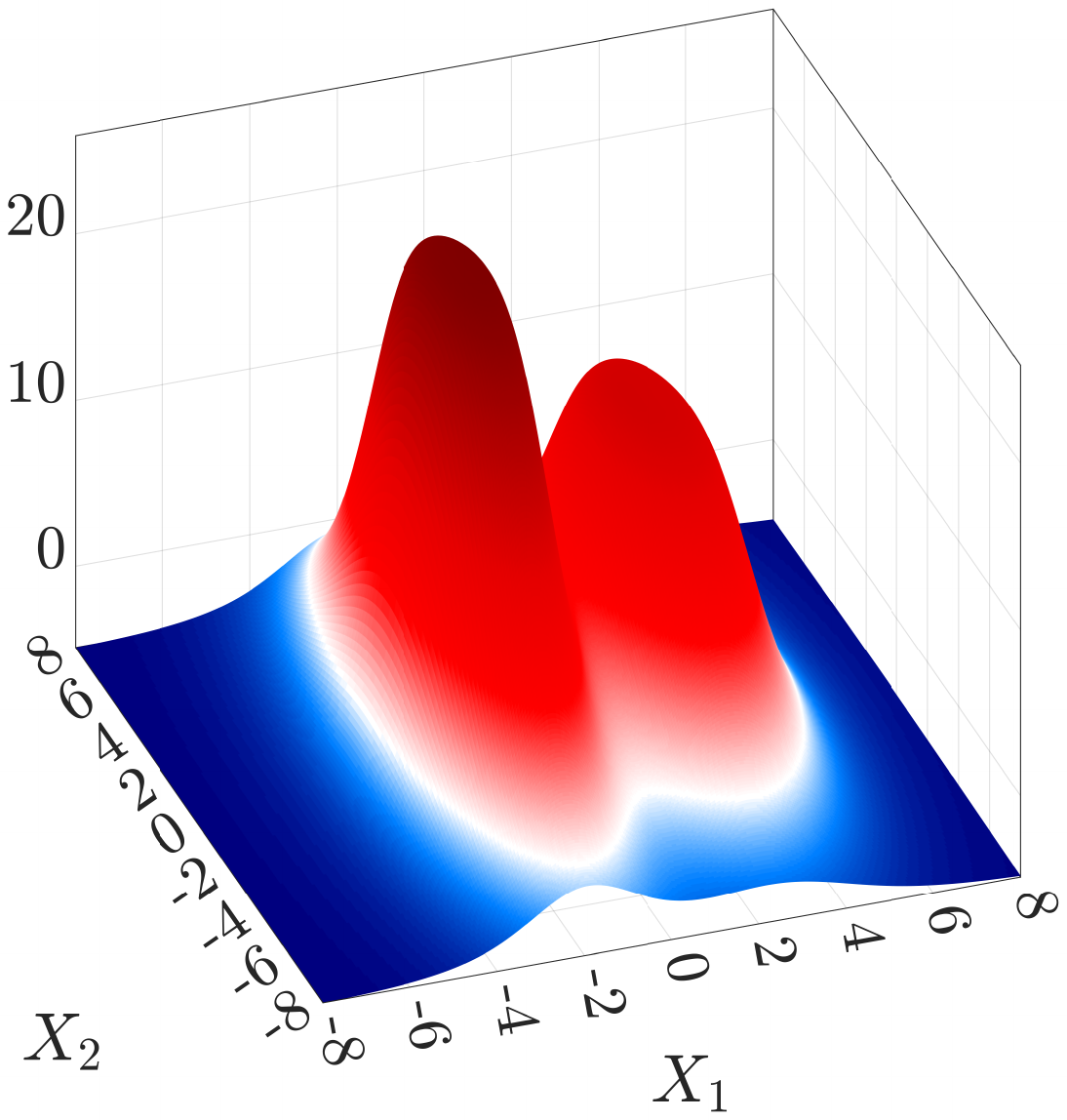}}\hspace{-0.2in}
\quad
\subfigure[]{\includegraphics[trim=0cm 3.5cm 0cm 3cm,width=0.3\textwidth]{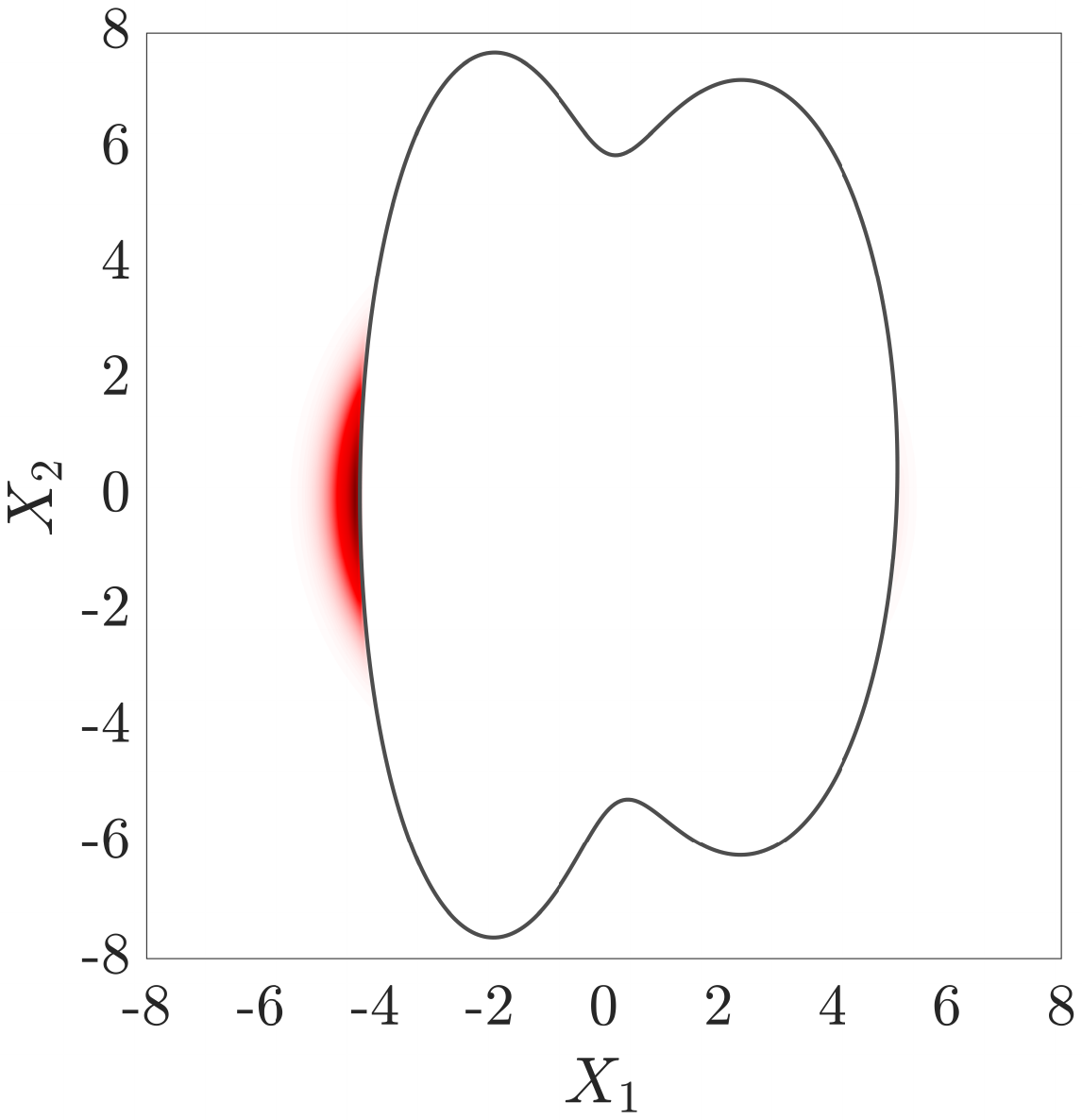}}\hspace{-0.2in}
\quad
\subfigure[]{\includegraphics[trim=0cm 3.5cm 0cm 3cm,width=0.3\textwidth]{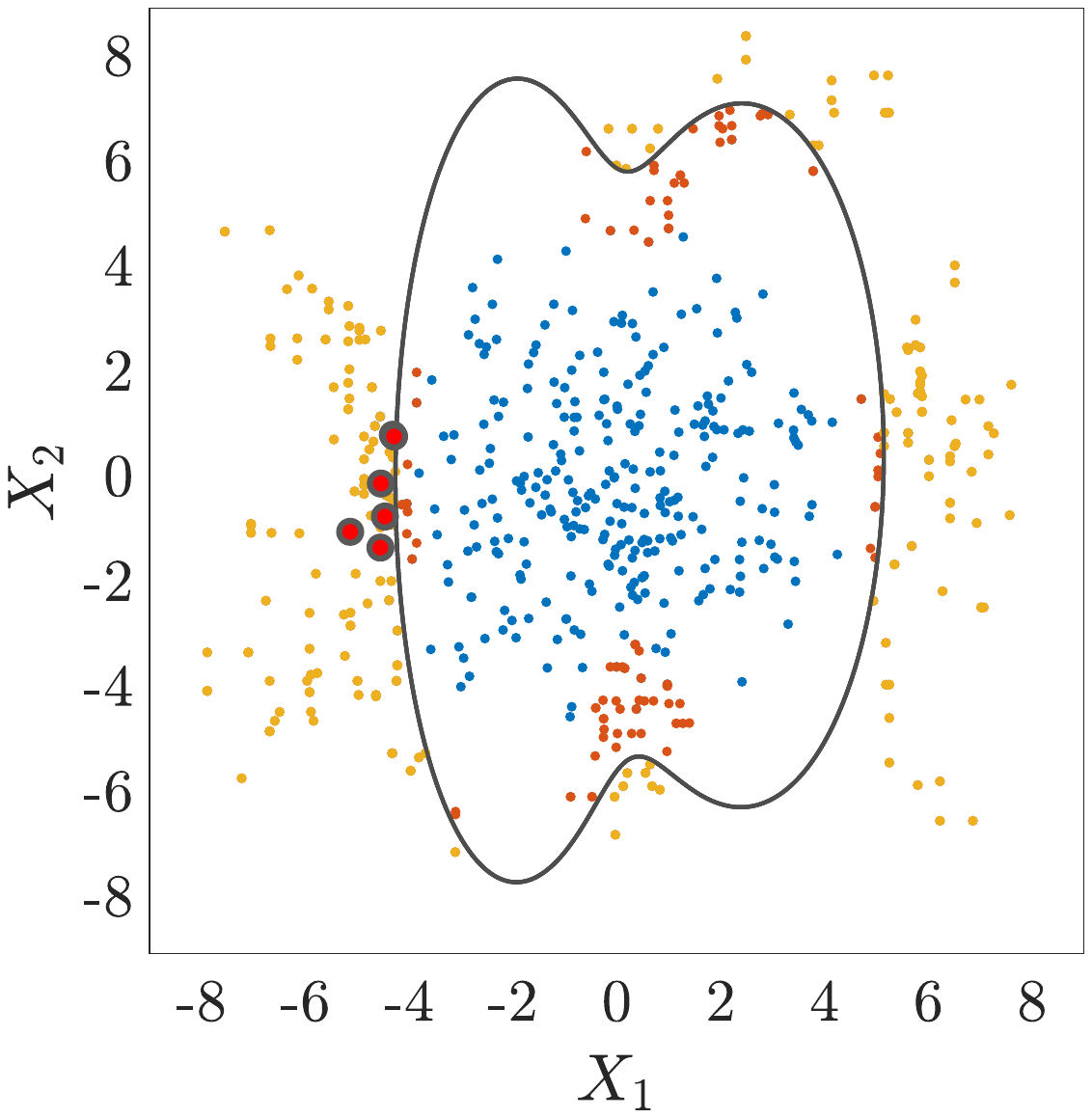}}\hspace{-0.2in}}
\vspace{-0.1in}
\centerline{\subfigure[]{\includegraphics[trim=0cm 3.5cm 0cm 3cm,width=0.3\textwidth]{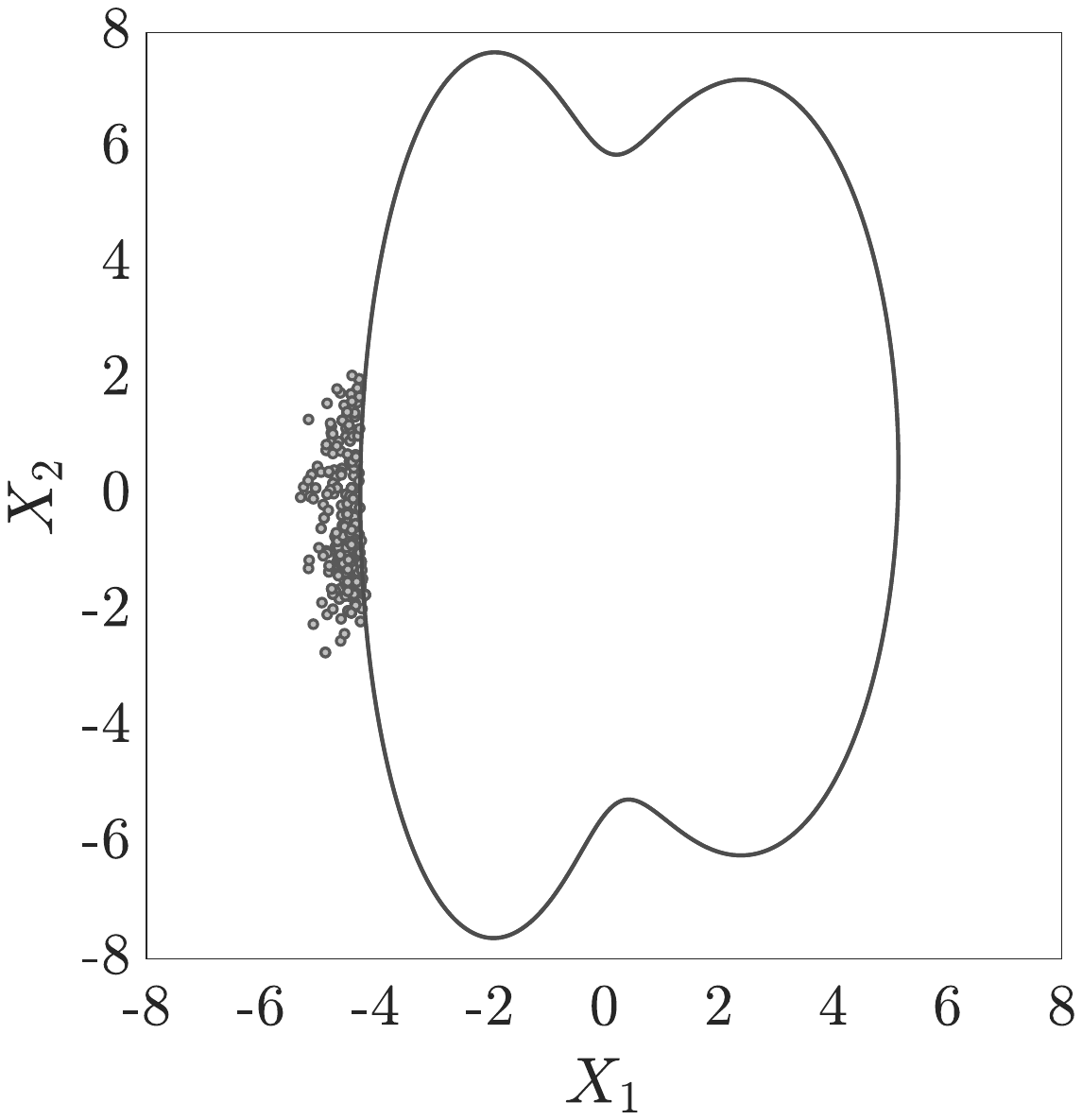}}\hspace{-0.2in}
\quad
\subfigure[]{\includegraphics[trim=0cm 3.5cm 0cm 3cm,width=0.3\textwidth]{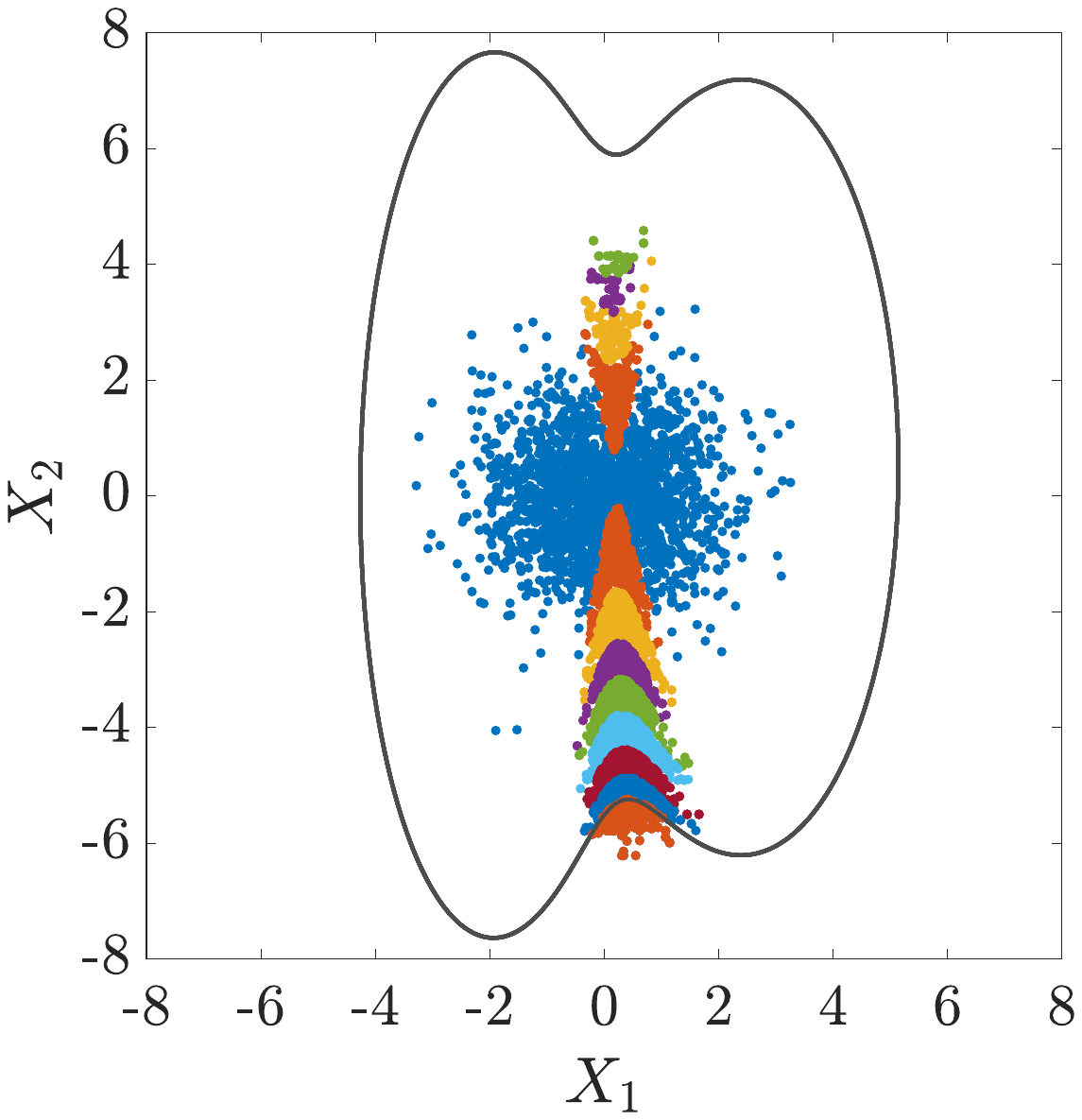}}}
  \captionsetup{labelfont={color=Black}}
\caption{Example 4: (a) The surface of the limit-state in \Cref{eq: Chang_topo}, showing changing topology, (b) the constructed ASTPA sampling target $\tilde{h}$, capturing the influential rare event mode, (c) the sampling evolution of the rare event domain discovery stage, with red points representing the selected pCN seeds, (d) samples generated from $\tilde{h}$, and (e) the sampling evolution of SuS, which fails to identify the rare event mode with a competitive number of model calls, demonstrating the limitations of sequential methods in such scenarios.}\label{fig: Chang_topo}
\vspace*{-0.2in}\end{figure}

This example examines a limit-state function with significant topological changes, making it a challenging scenario for rare event sampling methods, as described in \citep{breitung2019geometry}. The function is defined as:
\begin{equation}
    g(\bm{X}) =30\left[\bigg(\dfrac{4(X_1+2)^2}{9}+\dfrac{X_2^2}{25}\bigg)^2+1 \right]^{-1}
+20\left[\bigg(\dfrac{(X_1-2.5)^2}{4}+\dfrac{(X_2-0.5)^2}{25}\bigg)^2+1 \right]^{-1}-5
\label{eq: Chang_topo}
\end{equation}
where $X_1$ and $X_2$ are independent standard normal random variables. This limit-state surface is particularly challenging due to its abrupt changes in topology, as shown in \Cref{fig: Chang_topo}(a),  requiring a robust sampling framework to accurately capture the rare event domain.

The guided pCN-ASTPA algorithm is applied with a likelihood dispersion factor of $\sigma = 0.1$ and a scaling constant $g_c = \nicefrac{g(\mathbf{0})}{5}$, as outlined in \Cref{sec: Targ_form}. The discovery stage uses $N_{\text{level}} = 300$ and $p_0 = 0.1$, generating 6 rare event seeds for the pCN sampler, with each chain running for $L_{\text{chain}} = 100$ iterations. Inverse importance sampling (IIS) is employed with $M = 200$ samples. \Cref{fig: Chang_topo} visualizes the limit-state surface, the constructed sampling target, and the evolution of samples in the rare event domain discovery process, highlighting the effectiveness of guided pCN-ASTPA in accurately capturing the influential rare event region. In contrast, \Cref{fig: Chang_topo}(e) demonstrates the limitations of sequential methods like SuS, which struggle to efficiently identify the rare event mode with a competitive number of model evaluations. These rare event domain discovery limitations are more pronounced in iCE-IS in this example, which exhibits a significant error in the estimated mean of the rare event probability. As presented in \Cref{table: Changing_topo}, the guided pCN-ASTPA achieves an accurate rare event probability estimate, $\mathop{\mathbb{E}}[\hat{p}_{\mathcal{F}}]$, with significantly fewer model evaluations and a lower Coefficient of Variation (C.o.V) compared to the other methods. This underscores the effectiveness of the pCN-ASTPA framework in tackling problems with complex topological changes, where traditional sequential methods often underperform.

\begin{table}[t!]
\captionsetup{justification=centering}
\caption{Example 4: Performance of various methods for the limit-state surface with changing topological parameter space.}
\centering
\footnotesize
\setlength\tabcolsep{4pt}
\begin{tabular}{p{1.0cm}cccccccccc}
  \toprule[1.5pt]
  \multirow{2}{*}{\Cref{eq: Chang_topo}}& \multirow{2}{*}{\textbf{500 Simulations}} & \multirow{2}{*}{\textbf{MCS}}& \textbf{pCN-ASTPA}& \multirow{2}{*}{\textbf{iCE-IS}}& \multirow{2}{*}{\textbf{SIS}}&\multirow{2}{*}{\textbf{aCS-SuS}}\\ 
  \cline{4-4}
    \addlinespace[2pt]
  & && \multicolumn{1}{c}{($\sigma = 0.1, \, q =5$)} &&  &\\ 
 \cmidrule(lr){1-7}
         \multirow{3}{*}{\shortstack[l]{$d=2$}}\rule{0pt}{2.5ex}    &$\mathop{\mathbb{E}}[N_{\text{total}}]$ &1.00E9 & 1,370 & 14,744 & 13,542 &14,597\\
      &C.o.V &0.01& 0.11$\color{ForestGreen}($0.13$\color{ForestGreen})$& 0.09 & 3.33 & 1.92 \\
       &$\mathop{\mathbb{E}}[\hat{p}_{\mathcal{F}}]$    & 1.13E-5 & 1.10E-5  &1.92E-8& 1.32E-5& 1.26E-5\\
      \bottomrule[1.5pt]
\end{tabular}\label{table: Changing_topo}
\end{table}

\subsection{Example 5: A thirty four-story structural example}

\begin{figure}[t!]
\vspace*{-0.5in}
 \centering
 \includegraphics[width=0.59\textwidth]{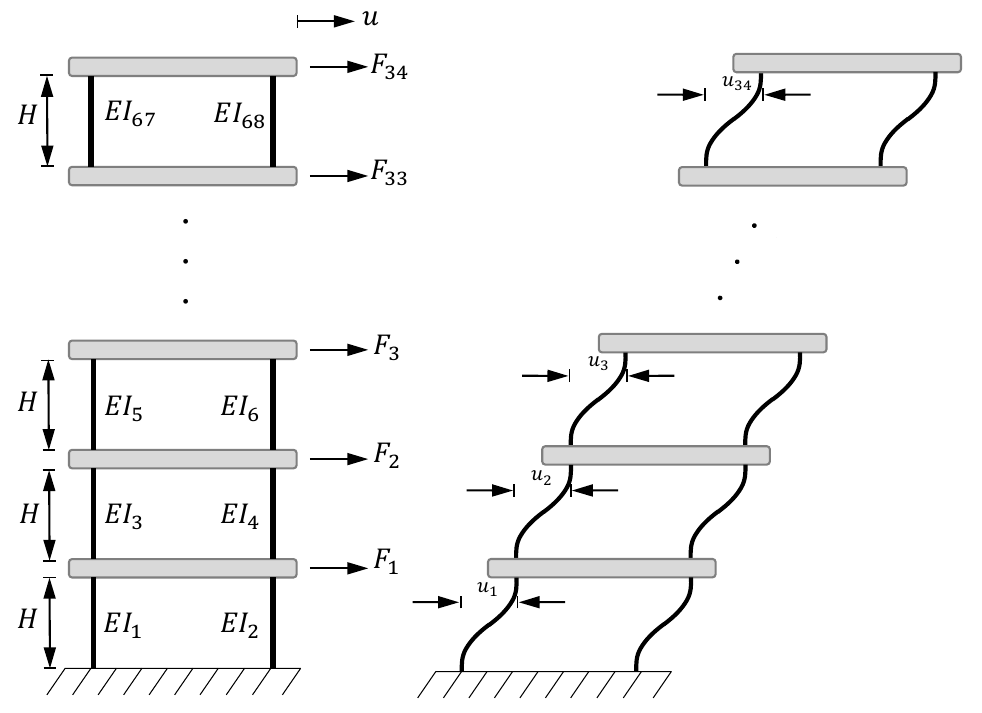}
 \caption{Example 5: A thirty four-story structure under static loads.}
 \label{fig: multistory_structure}
 \end{figure}

This example analyzes a thirty-four-story structure, adapted from \citep{bucher2009computational}, as represented in \Cref{fig: multistory_structure}. The structure is subjected to thirty four static loads $F_{i}$, for $i = 1, 2, \dots, 34$. The floor slabs/beams are assumed to be rigid, and all columns have the same height, $H = 4$ m, but different flexural stiffnesses $EI_{k}$, for $k = 1, 2, \dots, 68$. Both the loads and stiffnesses are treated as random variables, resulting in a total of $d = 102$ random variables. The loads are modeled as normally distributed with a mean of 2 kN and a C.o.V. of 0.4. Similarly, the stiffnesses are normally distributed with a mean of 20 MNm² and a C.o.V. of 0.2. The top story displacement, $u$, can be determined by summing the interstory displacements $u_{i}$, expressed as:
\begin{equation}
u_i = \dfrac{\left(\sum_{j=i}^{34} F_{j}\right) H^{3}}{12\left(\sum_{k=2i-1}^{2i} EI_{k}\right)}, \quad
g = Y_{0} - \sum_{i=1}^{34} u_{i}
\label{eq: multistory}
\end{equation}  
The limit-state function $g$ indicates failure (a rare event) when the top story displacement $u$ exceeds a threshold value $Y_{0}$, set to 0.22, 0.23, and 0.235 m for different cases, to examine different levels of failure probability. All random variables are transformed into the independent standard Gaussian space for the analysis. Since $g(\bm{0}) < 2$, as discussed in \Cref{sec: Targ_form}, scaling is performed with $g_c = \nicefrac{g(\mathbf{0})}{4}$. The discovery stage is performed with $N_{\text{level}} = 300$ and $p_0 = 0.2$ to effectively identify the rare event seeds for 5 pCN chains ($N_{\text{chains}} = 5$), each with a chain length of $1,000$ samples ($L_{\text{chain}} = 1,000$). The inverse importance sampling (IIS) is performed with $2,000$ resampled points ($M = 2,000$). \Cref{Table: multistory} highlights the outstanding performance of the guided pCN-ASTPA in this high-dimensional problem. Our framework consistently provides accurate failure probability estimates with a low Coefficient of Variation (C.o.V.), while requiring significantly fewer total model evaluations compared to the other methods considered.

\begin{table}[t!]
\captionsetup{justification=centering}
\caption{Example 5: Performance of various methods for the thirty four-story structure.}
\centering
\footnotesize
\setlength\tabcolsep{4pt}
\begin{tabular}{p{1.0cm}cccccccccc}
  \toprule[1.5pt]
  \multirow{2}{*}{\Cref{eq: multistory}}& \multirow{2}{*}{\textbf{500 Simulations}} & \multirow{2}{*}{\textbf{MCS}}& \textbf{pCN-ASTPA}& \multirow{2}{*}{\textbf{iCE-IS}}& \multirow{2}{*}{\textbf{SIS}}&  \multirow{2}{*}{\textbf{aCS-SuS}}\\ 
  \cline{4-4}
    \addlinespace[2pt]
  & && \multicolumn{1}{c}{($\sigma = 0.3, \, q = 4$)} & &\\ 
 \cmidrule(lr){1-7}
         \multirow{3}{*}{\shortstack[l]{$d=102$\\$Y_{0}=0.22$}}\rule{0pt}{2.5ex}    &$\mathop{\mathbb{E}}[N_{\text{total}}]$ &1.00E8& 7,540 & 7,570 &10,082 & 9,200\\
      &C.o.V &0.03& 0.14$\color{ForestGreen}($0.14$\color{ForestGreen})$& 0.55  &0.60 & 0.24 \\
       &$\mathop{\mathbb{E}}[\hat{p}_{\mathcal{F}}]$    & 2.41E-5 & 2.44E-5&  1.93E-5& 2.75E-5  & 2.42E-5\\
             \cmidrule(lr){1-7}
          
    \multirow{3}{*}{\shortstack[l]{$d=102$\\$Y_{0}=0.23$}}\rule{0pt}{2.5ex}  &$\mathop{\mathbb{E}}[N_{\text{total}}]$ &1.00E8& 7,540& 10,851 &11,504  & 11,470 \\
       & C.o.V  & 0.09 &  0.22$\color{ForestGreen}($0.20$\color{ForestGreen})$& 0.81 &0.63&  0.30\\
       &$\mathop{\mathbb{E}}[\hat{p}_{\mathcal{F}}]$   &1.22E-6 &1.22E-6& 0.83E-6&1.25E-6  &1.22E-6  \\
             \cmidrule(lr){1-7}
  
    \multirow{3}{*}{\shortstack[l]{$d=102$\\$Y_{0}=0.235$}}\rule{0pt}{2.5ex}  &$\mathop{\mathbb{E}}[N_{\text{total}}]$ &1.00E8&7,540& 11,833 &  12,782 & 12,815 \\
       & C.o.V  & 0.20&  0.27$\color{ForestGreen}($0.28$\color{ForestGreen})$& 0.85 &1.32&  0.35\\
       &$\mathop{\mathbb{E}}[\hat{p}_{\mathcal{F}}]$   &2.46E-7 &2.48E-7& 1.55E-7 &2.41E-7  &2.41E-7  \\
      \bottomrule[1.5pt]

\end{tabular}\label{Table: multistory}
\end{table}


\subsection{Example 6: Steel plate }

\begin{figure}[t!]
\vspace*{-0.5in}
 \centering
  \begin{tabular}{ccccc}
   \includegraphics[trim=0cm 0cm 0cm 3.5cm,width=.3\textwidth,keepaspectratio]{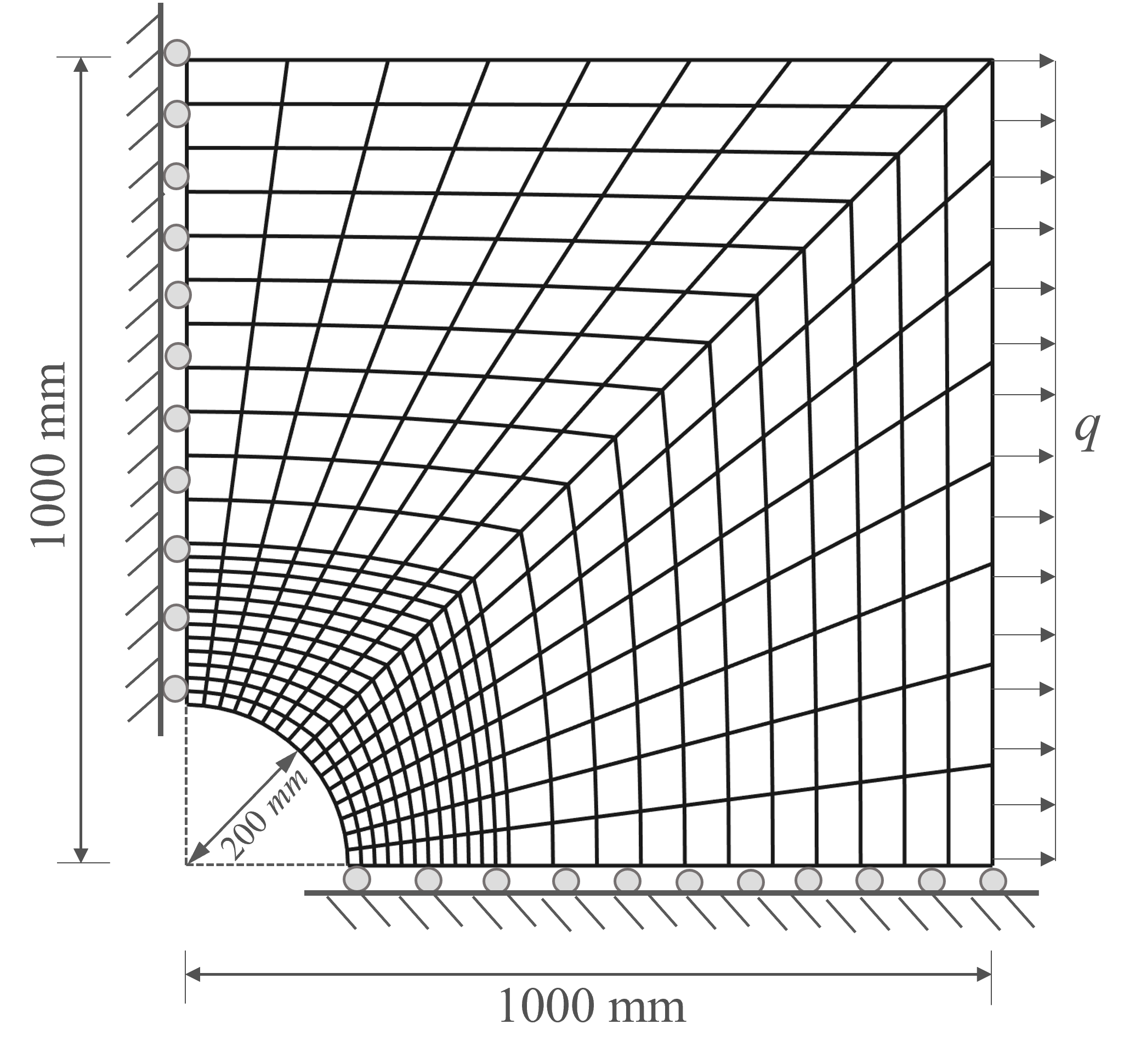}&
   \includegraphics[trim=0cm 3.3cm 0cm 3.5cm,width=.31\textwidth,keepaspectratio]{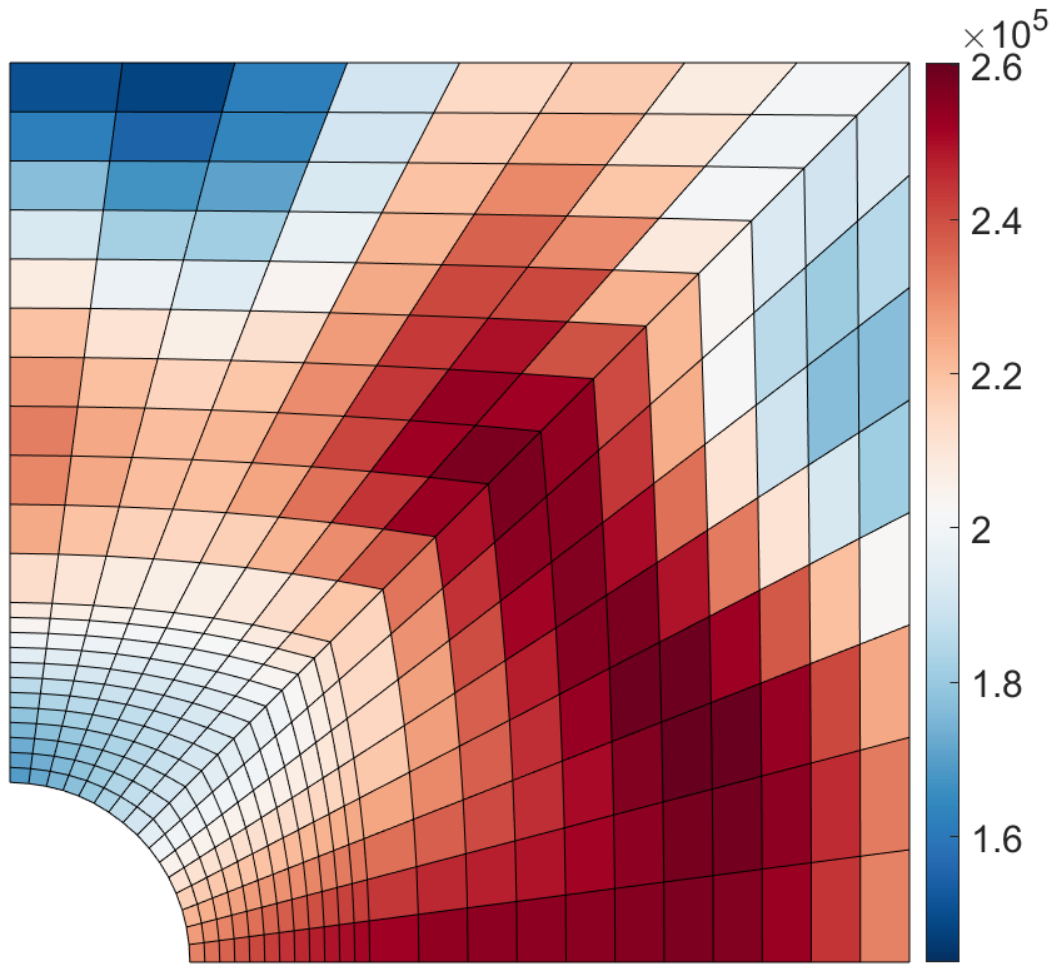}&\\
    {\fontsize{9.5pt}{9pt}\selectfont(a)}&
    {\fontsize{9.5pt}{9pt}\selectfont(b)}&
\\
  \vspace*{-0.1in}
 \end{tabular}
  \captionsetup{labelfont={color=Black}}
   \vspace*{-0.13in}
 \caption{ Example 6: A quarter of a symmetric 2D plate-with-a-hole under random elastic modulus. (a) shows the finite element mesh. (b) showcases a random field realization of the elastic modulus.}
 \label{fig: Plate}
\end{figure}

In this example, we consider a low-carbon steel plate, as shown in \Cref{fig: Plate} \citep{papaioannou2019pls, ko2024quadratic}. The plate has dimensions of $2,000$ mm × $2,000$ mm, with a circular hole of radius 200 mm located at the center. Due to symmetry, only a quarter of the plate is analyzed. The Poisson ratio is assumed to be $\nu =0.3$ and the plate thickness is $t=10$ mm. A deterministic uniform load of $q=100 \,\text{N/mm}^2$ is applied on the right edge of the plate. The elastic modulus $E(x, y)$ of the plate is modeled as a two-dimensional homogeneous normal random field with a mean value of $\mu_{E}=2 \times 10^5 \,\text{N/mm}^2$ and a standard deviation of $\sigma_{E}=2 \times 10^4 \,\text{N/mm}^2$. The autocorrelation function of the random field $E(x, y)$ is given as:
\begin{equation}
\begin{aligned}
 \rho_{E}(\Delta x, \Delta y)= \exp \left[-\left(\dfrac{\Delta x}{l_x}\right)^2 - \left(\dfrac{\Delta y}{l_y}\right)^2 \right]
\end{aligned}
\end{equation}
where $l_x$ and $l_y$ are correlation lengths, both set to 300 mm in this example. The stochastic field is described using the Karhunen-Loève (KL) expansion with 100 terms, implying 100 standard Gaussian random variables utilized within the KL expansion ($d = 100$). 

The stress, strain, and displacement fields of the plate are governed by the Navier-Cauchy equilibrium equation \citep{johnson2012numerical}, simplified under the plane stress assumption as:
\begin{equation}
 G(x, y)\nabla^2 \bm{u}(x, y) + \dfrac{E(x, y)}{2(1 - \nu)} \nabla(\nabla \cdot \bm{u}(x,y)) + \mathbf{f} = \mathbf{0}
\end{equation}
where $G(x, y)$ is the shear modulus, defined as $E(x, y)/[2(1 + \nu)]$, $\bm{u}$ is the displacement vector, and $\mathbf{f}$ is the body force acting on the plate. This equation is solved using Finite Element Analysis (FEA) with a mesh of 368 four-node quadrilateral elements, as shown in \Cref{fig: Plate}. The limit-state function is defined as:
\begin{equation}
 g(\bm{X})=410 - \max(\sigma_1)
\label{eq: plate_lim}
\end{equation}
where $\max(\sigma_1)$ is the maximum first principal stress, with failure occurring when this stress exceeds the threshold of 410 MPa.\par

For the guided pCN-ASTPA implementation, 5 chains ($N_{\text{chains}} = 5$) are used, each with a chain length of $1,000$ iterations ($L_{\text{chain}} = 1,000$). The IIS step is performed with $2,000$ resampled points ($M= 2,000$). The likelihood dispersion factor is set to $\sigma = 0.2$, and the scaling constant $g_c = \nicefrac{g(\mathbf{0})}{4}$ is applied. The discovery stage utilizes $N_{\text{level}} = 100$ and $p_0 = 0.2$. The results in \Cref{Table: Plate} demonstrate that the guided pCN-ASTPA and iCE-IS methods provide comparable performance in this example, with pCN-ASTPA being slightly better, and with both significantly outperforming SIS and aCS-SuS.

\begin{table}[t!]
\captionsetup{justification=centering}
\caption{Example 6: Performance of various methods for the plate example.}
\centering
\footnotesize
\setlength\tabcolsep{4pt}
\begin{tabular}{p{1.0cm}cccccccccc}
  \toprule[1.5pt]
  \multirow{2}{*}{\Cref{eq: plate_lim}}& \multirow{2}{*}{\textbf{500 Simulations}} & \multirow{2}{*}{\textbf{MCS}}& \textbf{pCN-ASTPA}& \multirow{2}{*}{\textbf{iCE-IS}}& \multirow{2}{*}{\textbf{SIS}}&\multirow{2}{*}{\textbf{aCS-SuS}}\\ 
  \cline{4-4}
    \addlinespace[2pt]
  & && \multicolumn{1}{c}{($\sigma = 0.2, \, q =4$)}  &  &\\ 
 \cmidrule(lr){1-7}
         \multirow{3}{*}{\shortstack[l]{$d=100$}}\rule{0pt}{2.5ex}    &$\mathop{\mathbb{E}}[N_{\text{total}}]$ & 1.00E7 &  7,247& 7,611 & 11,605  &12,996\\
      &C.o.V &0.29& 0.16$\color{ForestGreen}($0.17$\color{ForestGreen})$&  0.21 & 1.22 & 0.31 \\
       &$\mathop{\mathbb{E}}[\hat{p}_{\mathcal{F}}]$    & 1.04E-6 & 1.01E-6 &1.02E-6& 0.96E-6& 1.01E-6\\
      \bottomrule[1.5pt]

\end{tabular}\label{Table: Plate}
\end{table}

\subsection{Example 7: Bimodal high-dimensional decic problem}
To further investigate the performance of the proposed method in high-dimensional spaces with very significant nonlinearities, we consider the following challenging decic limit-state function in the standard normal space: 
\begin{equation}
    g(\bm{X}) = \min\begin{cases} 
    2.80 - \frac{1}{\sqrt{d}}\ \sum_{i=1}^{200} X_{i}  +f(\bm{X})\\[12pt]
    2.80 + \frac{1}{\sqrt{d}}\ \sum_{i=1}^{200} X_{i} +f(\bm{X})
\end{cases}, \text{where}\, f(\bm{X})= \big (\sum_{j=1}^{\gamma} X_{j} \big )^{2}+\exp\big[\big (\sum_{k=1}^{\gamma} X_{k} \big)^{7}\big] +\big (\sum_{l=1}^{\gamma} X_{l} \big)^{10}
\label{eq: decic_lim}
\end{equation}
where $\gamma$ controls the number of nonlinear terms, and hence the problem's complexity. To construct the sampling target in ASTPA, the likelihood dispersion factor is set to $\sigma = 0.5$, and the scaling constant to $g_c = 1$. The discovery stage of the proposed framework employs $N_{\text{level}} = 500$ and $p_0 = 0.2$. Following this stage, $20$ to $22$ pCN chains are used, each with a chain length ranging from $1,000$ to $1,200$ iterations. Higher values within this range are utilized in scenarios with greater nonlinearities to account for the increased sampling complexity. The inverse importance sampling (IIS) step employs $4,000$ to $7,000$ samples, with larger numbers chosen for more complex scenarios. As shown in \Cref{Table: decic}, the guided pCN-ASTPA framework demonstrates remarkable adaptability to increasing nonlinearity, achieving the lowest coefficient of variation (C.o.V.) while requiring fewer model evaluations compared to other methods. In contrast, the iCE-IS method fails to converge with a comparable number of model calls in this example. The ability of the guided pCN-ASTPA to efficiently manage both high dimensionality and strong nonlinearity highlights its robustness and computational efficiency, making it particularly well-suited for complex reliability estimation problems involving intricate variable interactions.

\begin{table}[t!]
\captionsetup{justification=centering}
\caption{Example 7: Performance of various methods for the bimodal high-dimensional decic problem.}
\centering
\footnotesize
\setlength\tabcolsep{4pt}
\begin{tabular}{p{1.0cm}cccccccc}
  \toprule[1.5pt]
  \multirow{2}{*}{\Cref{eq: decic_lim}}& \multirow{2}{*}{\textbf{500 Simulations}} & \multirow{2}{*}{\textbf{MCS}}& \textbf{pCN-ASTPA}& \multirow{2}{*}{\textbf{SIS}}& \multirow{2}{*}{\textbf{aCS-SuS}}\\ 
  \cline{4-4}
    \addlinespace[2pt]
  & && \multicolumn{1}{c}{($\sigma = 0.6, \, q = \,$n.a.)}  &  &\\ 
 \cmidrule(lr){1-6}
         \multirow{3}{*}{\shortstack[l]{$d=200$\\$\gamma=10$}}\rule{0pt}{2.5ex}    &$\mathop{\mathbb{E}}[N_{\text{total}}]$ &1.00E7 &25,434& 38,488& 28,526\\
      &C.o.V &0.08& 0.26$\color{ForestGreen}($0.23$\color{ForestGreen})$  &3.39 & 0.76 \\
       &$\mathop{\mathbb{E}}[\hat{p}_{\mathcal{F}}]$  &  1.02E-5&1.01E-5 & 1.20E-5 & 1.06E-5\\
       
             \cmidrule(lr){1-6}
         \multirow{3}{*}{\shortstack[l]{$d=200$\\$\gamma=15$}}\rule{0pt}{2.5ex}    &$\mathop{\mathbb{E}}[N_{\text{total}}]$ &1.00E7 &26,568& 41,921&29,946\\
      &C.o.V &0.12 & 0.33$\color{ForestGreen}($0.30$\color{ForestGreen})$&1.66& 1.26\\
       &$\mathop{\mathbb{E}}[\hat{p}_{\mathcal{F}}]$  &  6.66E-6&6.58E-6& 6.42E-6 & 6.83E-6   \\

                          \cmidrule(lr){1-6}
         \multirow{3}{*}{\shortstack[l]{$d=200$\\$\gamma=20$}}\rule{0pt}{2.5ex}    &$\mathop{\mathbb{E}}[N_{\text{total}}]$ &1.00E7 &29,630 & 43,886& 31,007\\
      &C.o.V &0.16&0.34$\color{ForestGreen}($0.31$\color{ForestGreen})$ &1.86 & 1.56 \\
       &$\mathop{\mathbb{E}}[\hat{p}_{\mathcal{F}}]$  &  4.51E-6&4.46E-6 & 4.41E-6 & 4.58E-6  \\
       
           \cmidrule(lr){1-6}
    \multirow{3}{*}{\shortstack[l]{$d=200$\\$\gamma=25$}}\rule{0pt}{2.5ex}  &$\mathop{\mathbb{E}}[N_{\text{total}}]$ &1.00E7&35,072& 45,830& 38,520\\
       & C.o.V & 0.18&0.33$\color{ForestGreen}($0.33$\color{ForestGreen})$  & 2.89 & 2.07\\
       &$\mathop{\mathbb{E}}[\hat{p}_{\mathcal{F}}]$   & 3.12E-6 &3.15E-6& 3.24E-6&3.42E-6 \\
      \bottomrule[1.5pt]

\end{tabular}\label{Table: decic}
\end{table}
\section{Conclusions}\label{sec: Conclusion}

In this work, we introduce a novel gradient-free framework for reliability and rare event probabilities estimation in complex, high-dimensional Gaussian spaces, particularly suited for black-box computational models where analytical gradients are not available. At the core of this framework is the foundational Approximate Sampling Target with Post-processing Adjustment (ASTPA) approach, which efficiently estimates rare event (failure) probabilities by constructing and sampling from an unnormalized target distribution, relaxing the optimal importance sampling density (ISD) \citep{Papakon2023HMCMC}. This relaxation enables efficient sampling, especially in high-dimensional and highly nonlinear problem settings, with the generated samples employed to compute a shifted estimate of the sought probability. The inverse importance sampling (IIS) procedure subsequently refines the probability estimation by employing a fitted ISD using the already obtained samples to compute the target's normalizing constant, avoiding scalability challenges.

While previous work has shown that integrating gradient-based samplers with ASTPA yields outstanding performance in both Gaussian and non-Gaussian spaces \citep{eshra2024direct, Papakon2023HMCMC}, the absence of analytical gradients hinders the efficient applicability of such approaches. The proposed gradient-free framework addresses this limitation by advancing the preconditioned Crank-Nicolson (pCN) algorithm, incorporating a rare event domain discovery stage to significantly enhance the sampling process. The scale parameter of the pCN algorithm is adapted using the Robbins–Monro stochastic approximation algorithm to achieve a $20$--$40\%$ average acceptance probability, which was proven effective in our numerical examples. Despite the dimension-robustness of pCN, its effectiveness depends on carefully initializing its chains in relation to the regions of interest, particularly in high-dimensional multimodal distributions. To address this, we developed a computationally efficient rare event domain discovery technique, conditionally sampling well-designed synthetic variants of the original distribution, thereby facilitating rapid diffusion toward relevant regions in the random variable space. Our investigation into selecting rare event (pCN) seeds (from the obtained rare event sample set) shows that weighted sampling is effective in low-dimensional cases, whereas uniform sampling is more suitable for initializing the pCN chains in high-dimensional, multimodal cases, where the target distributions become sensitive to even small perturbations. This approach significantly enhances the performance of the pCN algorithm in sampling the approximate target in ASTPA.

The guided pCN-based ASTPA methodology has been tested across a range of challenging Gaussian problems, including mathematically formulated test functions and engineering examples, such as a 34-story frame structure and a steel plate, with rare event probabilities as low as $10^{-8}$. These mathematical limit-state functions are carefully designed to involve multimodality and up to decic nonlinear terms, representing challenging benchmarks. Our method consistently outperformed state-of-the-art gradient-free approaches, including improved Cross-Entropy Importance Sampling (iCE-IS), Sequential Importance Sampling (SIS), and adaptive Conditional Sampling Subset Simulation (aCS-SuS), both in computational cost and accuracy. Moreover, the ASTPA analytical coefficient of variation (C.o.V) demonstrated close agreement with empirical results, further validating the accuracy of this theoretical estimate.

This guided pCN-based ASTPA framework sets a new performance benchmark for gradient-free rare event probability estimation, offering a scalable and computationally efficient solution for high-dimensional, nonlinear, and multimodal problems.  Future work will explore further extensions of the framework, including the development of gradient-free variants suited for direct applicability in non-Gaussian spaces and estimating high-dimensional first-passage problems under various settings \citep{PapakonICASP2023}.

\bibliographystyle{ieeetr}
\bibliography{references}  
\end{document}